\documentclass[elsarticle-num-names]{elsarticle}

\usepackage{diagbox}

\usepackage{graphicx}%
\usepackage{multirow}%
\usepackage{amsmath,amssymb,amsfonts}%
\usepackage{amsthm}%
\usepackage{mathrsfs}%
\usepackage[title]{appendix}%
\usepackage{xcolor}%
\usepackage{textcomp}%
\usepackage{manyfoot}%
\usepackage{booktabs}%
\usepackage{algorithm}%
\usepackage{algorithmicx}%
\usepackage{algpseudocode}%
\usepackage{listings}%

\usepackage{ dsfont }
\usepackage{enumitem}            
\usepackage{faktor}

\usepackage{academicons}
\usepackage{accents}
\definecolor{orcidlogocol}{HTML}{A6CE39}

\usepackage[normalem]{ulem}

\usepackage{aliascnt}   
\usepackage{marvosym} 
\newcommand{\dardag}{\text{\Ankh}}
\usepackage{tikz-cd}
\newtheorem{theorem}{Theorem}[section]
\newaliascnt{proposition}{theorem}
\newaliascnt{corollary}{theorem}
\newcommand{\utilde}[1]{\underaccent{\tilde}{#1}}
\newcommand{\ubreve}[1]{\underaccent{\breve}{#1}}

\aliascntresetthe{proposition}
\aliascntresetthe{corollary}

\usepackage{comment}
\usepackage{lscape}

\usepackage{hyperref}
\usepackage{float}

\begin{document} 

\title{Geometric flavours of Quantum Field theory on a Cauchy hypersurface. Part II: Methods of quantization and evolution}


\author[1,2,3]{Jos\'e Luis Alonso}

\author[1,2,3]{Carlos Bouthelier-Madre}
\author[1,2,3]{Jes\'us Clemente-Gallardo}

\author[1,3]{David Mart\'inez-Crespo} \ead{dmcrespo@unizar.es}

\affiliation[1]{organization={Department of Theoretical Physics, {University of Zaragoza}},
            addressline={{Facultad de Ciencias, Campus San Francisco}}, 
            city={{Zaragoza}},
            postcode={{50009}}, 
            country={{Spain}}}
\affiliation[2]{organization={ {Institute for Biocomputation and Physics of Complex Systems BIFI}},
            addressline={{Edificio I+D-Campus Río Ebro}, {University of Zaragoza}}, 
            city={{Zaragoza}},
            postcode={{50018}}, 
            country={{Spain}}}

\affiliation[3]{organization={{Center for Astroparticles and High Energy Physics CAPA}},
            addressline={{Facultad de Ciencias, Campus San Francisco, {University of Zaragoza}}}, 
            city={{Zaragoza}},
            postcode={{50009}}, 
            country={{Spain}}}


\begin{abstract}
    In this series of papers we aim to provide a mathematically comprehensive framework to the Hamiltonian pictures of quantum field theory in curved spacetimes. Our final goal is to study the kinematics and the dynamics of the theory from the point of differential geometry in infinite dimensions. In this second part we use the  tools of Gaussian analysis in infinite dimensional spaces introduced in the first part to describe rigorously the procedures of geometric quantization in the space of Cauchy data of a scalar theory. This leads us to discuss and establish relations between different pictures of QFT. We also apply these tools to describe the geometrization of the space of pure states of quantum field theory as a Khäler manifold. We use this to derive an evolution equation that preserves the geometric structure and avoids norm losses in the evolution. This leads us to a modification of the Schrödinger equation via a quantum connection that we discuss and exemplify in a simple case.
\end{abstract}



\maketitle
\tableofcontents

\section{Introduction}

In the first paper of this series \cite{alonsoGeometricFlavours2023} we presented a set of mathematical tools to develop Gaussian integration theory on infinite dimensional spaces. Those tools are important to characterize geometrically the Hilbert space of pure states of a particular quantum field theory. In this paper we aim to clarify the physical interpretation of the canonical quantum field theory of a scalar field described over a Cauchy hypersurface of any generic spacetime. These tools will be used and enlarged in \cite{alonsobujHybridGeometrodynamics2024} to describe a dynamical coupling of gravity and the quantum field theory of a scalar field.

 The fields of Gaussian and White noise analysis \cite{gelfandGeneralizedFunctions1964,hidaBrownianMotion1980,hidaWhiteNoise1993,kuoWhiteNoise1996,obataWhiteNoise1994,huAnalysisGaussian2016,kondratievGeneralizedFunctionals1996,sampedroSpaceInfinite2020} have been widely developed in relation to the modelling of financial models \cite{nunnoMalliavinCalculus2009}. Nonetheless,   their deep connection with the informally defined Feynman integrals and their applications to Quantum Field Theory (QFT) was noticed and studied by many authors from a rigorous mathematical point of view \cite{glimmQuantumPhysics1987,
 cartierFunctionalIntegration2006,
 cartierRigorousMathematical1997,westerkampRecentResults2003}. Following these works, in these article we will make systematic use of Gaussian analytic methods to interpret physical aspects  of QFT with a strong focus on geometry rather than on analytical or algebraic \cite{brunettiAdvancesAlgebraic2015} techniques.

However, not every physical aspect of the theory relies solely on the mathematics of Gaussian integration. In order to fully understand the physics we focus on the scalar field. In that case must choose a Kähler structure $(\omega,\mu,J)$ on the set of classical fields \cite{ashtekarQuantumFields1975,corichiSchrodingerFock2004,muchComplexStructures2021}.  The corresponding Hilbert space of pure states  can be then defined by a Gaussian measure associated to this structure. Thus, we also investigate  how the choice of Kähler structure determines very relevant features of the quantum theory.

In particular, in this work we try to answer two central questions. Firstly: what does it mean to quantize a classical field theory? And, secondly, what is the most efficient way of describing a QFT in geometric terms  to couple it to gravity  on equal footing?

To answer the first question we thoroughly develop a quantization program,  that we summarize as follows. In the first place, we investigate how to rigorously quantize canonical Cauchy data of a scalar field using the well known procedure of  geometric quantization  \cite{woodhouseGeometricQuantization1997,hallQuantumTheory2013,tuynmanMetaplecticCorrection2016,oecklSchrodingerRepresentation2012,oecklAffineHolomorphic2012,oecklHolomorphicQuantization2012}. In relation to it, we address the ordering problems \cite{ditoStarproductApproach1990} of the algebra of observables in this context of canonical quantization. Lastly, we provide
physical interpretation of each one of the mathematical tools introduced in \cite{alonsoGeometricFlavours2023}, reflecting different aspects of the structure of Gaussian integration through the lense of this quantization program. Other approaches, as the field of stochastic quantization, are not considered in this work. However the connection with stochastic calculus may provide a way to extend the formalism to fermions \cite{albeverioGrassmannianStochastic2020} or gauge fields \cite{masujimaPathIntegral2008}.  

Regarding the second question, we argue that one needs to detach the particular physical features of a theory from the general geometric structures underlying it. In order to do so, the concept of  {\it second quantized} test function, which was reviewed  in the first part of this series as the algebra of Hida test functions \cite{hidaBrownianMotion1980,hidaWhiteNoise1993,nunnoMalliavinCalculus2009}, proves to be particularly useful. These functions, whose domain is a space of distributions, will be used to model a common dense subset of any Hilbert space  that we may consider as the space of pure quantum states of any particular theory. The main advantage of this procedure is that we can describe a generic quantum theory modelling the manifold  $\mathscr{P}$ of pure states regardless of any particular choice of Kähler structure at the classical level and encode the particular features of each theory in a {\it second quantized} Kähler structure  $(\mathcal{G}, \Omega, \mathcal{J})_\mathscr{P}$.

In this work we take the gravitational degrees of freedom as a given background, whereas the case of a  spacetime that evolves subject to the \textit{backreaction} of QFT matter is studied in \cite{alonsobujHybridGeometrodynamics2024}.
We assume space-time to be globally hyperbolic and consider it as a foliation of spatial hypersurfaces, labelled by a time-like parameter $t$. On each hypersurface, the background metric induces a 3-metric $h_{ij}$ and the corresponding field-momenta $\pi_{ij}$, which encode the extrinsic curvature of each leaf. The transformations  defining the evolution from one leaf to the infinitesimally following one in the foliation are encoded in the lapse and shift vector fields $(N, N^i)$. 
The spacetime foliation is thus represented by the $t$-parametric family of geometrical objects $(h_{ij}(t), \pi_{ij}(t), N(t), N^i(t))$.

On such a foliation, a Hamiltonian description of a classical scalar field theory of matter is considered, with its fields and field-momenta having domain on the spatial hypersurfaces. These fields are then quantized on each hypersurface, the quantization depending on a suitably chosen complex structure, which is in turn dependent on the geometry of the leaf. Thus, quantization becomes $t$-dependent, as the dependence on the geometrical objects of the leaf is composed with their intrinsic time dependence, given the non-constant nature of $(h_{ij}(t), \pi_{ij}(t), N(t), N^i(t))$ due to the dynamical nature of generic background spacetimes. This additional  $t-$dependence is inherited by quantum states and operators, and also by the K\"ahler structure at the second quantized level.

Our final goal is to build a description of the Hamiltonian dynamics of QFT as a Hamiltonian system with respect to a canonical Poisson structure, following the well known construction of Kibble  (and, later, its generalization by Ashtekar and Schilling) for geometrical quantum mechanics \cite{kibbleGeometrizationQuantum1979,ashtekarGeometricalFormulation1999}. The geometric quantization procedure based on the geometric structures defined on the set of classical fields, as well as their $t$-dependence, will determine crucial aspects of the resulting dynamics.  In particular, the {\it second quantized} structure will provide an evolution equation that modifies the Schrödinger equation and preserves norm in the evolution  in a similar way suggested in other works \cite{agulloUnitarityUltraviolet2015,kozhikkalBogoliubovTransformation2023}. This preservation of norm is not    obtained as a property  in  the usual prescription as it is explored in \cite{hofmannClassicalQuantum2015,hofmannNonGaussianGroundstate2017,hofmannQuantumComplete2019,eglseerQuantumPopulations2021}.  This kind of geometric structure is key to build a mathematically consistent theory of classical gravity coupled to quantum field, as we showed in \cite{alonsobujHybridGeometrodynamics2024}, where the space-time is no longer treated as a background, but instead it suffers the \textit{backreaction} of the quantum fields.

  The structure of the paper is the following:
On the remainder of this section we recall some mathematical features and conventions introduced in the first part of this series \cite{alonsoGeometricFlavours2023}.
  Then, Section \ref{sec:QFTCauchy} introduces the first step towards geometric quantization, prequatization theory. We introduce the prequantum Hilbert space and the prequantization of linear observables that will serve as basis for geometric quantization.  Continuing  with this program, Section \ref{sec:holom} introduces the set of quantum states for different relevant cases and presents a summary of the geometric quantization of linear observables (on fields).   In particular, we consider  holomorphic, antiholomorphic, Schrödinger and momentum-space quantizations; and we also provide isomorphisms and quantization preserving mappings relating them. Then, in Section \ref{sec:QuantizHolom} we study  the quantization of arbitrary observables in these cases and, again, we provide a way to relate them.  This is followed by Section \ref{sec:GQFT} presenting the construction of geometric quantum field theory analogously to Kibble's geometric quantum mechanics. We identify the canonical Kähler structure of the set of quantum states built previously and study its relation with the one defined on the set of classical fields
proper to geometric quantization.  Then, in Section \ref{sec:timeevolution} we use the geometric formalism to describe the solutions of Schrödinger equation as the integral curves of a Hamiltonian vector field associated with the canonical symplectic structure. One key novelty presented in this section is the fact that the geometric construction requires the introduction of a connection to take into account the dependence of the quantization procedure  on the geometrical objects of the hypersurface, as in \cite{alonsobujHybridGeometrodynamics2024}. In this case, these structures are in turn $t$-dependent, which turns the whole construction time-dependent and is reflected in the time-evolution equation. Finally, Section \ref{sec:KleinGordonExample} presents an application of all the ingredients introduced so far to the case of the dynamics of a free quantum scalar field on a flat Friedman-Lemaitre-Robertson-Walker (FLRW) spacetime. 
We finish this work with Section \ref{sec:conclusions} presenting the main conclusions of the two papers and describing its potential applications.

\paragraph{Mathematical summary}
\

Based on infinite-dimensional integration theory, in \cite{alonsoGeometricFlavours2023} we argued that the natural model space for quantum fields is a space of distributions regarded as the strong dual (DNF) of a Nuclear-Fréchet (NF) space. Regarding spacetime structure, we remind our restriction to globally hyperbolic spacetimes such that admit a compact space-like Cauchy hypersurfaces $\Sigma$. In that case, the usual model space for classical fields used in the literature is the space of smooth compactly supported functions $\mathcal{N}=C^{\infty}_c(\Sigma)$. Consequently, in \cite{alonsoGeometricFlavours2023} we made the choice  of modelling quantum fields over its strong dual $\mathcal{N}'=D'(\Sigma)$. The ultimate reason for that choice was to develop a rigorous quantization program, as explained in the forthcoming sections. Throughout this work $\mathcal{N}$ will be assumed to be real while $\mathcal{N}_\mathbb{C}$ is its complex counterpart. It is important to notice that, even though we will limit our study to compact $\Sigma$ manifolds, the only ingredient needed to generalize our analysis is the NF topology of $\mathcal{N}$, which amounts to proper boundary conditions. For this reason, we may use Minkowski spacetime as an example of our analysis with the space of tempered functions $\mathcal{S}(\mathbb{R}^3)$ and distributions (its dual) as model spaces. To emphasize the generalizability of the framework, we will use the notation  $\mathcal{N}$ and  $\mathcal{N}'$ for the NF space and its dual instead of any particular functional space, such as $C^{\infty}_c(\Sigma)$.  Let us also recall  from \cite{alonsoGeometricFlavours2023} the notation that we will employ in this work

\begin{align}
    \label{eq:dualityNotationDistributions}
    \xi_{\mathbf{x}}\in\mathcal{N}, \varphi^{\mathbf{x}}\in\mathcal{N}' \textrm{ paired by }& \xi_x\varphi^x :=\langle\xi_{\mathbf{x}}, \varphi^{\mathbf{x}}\rangle \in \mathbb{R}\nonumber\\
    \rho_{\mathbf{x}}\in\mathcal{N}_{\mathbb{C}}, \phi^{\mathbf{x}}\in\mathcal{N}^{\prime}_{\mathbb{C}}\textrm{ paired by }& \rho_x\phi^x :=\langle \rho_{\mathbf{x}},\phi^{\mathbf{x}}\rangle\in \mathbb{C}
\end{align}
Notice that we do not conjugate any of the variables in the complex pairing. Instead we choose a dual coordinate convention for holomorphic coordinates of functions and distributions
\begin{equation}
    \label{eq:coordinateConventions}
\phi^{\mathbf{x}}=\frac{1}{\sqrt{2}}(\varphi^{\mathbf{x}}-i\pi^{\mathbf{x}}),\hspace{2 em}\rho_\mathbf{x}=\frac{1}{\sqrt{2}}(\xi_\mathbf{x}+i\eta_\mathbf{x}).
\end{equation}
for  $(\varphi^{\mathbf{x}},\pi^\mathbf{x})\in \mathcal{N}'\times \mathcal{N}'$ and $(\xi_{\mathbf{x}},\eta_{\mathbf{x}})\in \mathcal{N}\times\mathcal{N}$. To fix notation we indicate that, throughout  this work, the strong dual will be denoted  by $'$ while $*$ indicates only  complex conjugation. 

Operators and bilinears are denoted with abstract indices as $A^{\mathbf{x}}_{\mathbf y}, A^{\mathbf{xy}}$ or $A_{\mathbf{xy}}$ and their distinction gives rise to three notions of Dirac delta. Thus,  $\delta^{\mathbf x}_{\mathbf y}$ represents point evaluation while $\delta^{\mathbf{xy}}$ and $\delta_{\mathbf{xy}}$ is the dual pairing for test functions and distributions respectively. See \cite{alonsoGeometricFlavours2023} for further information. We define operator composition and raise and lower indices following the conventions 
\begin{equation}
    \label{eq:raiseandlowerconvention}
    (AB)^{\mathbf{x}}_{\mathbf{y}}=A^{\mathbf{x}}_zB^z_{\mathbf{y}},\ (A^t)^{{\mathbf{x}}}_{{\mathbf{y}}}= \delta_{{\mathbf{y}}u} A^u_v\delta^{v{{\mathbf{x}}}},\ A_{{\mathbf{x}}{\mathbf{y}}}=\delta_{{\mathbf{x}}z}A^z_{\mathbf{y}}\textrm{ and }A^{{\mathbf{x}}{\mathbf{y}}}=A^{\mathbf{x}}_z\delta^{z{{{\mathbf{y}}}}}
\end{equation}

Using these conventions we have that $\delta^{\mathbf{x}z}\delta_z{\mathbf y}=\delta^{\mathbf x}_{\mathbf y}$.
A particular tool of huge importance in our work is  the concept of  Hida test function. The domain of those functions is the space of distributions $\mathcal{N}'$, which possesses a NF topology that relies solely on the structure of $\mathcal{N}$. For this reason, those functions are denoted by a parenthesized expression $(\mathcal{N})$. We will use Hida test functions to model a subset of pure states that is common to and dense in any particular Hilbert space representing a particular theory. The main advantage of this procedure is that we can describe a generic quantum theory modelling the manifold  $\mathscr{P}$ of pure states over $(\mathcal{N})$ regardless of any particular choice of Kähler structure at the classical level. Moreover, $(\mathcal{N})$ is a NF space and therefore a convenient model space for geometry  \cite{krieglConvenientSetting1997,dodsonGeometryFrechet2016}.  In turn, as we explained above, this allows us to describe a particular theory with its {\it second quantized} Kähler structure  $(\mathcal{G}, \Omega, \mathcal{J})_\mathscr{P}$.  As we will see, the ability to characterize the space of states independently of the choice of classical complex structure is of crucial importance in curved spacetimes, as $(\mathcal{N})$ will be  a common subset to all the Hilbert spaces arising from the parametric family of complex structures similarly to the construction  considered in \cite{agulloUnitarityUltraviolet2015}.

\section{ Geometric quantization I: the general framework}
\label{sec:QFTCauchy}

The goal of this section is to build the quantum field theory corresponding to the classical model presented in \cite{alonsoGeometricFlavours2023}. As our goal is a geometric description of the resulting quantum model, we will consider a geometric quantization framework adapted to our infinite dimensional manifold. Our presentation in this section is similar to others in the literature.  Examples of this construction for field theories can be found in Oeckl's work \cite{oecklHolomorphicQuantization2012, oecklAffineHolomorphic2012,oecklSchrodingerRepresentation2012}, while, for finite dimensional systems, \cite{woodhouseGeometricQuantization1997,hallQuantumTheory2013} contain detailed discussions of the aspects discussed below.  Our aim is to approach the discussion with the mathematical tools of \cite{alonsoGeometricFlavours2023}. The procedure is, roughly speaking, similar to the usual construction on finite dimensional symplectic manifolds (see \cite{woodhouseGeometricQuantization1997,hallQuantumTheory2013,tuynmanMetaplecticCorrection2016}) and  it is based in two steps.

\begin{itemize}
	\item Firstly, a pre-quantization is defined, which builds an initial Hilbert space. This is constructed from  the set of sections of a complex line bundle defined on the classical phase space. Also, the prequantization procedure provides a representation of classical functions as operators acting on that prequantum Hilbert space. 
	\item Secondly, the definition of a polarization completes the quantization procedure. This is the  choice of a Lagrangian submanifold on the phase space of fields. The quantum Hilbert space is obtained by restriction of prequantum sections to this submanifold. Quantum operators are also restricted to those preserving the polarization in a suitable way.
\end{itemize}

The first step is straightforward and relatively simple to define for a general symplectic classical manifold. The definition of the polarization is a more subtle task and introduces arbitrary choices which restrict the set of classical magnitudes that can be quantized.  We will firstly present the general construction adapted to our manifold of fields $\mathcal{M}_F$. For simplicity we will assume that this manifold is linear and modelled over the DNF space $\mathcal{N}'$. Then, we will exploit a Kähler structure defined on its complexification $\mathcal{M}_C$ to consider one particular type of polarization, the holomorphic polarization, which exhibits several nice properties and largely simplifies the technicalities of the construction. This structure was introduced in \cite{alonsoGeometricFlavours2023}. Finally, we will compare the resulting construction with the usual approach based on a real polarization.

\subsection{Prequantization I: The definition of the bundle and the measure}

The first step towards quantization is the complex prequantum line bundle $B$ over the classical field phase space. This is an Hermitian line bundle $\pi_{\mathcal{M}_C,B}:B\rightarrow \mathcal{M}_C$, associated with a $U(1)$--principal bundle on $\mathcal{M}_C$. On this bundle we define a principal connection which is required to have, as local curvature form, the symplectic form of $\mathcal{M}_C$. This can be done if we consider the local connection one form to be defined by the corresponding symplectic potential $\theta$ (as our field manifold is a linear space, this choice is well defined globally). By using this connection, we can define a covariant derivative on the associated bundle $B$.

The next step is to find a prequantized Hilbert space $\mathscr{H}_P$. To define this space we consider the set of square integrable sections $B$ with respect to a measure $\mu_c$ invariant under symplectic transformations. Let us consider for simplicity the case where $\mathcal{M}_C$ can be covered with a single chart in $\mathcal{N}'_\mathbb{C}$, the complexified space of distributions. In this case we will consider the measure $\mu_c$ whose characteristic functional  $C$ obtained as  
\begin{equation}
\label{eq:gaussianmeasure}
C(\rho_{\mathbf{x}}, \bar \rho_{\mathbf{x}})=\int_{\mathcal{N}^\prime_\mathbb{C}}D\mu_c(\phi^\mathbf{x} )e^{i (\overline{\rho_x\phi^x}+\rho_x \phi^x) },
\end{equation}
satisfies:
\begin{itemize}
    \item $C(0,0)=1$
    \item $C$ is positive definite,
    \item it is continuous in the Fréchet topology of $\mathcal{N}_{\mathbb{C}}$.
\end{itemize}
Hence,  Bochner-Minlos theorem \cite{gelfandGeneralizedFunctions1964,hidaBrownianMotion1980,hidaWhiteNoise1993} assigns a unique Borel probability measure $\mu_c$ on $\mathcal{M}_C$ to each choice of functional $C$ satisfying the  conditions above.  Let us see how there are natural choices of functionals based on physical arguments.

Along this and the next section, it is convenient to distinguish between $\phi^\mathbf{x}\in \mathcal{N}^\prime_\mathbb{C}$ and the holomorphic coordinate of $\mathcal{M}_C$ that we denote $\utilde\phi^\mathbf{x}\in \mathcal{M}_C$. Recall from \cite{alonsoGeometricFlavours2023} that, in holomorphic coordinates, the Kähler structure $(\omega,\mu,J)_{\mathcal{M}_C}$ is derived from the Hermitian form

\begin{equation}
    \label{eq:kahlerstr}
    h_{\mathcal{M}_C}= \frac{\mu_{\mathcal{M}_C}-i\omega_{\mathcal{M}_C}}{2}= K_{xy}d\utilde\phi^x\otimes d\bar{\utilde\phi}^y,
    \end{equation}
    with trivial complex structure and  inverse  $h^{-1}_{\mathcal{M}_C}=\Delta^{xy}\partial_{\utilde{\bar\phi}^x}\otimes\partial_{\utilde{\phi}^y}$. From a physical point of view we should ask for the measure to be invariant under symplectic transformations.  Thus we select $C$ to correspond to a function of the Hermitian form   $h^{-1}_{\mathcal{M}_C}$, acting on  $\rho_x d\utilde\phi^x$ as covectors. This provides the measure whose functional $C$ is equal to

\begin{equation}
\label{eq:C}
C(\rho_\mathbf{x}, \bar \rho_\mathbf{x})=e^{-h^{-1}_{\mathcal{M}_c}(\bar \rho_\mathbf{x}, \rho_\mathbf{x})}= e^{-\bar \rho_{x}\Delta^{xy} \rho_{y}}.
\end{equation}

This  defines a gaussian  measure $\mu_c$ that can be fully understood from the Gel'fand triple provided by its Cameron-Martin Hilbert space 

\begin{equation}
    \label{eq:holomorphicTriple}
    \mathcal{N}_{\mathbb{C}}\subset \mathcal{H}_\Delta\cong \mathcal{H}_{K} \subset {\mathcal{N}_{\mathbb{C}}'}
\end{equation}
where $\mathcal{H}_\Delta=\overline{(\mathcal{N}_{\mathbb{C}},\Delta^{\mathbf{xy}})}$. In this triple we identify $\mathcal{H}_\Delta$ with its dual $\mathcal{H}_{K}$. The latter is obtained restricting ${\mathcal{N}_{\mathbb{C}}'}$ to the subset in which the norm computed with \eqref{eq:kahlerstr} converges. This space represents the allowed directions of translation that leave invariant the spaces $L^p(\mathcal{N}'_{\mathbb{C}},D\mu_c)$\footnote{See Proposition 6.9 of \cite{huAnalysisGaussian2016}.} and has the physical interpretation of a \textit{one particle state structure}. It is also the main ingredient of the Wiener-Ito decomposition theorem that provides the particle interpretation of a QFT written as a gaussian integral. For further details see \cite{alonsoGeometricFlavours2023}. See also \cite{oecklHolomorphicQuantization2012,cartierRigorousMathematical1997,sampedroSpaceInfinite2020} for a different construction of the measure using projective limits.

With this choice, the measure exists and then the scalar product of 
$\Psi^s,\Phi^s\in \Gamma(\mathcal{M}_C,B)$, sections of the prequantum bundle, can be written as:
\begin{equation}
\label{eq:scalarproductprequantum}
\langle\Phi^s,\Psi^s \rangle=\int_{\mathcal{M}_C}D\mu_c\langle \Phi^s,\Psi^s\rangle. 
\end{equation}

\subsection{Prequantization II. The states: introducing the vacuum}

\subsubsection{A trivializing section}
In this way we have introduced a measure on the set of sections of the prequantum bundle and we can define our pure quantum states to be the square-integrable sections with respect to the measure $\mu_c$, or equivalently, those sections with finite norm. Now we want to trivialize the bundle and factorize those sections with respect to a preferred one $\Psi_r$. This section will be our reference and will be related with a phase factor with the physical vacuum state of the theory $\Psi_0$ as we will discuss in the next section. This preference will be stated in terms of a particularly simple expression for its covariant derivative given by \eqref{eq:vacuum}. We will ask this section not only to have norm 1 for this measure, but to have its local-on-$\utilde\phi^\mathbf{x}$ complex Hermitian product constant, i.e.,
\begin{equation}
\label{eq:vacuum_norm}
\langle \Psi_0( \bar{\utilde\phi}^\mathbf{x}, \utilde\phi^\mathbf{x}) \Psi_0( \bar{\utilde\phi}^\mathbf{x}, \utilde\phi^\mathbf{x})  \rangle=1, \qquad \forall ( \bar{\utilde\phi}^\mathbf{x}, \utilde\phi^\mathbf{x}) \in \mathcal{M}_C.
\end{equation}

All other sections can then be represented as simple functions, which will be square integrable with respect to a measure defined by using $\Psi_r$. As sections are also functions on $\mathcal{M}_C\cong \mathcal{N}_{\mathbb{C}}^\prime$, we can write our pre-quantum Hilbert space as
\begin{equation}
\label{eq:preHilbert}
\mathscr{H}_P=\left \{ \Phi: \mathcal{M}_C, \to \mathbb{C} \text{ s. t. } \Phi\in L^2(\mathcal{N}_{\mathbb{C}}^\prime,D\mu_c) \text{ as function } \right \}
\end{equation}

The local expression of the reference section can be further determined by the covariant derivative of the bundle. We can define the reference section $\Psi_r$ adapted to the symplectic potential (hence, to the complex structure \cite{alonsoGeometricFlavours2023}) requiring that
\begin{equation}
\label{eq:vacuum}
\nabla_{X}\Psi_r=-i\theta(X)\Psi_r.
\end{equation}
Notice that while condition \eqref{eq:vacuum_norm} fixes the modulus of the reference state, this condition fixes the phase.  We shall get back to this point below, when considering the two types of polarization.  In any case, we can already remark that clearly the local expression of the state (or equivalently, the connection) depends on the choice of the complex structure on $\mathcal{M}_C$. 
Notice also that  $\mu$ includes in its definition the expression of the modulus of the reference $\Psi_r$ which usually appears in the literature as the vacuum of the theory \cite{hofmannClassicalQuantum2015,hofmannNonGaussianGroundstate2017,hofmannQuantumComplete2019,eglseerQuantumPopulations2021,longSchrodingerWave1998,longSchrodingerWave1996,corichiSchrodingerFock2004}. In those works the reference is referred to the informally defined Lebesgue measure $D\phi$. In this way,  $\Psi_0'$ is an informally defined function representing the vacuum and its modulus is such that $D\phi\vert \Psi_0'\vert^2\simeq D\mu_c$ in a heuristic way.    This is the reason for the necessity of  Equation \eqref{eq:vacuum_norm}, which only leaves free the  phase of the reference state for a chosen measure.

Using this section $\Psi_r$ as a trivialization of the bundle $B$ seen as a principal bundle of fiber $\mathbb{C}\backslash\{0\}$, any other section  $\Phi^s$ is given by $\Phi^s=\Phi\Psi_r$. Here $\Phi:\mathcal{M}_C\to \mathbb{C}$ is identified with a regular smooth function of $(\mathcal{N}_{\mathbb{C}})\subset L^2(\mathcal{N}'_\mathbb{C},D\mu_c)$. Our notion of smooth function in this case is provided by the set of Hida test functions $C^\infty(\mathcal{M}_C)\cong (\mathcal{N}_{\mathbb{C}})\otimes (\mathcal{N}_{\mathbb{C}})^*$ defined in \cite{alonsoGeometricFlavours2023}. This notion also includes regularity  properties under integration that we studied in that paper.  Hence, this decomposition is  only valid for the subset of smooth sections in \eqref{eq:preHilbert}. However, this subset is dense  and, as such, is enough to describe the quantization procedure using it. The action of the covariant derivative is translated to the function $\Phi$ as:
\begin{equation}
\label{eq:covariantDerivative}
\nabla_X(\Phi\Psi_r)=\left[ X(\Phi)-i\theta(X)\Phi\right]\Psi_r.
\end{equation}

Our pre-quantum Hilbert space  is then identified with
\begin{equation}
\label{eq:wholePreq}
    \mathscr{H}_P\simeq L^2(\mathcal{N}'_\mathbb{C},D\mu_c),
\end{equation}
when referred to a reference section $\Psi_r$. See the discussion at the end of  \autoref{ssec:linearOp} for further details about the relation of  $\Psi_r$ with the vacuum of the theory $\Psi_0$. The effect of the reference on \eqref{eq:wholePreq} is that covariant derivatives incorporate an  extra multiplicative factor $-i\theta(X)$ provided by the symplectic potential to the directional derivative.   

Another key factor of this construction is that in the quantum theory every magnitude is obtained by integration of the classical degrees of freedom. This turns the classical manifold $\mathcal{M}_C$ into a rather auxiliary object. Let $\Psi$ be a pre-quantum state, and consider a translation by $\Psi(\phi^\mathbf{x})\mapsto \Psi(\phi^\mathbf{x}+h^\mathbf{x})$. Only when $h^\mathbf{x}\in \mathcal{H}_K\simeq \mathcal{H}_\Delta$, the Cameron-Martin Hilbert space, we can reabsorb this into the Gaussian measure \eqref{eq:gaussianmeasure} with an integrable Radon-Nikodym factor \cite{hidaWhiteNoise1993,nunnoMalliavinCalculus2009}. Hence, we can consider $\mathcal{H}_\Delta$ as the only well-defined tangent vectors on the linear manifold $\mathcal{M}_C\simeq \mathcal{N}'_{\mathbb{C}}$ for the quantum counterpart, since they are the only ones that leave the Hilbert space invariant. Nonetheless, in general curved spacetimes this is not the case, and evolution will not preserve  $L^2(\mathcal{N}'_\mathbb{C},D\mu_c)$ unless the spacetime is static or stationary \cite{ashtekarQuantumFields1975,agulloUnitarityUltraviolet2015,corichiSchrodingerFock2004}.  We will deal with these issues in Section \ref{sec:timeevolution}.

\subsubsection{Ambiguity of the construction}
\label{sec:Ambiguity}

In the following section we will relate the reference sections of different representations through a phase factor. This relation is provided by an ambiguity in the construction above that we explain below. For a similar discussion in finite dimensions follow \cite{woodhouseGeometricQuantization1997}. We may consider the existence of another connection on $B$ and the corresponding reference section $\tilde \Psi_r$ it is related to. As we want the section to be square integrable, we can consider the existence of a function $\alpha\in C^\infty(\mathcal{M}_C)$ satisfying

\begin{equation}
\label{eq:newvacuum}
\tilde \Psi_r=\alpha \Psi_r,
\end{equation}
for $\alpha:\mathcal{M}_C\to \mathbb{C}$ a function. Hence, sections of $B$ should satisfy
\begin{equation}
\label{eq:change-alpha}
\Phi_s= 
\begin{cases} 
    \tilde \Phi \tilde \Psi_r=\tilde \Phi \alpha\Psi_r \\
    \Phi \Psi_r
\end{cases}
\end{equation}

For the two covariant derivatives to coincide on a section of $B$, we obtain that
\begin{equation}
\label{eq:vacua}
\nabla_X \tilde \Psi_r=
\begin{cases} 
  i\tilde \theta (X) \tilde \Psi_r = i\tilde \theta (X)  \alpha \Psi_r  \\ 
  X(\alpha)\Psi_r+ \alpha(-i\theta(X) \Psi_r)
\end{cases}
\end{equation}
As this should happen for any vector field $X$, we conclude that the function $\alpha$ must satisfy
\begin{equation}
\label{eq:alphacondition}
d\alpha-i( \tilde \theta-\theta)\alpha=0
\end{equation}

Notice that as the modulus of the reference section is fixed to 1,  the function must be a phase $\alpha=e^{-if}$. This implies that the measure does not change but the symplectic potentials are required to satisfy the gauge condition
\begin{equation}
\label{eq:gauge}
\theta'-\theta=df.
\end{equation}
This is the freedom left from Equation \eqref{eq:vacuum_norm} once we fixed the measure $\mu_c$. 
Notice that if the modulus of the function is equal to one the measure is not changed, since we are just choosing among the different phases of the reference state. These kind of transformations that preserve the measure but change the reference phase are the usual ones for geometric (pre)quantization. We anticipate, however, that during the evolution in the context of curved spacetime, changes of measure will also occur \cite{alonsobujHybridGeometrodynamics2024, agulloUnitarityUltraviolet2015}

\subsection{Prequantization III: The prequantization of observables}
The covariant differentiation is precisely the basic tool to build a quantization mapping $\mathcal{Q}$ for the classical functions $f\in C^\infty(\mathcal{M}_C)$ acting on the set of sections of $B$. Thus, the operator $\mathcal{Q}(f)$  becomes
\begin{equation}
\label{eq:quantization}
\mathcal{Q}(f)=-i\hbar \nabla_{X_f} +f,
\end{equation}
where $X_f$ is the Hamiltonian vector field associated to $f$. For the sake of simplicity we will take $\hbar=1$ in the rest of the paper.

When we consider the section $\Psi_r$, and the corresponding function $\Phi$, we can adapt $\mathcal{Q}$ to the new setting described by \eqref{eq:wholePreq}. Thus,  $\nabla_{X_f}$ must become a self-adjoint operator on the Hilbert space of square integrable functions, where  $\Phi$ is contained. 

The classical observables are chosen as functions on the classical phase space $F:\mathcal{M}_C \to \mathbb{R}$. 
\begin{equation}
\label{eq:Hamiltonina}
X_{F}\lrcorner \omega =-dF.
\end{equation} 
Remember that, being weakly symplectic, the equation above may not have a solution in general. 
But we will consider also that the first Malliavin derivative,  see \cite{alonsoGeometricFlavours2023}, is defined for them and hence that $F\in\mathbb{D}_{\mu}^{2,1}$.   Therefore the expression
$$dF= \partial_{\utilde{\phi}^{x}}Fd\utilde\phi^x+\partial_{\bar{\utilde\phi}^{x}}Fd\bar{\utilde\phi}^x,$$
ensures that $dF\in \mathcal{H}_\Delta\otimes L^2(\mathcal{N}'_{\mathbb{C}},D\mu_c) $ and $\mathcal{H}_\Delta$ is the natural domain of $\omega$ so we can always find $X_F\in \mathcal{H}_\Delta\otimes L^2(\mathcal{N}'_{\mathbb{C}},D\mu_c)$. 
Then, its prequantum counterpart is an operator $\hat{F}:=\mathcal{Q}(F):\mathscr{H}_P\to \mathscr{H}_P$ given by 
\begin{equation}
\label{eq:actionPrecuantumOperator}
\mathcal{Q}(F) \Phi= F\Phi-i\nabla_{X_{F}}\Phi, \qquad \Phi\in \mathscr{H}_P.
\end{equation}

It follows that this (pre)quantization procedure $\mathcal{Q}$  meets the Dirac quantization (pre)conditions \cite{hallQuantumTheory2013}
\begin{enumerate}[label=Q\arabic*)]
    \item $\mathcal{Q}(F+G)=\mathcal{Q}({F})+\mathcal{Q}({G})$ 
    \item $\mathcal{Q}({F})\mathcal{Q}({G})-\mathcal{Q}({G})\mathcal{Q}({F})=-i\mathcal{Q}(\{F,G\}_{\mathcal{M}_F})$
    \item $\mathcal{Q}(F)=F\mathds{1}$ if $F$ is constant
\end{enumerate}
for any $F, G\in C^\infty(\mathcal{M}_C)$.

\subsection{The polarization}

The prequantization provides a Hilbert space $\mathscr{H}_P$  too big to realize the quantum states since, in principle, $\Phi\in \mathscr{H}_P$ are functions with domain on the whole manifold while, in regular quantum theories, the domain must have half the degrees of freedom of $\mathcal{M}_C$. 

To deal with this discrepancy we must restrict the domain of the functions $\Phi$ to a Lagrangian submanifold of $\mathcal{M}_{C}$. Notice, though, that in this setting the concept of Lagrangian submanifold is not as straightforward as in the finite-dimensional  case treated in \cite{woodhouseGeometricQuantization1997}. First, we must remember that $\omega_{\mathcal{M}_C}$ is defined, at most, on a dense subset of the vector fields of  $\mathcal{M}_{C}\cong \mathcal{N}'_{\mathbb{C}}$. Therefore that the condition of (co)isotropy must take into account this fact. 
 To bypass this difficulty we will introduce a polarization at the level of the \textit{one particle state structure} introduced in the complex description of the classical phase space \eqref{eq:holomorphicTriple}. Thus, we will choose a Lagrangian subspace $P$ of the complex Hilbert space $P\subset \mathcal{H}_{\Delta}$ on which the  symplectic form $\omega_{\mathcal{M}_C}$ is well defined. The quantum Hilbert space is given by:
 
 \begin{equation}
 \label{eq:ponarizedHilbert}
 \mathscr{H}=\overline{\left\{ \Psi^s \in \mathscr{H}_P \textrm{ s.t. } \nabla_{\bar{X}}\Psi^s =0 \ \  \forall\  X\in P\right\}}
 \end{equation}

As, again, the space is linear, we can choose a symplectic potential $-d\Theta=\omega_{\mathcal{M}_C}$ adapted to the polarization such that 

 \begin{equation}
 \label{eq:adaptedPolar}
 \Theta(\bar{X})=0\ \forall \ X\in P
 \end{equation}

Under this assumption the definition of the polarization simplifies enormously since, due to \eqref{eq:covariantDerivative} we can identify the polarized Hilbert space as:

\begin{equation}
\label{eq:quantumFunctions}
\mathscr{H}=\overline{\left\{ \Psi \in C^\infty (\mathcal{M}_C) \textrm{ s.t. } d\Psi(\bar{X}) =0 \ \  \forall\  X\in P\right\}}
\end{equation}

Unitary dynamics is trickier to implement due to the fact that the quantization rule \eqref{eq:actionPrecuantumOperator} may produce operators that do not respect the polarization. Moreover the situation is even more complicated if we ask the representation of the quantum operators to be irreducible adding a fourth condition to Q1-Q3 above:

\begin{enumerate}[label=Q4)]
    \item {\bf (Irreducibility condition)} For a given set of classical observables $\{f_i\}_{\mathcal{I}}$ such that $\{f_i,g\}=0\ \forall i\in \mathcal{I}$ implies $g$ constant then if an operator $A$ commutes with every $Q(f_i)$ then $A$ is a multiple of the identity.
\end{enumerate}

Then it is known that no quantization satisfying Q1-Q4 exists. We can partially escape from those two problems choosing to quantize only a subset of classical observables such as the linear ones in Darboux coordinates $(f_x\varphi^x+g_x\pi^x)$. More complicated functions  exhibit the usual ordering problem of quantization.  We will deal with this issue in \autoref{sec:QuantizHolom}. 

\section{Geometric quantization II:  types of quantization}
\label{sec:holom}  
Roughly speaking, we can consider two kind of polarizations: real polarizations, where the linear space $P$ is characterized by $P=P^*$ (which define what we call Schrödinger quantizations) and holomorphic ones for which $P\cap P^* =0$ (which define what we call holomorphic quantizations). Intermediate situations are combinations of those two extreme cases, at least in the finite dimensional setting, see \cite{woodhouseGeometricQuantization1997} for further details. As the holomorphic case is easier to handle in many ways and possesses better analytical properties \cite{alonsoGeometricFlavours2023}, we will begin our discussion with it. 

\subsection{Holomorphic quantization}

\subsubsection{Choosing holomorphic tensors}

For the holomorphic case the complex structure $J_{\mathcal{M}_C}$  provides a projection $X\to \frac12(\mathds{1}+iJ_{\mathcal{M}_C})X$ for the elements of the Cameron Martin tangent space $X\in \mathcal{H}_{\Delta}$, i.e., those elements in the tangent space which define integrable translations for the Gaussian measure $D\mu_c$. As $\mathcal{M}_C$ is a linear space,  the polarization is taken to be $P_{Hol}=\frac12(\mathds{1}+iJ_{\mathcal{M}_C}) \mathcal{H}_{\Delta}$.  Identifying coordinates $\phi^{\mathbf{x}}\in \mathcal{N}'_\mathbb{C}$ with holomorphic coordinates of $\utilde{\phi}^{\mathbf{x}}\in  \mathcal{M}_C$,  the adapted symplectic potential   is obtained from the (non unique) Kähler potential $\mathcal{K}=\bar{\phi}^xK_{xy}{\phi}^y$ 
\begin{equation}
\label{eq:SymplecticPotential}
\Theta=-{i}\partial_{\phi^x} \mathcal{K}d\phi^x.
\end{equation}

The set of states of the quantum Hilbert space given by \eqref{eq:quantumFunctions} is therefore the space of holomorphic (square integrable) functions $\Phi:\mathcal{N}_{\mathbb{C}}'\to \mathbb{C}$. In this representation we identify the reference and the vacuum sections $\Psi^H_r=\Psi_0$ because, as we will see, it will be annihilated by the annihilation operator. These functions $\Phi$ represent the excitations with respect to the vacuum state $\Psi_0$ \eqref{eq:vacuum}, in such a way that the physical state corresponds to the section of the complex line bundle $\Phi\Psi_0$, which are square integrable with respect to the measure $\mu_c$, or, equivalently, to the set of square integrable holomorphic functions if we take the section $\Psi_0$ as a reference:
\begin{equation}
    \label{eq:states_holomorphic}
    \mathscr{H}_{Hol}=L^2_{Hol}(\mathcal{N}_{\mathbb{C}}', D\mu_c).
    \end{equation}
This is, by construction, a well-defined Hilbert space but it is not the most frequent quantum model in QFT. Some examples of the study of this representation of QFT can be found in \cite{oecklSchrodingerRepresentation2012,oecklHolomorphicQuantization2012,oecklAffineHolomorphic2012}. Instead, it is more common to represent the states as element of a bosonic (or fermionic) Fock space, see for instance \cite{waldQuantumField1994,brunettiAdvancesAlgebraic2015}.  We will discuss this point later.

\subsubsection{Segal isomorphism: defining Fock states}

From the properties of Gaussian integration, we know that the Segal isomorphism presented in \cite{alonsoGeometricFlavours2023} allows us to identify the vectors in $L^2(\mathcal{N}_{\mathbb{C}}', D\mu_c)$ with the vectors of the symmetric Fock space constructed on the space of one-particle states $\mathcal{H}_\Delta$. As the quantum states correspond to holomorphic functions $\Psi\in L_{Hol}^2(\mathcal{N}_{\mathbb{C}}', D\mu_c)$, we can write:
\begin{equation}
\label{eq:Segal_holo}
\mathcal{I}(\Psi)=\left (\sqrt{n!}\psi^{(n,0)}_{\vec x_n} \right )_{n=0}^\infty, 
\end{equation}
where $\psi^{(n,0)}_{\vec x_n } $ correspond to the coefficients of the state $\Psi$ with respect to the space of Wick complex monomials

    \begin{equation}
        \label{eq:wienerItoSegalwithTtransform2}
        \psi^{(n,0)}_{\vec{x}_n}=\frac{1}{n!}\prod_{i=1}^n K_{x_iu_i}\frac{\partial}{\partial \bar\rho_{u_i}}  {S_{\mu_c}[\Psi](\rho_\mathbf{x},\bar{\rho}_\mathbf{x})} \bigg\lvert_{\rho_\mathbf{x}=0}
        \end{equation}
 where $S_{\mu_c}$ the integral transform introduced in \cite{alonsoGeometricFlavours2023}. This expression allows us to recover the usual description of quantum states in the Literature 
 \cite{waldQuantumField1994,brunettiAdvancesAlgebraic2015}.

\subsubsection{Quantizing linear operators}
\label{ssec:linearOp}
At the same time, as the connection is also holomorphic, the quantization mapping $\mathcal{Q}$ is simple to compute and defines holomorphic linear operators on $\mathscr{H}_{Hol}$. In particular  using \eqref{eq:actionPrecuantumOperator} for \eqref{eq:SymplecticPotential}, we  verify that
\begin{align}
\mathcal{Q}(\utilde\phi^\mathbf{y}) \Psi(\phi^\mathbf{x})=&\phi^\mathbf{y} \Psi(\phi^\mathbf{x}), & \nonumber\\
\label{eq:quantizationholomorphic}
    \mathcal{Q}(\bar{\utilde \phi}^\mathbf{y}) \Psi(\phi^\mathbf{x})=&\Delta^{\mathbf{y}z}\partial_{\phi^z }\Psi(\phi^\mathbf{x}), & \forall\, \Psi&\in \mathscr{H}_{Hol}.
\end{align}

Here we distinguish again $\phi^\mathbf{x}\in \mathcal{N}^\prime_\mathbb{C}$ and the holomorphic coordinate $\utilde\phi^\mathbf{x}\in \mathcal{M}_C$.  In this work we will consider bare coordinates $\phi^\mathbf{x}$ as simple placeholders of integration. In particular, they do not depend on time or any other structure built over the Cauchy hypersurface $\Sigma$. This is done to highlight the physical meaning of the coordinate $\utilde\phi^\mathbf{x}\in \mathcal{M}_C$. We will choose different systems of holomorphic coordinates of $\mathcal{M}_C$ with different interpretations that we will represent over the same $\mathcal{N}'_{\mathbb{C}}$ after quantization. This is crucial to relate different quantizations of the theory as we will see at the end of the section.

Notice that these operators correspond to the creation and annihilation operators, as it was to be expected  ${a^{\dagger \mathbf{x}}}=\mathcal{Q}(\utilde\phi^\mathbf{x})$ while $a^\mathbf{x}=\mathcal{Q}(\bar {\utilde\phi}^\mathbf{x})$.  Thus we obtain that  annihilation and creation operators correspond to the Malliavin derivative and Skorohod integral for holomorphic functions \cite{alonsoGeometricFlavours2023}. This interpretation that we obtain from geometric quantization was already noticed in the stochastic calculus literature from their algebraic relations  \cite{henry-labordereAnalysisGeometry2009}. 

Notice also that \eqref{eq:quantizationholomorphic} implies that $a^\mathbf{x} 1=0$. We identify  $\Psi_r^{H}=1$ as the representation as a holomorphic function $\Psi_r^{H}\in L^2_{Hol}(\mathcal{N}_{\mathbb{C}}', D\mu_c)$  of the reference section $\Psi_r^H$ in $\mathscr{H}_P$.   Thus, we conclude that $a^\mathbf{x}\Psi_r^H=0$. This implies that the reference section is annihilated by the annihilation operator and therefore we can identify the reference and the vacuum $\Psi_r^H=\Psi_0$ as we did above. 

\subsection{Schrödinger quantization}

Let us consider now the other most common example, the case of Schrödinger quantization as we can see in \cite{hofmannClassicalQuantum2015,hofmannNonGaussianGroundstate2017,hofmannQuantumComplete2019,eglseerQuantumPopulations2021,longSchrodingerWave1998,longSchrodingerWave1996,corichiSchrodingerFock2004}. Recall from  \cite{alonsoGeometricFlavours2023} that we can think of  $\mathcal{M}_C$ as a real manifold  $\mathcal{M}_F=\mathcal{N}'\times \mathcal{N}'$. We denote this process realification of the manifold.  The system of {realified}  conjugate coordinates  $(\varphi^{\mathbf{x}},\pi^{\mathbf{x}})$  is such that  

\begin{equation}
    \label{eq:symplecticFormlifted}
    \omega_{\mathcal{M}_F}=\delta_{xy} d\pi^x \wedge d\varphi^y =\int_\Sigma \frac{d^dx}{\sqrt{\lvert h\rvert}}\ [d\pi(x)\otimes d\varphi(x)-d\varphi(x)\otimes d\pi(x)].
\end{equation}

The fact that $\omega_{\mathcal{M}_F}$ is written in the canonical form above in turn defines the field momenta $\pi^{\mathbf{x}}$, since it must correspond to the conjugate variable to the field $\varphi^{\mathbf{x}}$. The particular physical meaning of these variables depends on the particular classical theory we want to quantize and is unimportant in this context.  Notice that the notion of conjugate acquires a dependence of Riemannian metric $h$ defined over the Cauchy hypersurface $\Sigma$. This is because \eqref{eq:symplecticFormlifted} is canonical only for that particular $h$.

In the Schrödinger representation  the quantum states are represented by the wave functional $\Psi(\varphi^{\mathbf{x}})$ , i.e., they must depend only on the field variables $\varphi^{\mathbf{x}}$, and not on the field momenta $\pi^{\mathbf{x}}$. This is derived from \eqref{eq:wholePreq} by applying \eqref{eq:ponarizedHilbert} with  a real polarization spanned by the momentum directions  in $\mathcal{M}_F$. See \cite{oecklSchrodingerRepresentation2012} for a similar construction and \cite{corichiSchrodingerFock2004} for an algebraic derivation of the representation.

Recall also that the
  complex structure on $\mathcal{M}_F$ is expressed in canonical coordinates as 
\begin{multline}
\label{eq:complexStructure}
-J_{\mathcal{M}_F} =
(\partial_{\varphi^y},\partial_{\pi^y})\left( \begin{array}{cc}
A^y_x & \Delta^y_x \\
D^y_x & -(A^t)^y_x
\end{array} \right)  \left( \begin{array}{c}
d\varphi^{x} \\
d\pi^{x}
\end{array} \right)=\\
=  \partial_{\varphi^y} \otimes[A^y_xd\varphi^x+\Delta^y_xd\pi^x]+ \partial_{\pi^x}\otimes [D^y_xd\varphi^x-(A^t)^y_xd\pi^y]
\end{multline}
With the conventions of \eqref{eq:raiseandlowerconvention} we have $A^2+\Delta D=-\mathds{1},\ \Delta^t=\Delta, D^t=D, A\Delta=\Delta A^t$ and $A^tD=DA$. Notice that $D$ is fixed once $\Delta$  and $A$ are known. Let $K^{\mathbf{x}}_{\mathbf{y}}$ be the inverse of $\Delta_\mathbf{y}^{\mathbf{x}}$, i.e. $\Delta^{x}_zK^z_{y}=\delta^x_y$ then $D=(iA^t+\mathds{1})K(iA-\mathds{1})$.

This complex structure, together with the symplectic form \eqref{eq:symplecticFormlifted} recovers a Riemannian structure $\mu_{J,\mathcal{M}_F}(\cdot,\cdot)=\omega_{\mathcal{M}_F}(\cdot,-J\cdot)$, which, in the coordinates above, reads: 
\begin{multline}
\label{eq:riemannianInduced}
\mu_{\mathcal{M}_F}=(d\varphi^{x},d\pi^{x}) \left( \begin{array}{cc}
    -D_{xy} & A^t_{xy} \\
    A_{xy} & \Delta_{xy}
    \end{array} \right) \left( \begin{array}{c}
        d\varphi^{y} \\
        d\pi^{y}
    \end{array} \right)=\\
    -D_{xy}d\varphi^x\otimes d\varphi^y+\Delta_{xy}d\pi^x\otimes d\pi^y+ A_{xy}(d\varphi^y\otimes d\pi^x+d\pi^x\otimes d\varphi^y)
\end{multline}
 Due to the fact that it is  Riemannian structure we derive  also $\Delta_{\mathbf{xy}}>0>D_{\mathbf{xy}}$. 

We recover holomorphic coordinates with  a change of coordinates, locally given by
\begin{align}
\label{eq:changeOfCoordinates}
d{\tilde{\varphi}^x}&=d{\varphi^x} &
d{\tilde{\pi}^x}&=A^x_yd{\varphi^y}+\Delta^x_yd{\pi^y} 
\\
\partial_{\tilde{\varphi}^x}&=\partial_{\varphi^x}- (K A)^y_x\partial_{\pi^y} &
\partial_{\tilde{\pi}^x}&=K^y_x\partial_{{\pi}^y}
\end{align}
Then $(\utilde\phi^{\mathbf{x}},\bar{\utilde\phi}^{\mathbf{y}})=\frac{1}{\sqrt{2}}(\tilde{\varphi}^{\mathbf{x}}-i\tilde{\pi}^{\mathbf{x}},\tilde{\varphi}^{\mathbf{y}}+i\tilde{\pi}^{\mathbf{y}})$ in accordance with the conventions \eqref{eq:coordinateConventions}.
Notice  that the identification $\mathcal{N}'\times \mathcal{N}'\simeq \mathcal{N}'_\mathbb{C}$ depends on the particular expression of the  complex structure introduced.

\subsubsection{The measure and the connection}
     The change from complex to real coordinates does not really affect the structure of the line bundle $B$ required to define the prequantization: we construct identical bundles on the two identical base manifolds $\mathcal{M}_C$ and $\mathcal{M}_F$ with different coordinates; and for each bundle, we can define a connection. These two connections need not to be identical (since we can consider several different connections for each case), but they must have a curvature proportional to the symplectic form. 

Identifying $\mathcal{M}_F$ and $\mathcal{M}_C$ also means that the measure on the space of fields can also be considered associated with the functional 
\begin{equation}
\label{eq:C_real}
C_S(\xi_\mathbf{x},\eta_\mathbf{x})=\int_{\mathcal{N}'\times \mathcal{N}'} D\mu(\varphi^\mathbf{x}, \pi^\mathbf{x}) e^{{i}\left(\xi_x \varphi^x+ \eta_x\pi^x\right)}.
\end{equation}
 Real polarizations will understood as taking only one copy of $\mathcal{N}'$ this is to restrict to the coordinate $\varphi^\mathbf{x}$. For this reason, in order to avoid the rescaling factor $\frac{1}{\sqrt{2}}$ in the quantization prescription we must rescale the covariance of the measure adapted to the holomorphic prescription $\mu_c$. Taking   $\rho_{\mathbf{x}}=\frac{1}{\sqrt{2}}(\xi_\mathbf{x}+i\eta_\mathbf{x})$ and $\phi^\mathbf{x}=\frac{1}{\sqrt2}(\varphi^{\mathbf{x}}-i \pi^{\mathbf{x}} )$, we must have that $\mu$ is defined by the characteristic functional
\begin{equation}
\label{eq:C_equiv}
C_S(\xi_\mathbf{x},\eta_\mathbf{x})=C_H\left(\frac{\bar\rho_{\mathbf{x}}}{\sqrt{2}}, \frac{\rho_{\mathbf{x}}}{\sqrt{2}}\right)=C_H\left(\frac{\xi_\mathbf{x}-i\eta_\mathbf{x}}{2}, \frac{\xi_\mathbf{x}+i\eta_\mathbf{x}}{2}\right).
\end{equation}

Again, we will consider a pre-quantum Hilbert space of sections of the bundle $B$, square-integrable with respect to the measure $\mu$. We can also identify a trivializing section $\Psi_r^S$ compatible with the connection in the sense of Equation \eqref{eq:vacuum}. With this trivialization, the space of states becomes the set of square-integrable functions on $\mathcal{N}'$ with respect to the measure $\mu$. 

From the isomorphism discussed above between $\mathcal{M}_C$ and $\mathcal{M}_F$, we can also conclude the equivalence of the connections on both spaces and the corresponding covariant derivatives. Strictly speaking, we should use different notation for the vector fields with respect to which we differentiate, since they are isomorphic and not identical, but we will use the same, to make the manuscript easier to read.  We shall use then the notation of Equation \eqref{eq:covariantDerivative} to represent the covariant derivative in the real polarization case, keeping in mind the fact that the only relevant vector fields are those tangent to the submanifold spanned by the field directions, as we will see now.

\subsubsection{The real polarization}
In the Schrödinger case the polarization is spanned by the momentum directions, i.e. the real subspace $P_{S}=Im(H_\Delta)$. We can consider an adapted symplectic potential with respect to $P_S$ satisfying
\begin{equation}
\label{eq:symlectic_potential_S}
\theta_S(X)=0 \qquad \forall X\in P_S.
\end{equation}

This means that $\theta_S$ must be a linear combination of the set $\{ d\tilde \varphi^\mathbf{x} \}$ of "basical" forms. The coefficients are read from the identification of the Gel'fand triple and the canonical Liouville 1-form of $T^*\mathcal{N}'$. Writing it in terms of the variables we introduced in the previous section we obtain: 

\begin{equation}
\label{eq:symplecticAdaptedSch}
\theta_S=-{i}(\tilde\varphi^x+i\tilde\pi^x)K_{xy}d\tilde\varphi^y=\pi^x\delta_{xy}d\varphi^y-i\varphi^xK_{xy} d\varphi^y+\varphi^x(K A)_{xy}d\varphi^y.
\end{equation}
Notice that we added the term $-i\tilde\varphi^xK_{xy} d\tilde\varphi^y$ (which is just an exact one form equal to the differential of the norm of the $\bar\varphi^x$) to the usual conventions because in this way we ensure that te quantization prescription provides hermitian operators in $L^2(\mathcal{N}',D\mu)$, as we will see below, and $2\Theta=\theta_S+\theta_M$ with \begin{equation}
\label{eq:symplecticmomentumnaive}
\theta_M= -(\tilde\varphi^x+i\tilde\pi^x)K_{xy}{d\tilde\pi^y}.
\end{equation} From this point of view $\theta_S$ is the restriction of $\Theta$ to $P_S$ directions with an scale factor of $2$ to keep up with the conventions in  changes of covariances. In this way we can establish the Schrödinger picture as a restriction of the holomorphic case to real directions. Nonetheless, this restriction must be  considered with care. 

The first conclusion of this analysis on the polarization is that functions on $\mathscr{H}_{S}$ (i.e., the quantum states) must be functions depending on the submanifold $\mathcal{N}'$ with coordinates $\varphi^{\mathbf{x}}$ rather on the whole $\mathcal{M}_F$, as it was to be expected.  Indeed, as for any $X\in P_S$,  $\theta_S(X)=0$, 
\begin{equation}
\label{eq:condition_polari}
\nabla_X \Psi=0 \Rightarrow d\Psi(X)=0, \forall X\in P.
\end{equation}
As the reference section $\Psi_r^S$ must satisfy this condition, any other section in the set of square-integrable ones will be obtained as a product by a function which is also annihilated by the momentum directions. Nonetheless, the definition of a measure for this set of functions is not immediate, since the measure in the original scheme above was defined for functions depending on both the fields and their momenta. However, we can repeat the construction for a different Gel'fand triple defined as
\begin{equation}
\label{eq:gelfand_sch}
\mathcal{N}\subset \mathcal{H}_\Delta^S \subset \mathcal{N}',
\end{equation}
where $\mathcal{H}_\Delta^S$ is the subspace of $\mathcal{H}_\Delta$ defined by the field states $ \xi_\mathbf{x}$. This triple defines a corresponding dual product $\langle \xi_\mathbf{x}, \varphi^\mathbf{x} \rangle= \xi_x\varphi^x$ which allows us to define a measure over $\mathcal{N}'$ that, with an slight abuse of notation, we denote with the same symbol $\mu$ by the functional
\begin{equation}
\label{eq:measure_sch}
C(\xi_{\mathbf{x}})=\int_{\mathcal{N}'} D\mu( \varphi^\mathbf{x}) e^{i \xi_x\varphi^x }.
\end{equation}
Notice that, by construction
\begin{equation}
\label{eq:measure_sch_C}
C(\xi_\mathbf{x})=C_S(\xi_\mathbf{x}, 0)= e^{-\frac14\xi_x\Delta^{xy}\xi_y}, 
\end{equation}
for $C_S$ the functional defined in Equation \eqref{eq:C_real}. Again, Bochner-Minlos theorem ensures that there exists a unique measure satisfying this condition and therefore we can define the set of (polarized) quantum states to be the space of square-integrable functions (with respect to the reference  section) $\mathscr{H}_S=L^2(\mathcal{N}', D\mu)$.

\subsubsection{Quantizing linear operators}
\label{sssec:SchLinearOperators}
We can readily quantize linear operators, such as the field operator $\varphi^x$ or the field momentum  $\pi^x$, whose action on the states $\Phi\in L^2(\mathcal{N}', D\mu)$ will be:
\begin{align}
\mathcal{Q}(\varphi^{\mathbf{y}}) \Phi (\varphi^\mathbf{x})&= \varphi^{\mathbf{y}} \Phi(\varphi^\mathbf{x}),\nonumber\\
 \mathcal{Q}(\pi^x\delta_{x{\mathbf{y}}})\Phi (\varphi^\mathbf{x})&=\left (-i\partial_{\varphi^{\mathbf{y}}}+i{\varphi^xK_{x{\mathbf{y}}}}-{\varphi^x(K A)_{x{\mathbf{y}}}}\right ) \Phi(\varphi^\mathbf{x}).
 \label{eq:quantizationOfSchOperators}
\end{align}

Analogously, we can also introduce creation and annihilation operators from the field and momentum operators as:

\begin{equation}
\label{eq:crationannihilation}
a^{\mathbf{x}}= \frac{\Delta^{{\mathbf{x}}y}\partial_{\varphi^y}-iA^{\mathbf{x}}_{y}\varphi^y}{\sqrt{2}}, \qquad a^{\dagger {\mathbf{x}}}=\sqrt{2}\varphi^{\mathbf{x}}-\frac{\Delta^{{\mathbf{x}}y}\partial_{\varphi^y}-iA^{\mathbf{x}}_{y}\varphi^y}{\sqrt{2}}.
\end{equation}

These operators satisfy that  $\mathcal{Q}(\varphi^{\mathbf{x}})= \frac{a^{\mathbf{x}}+a^{\dagger{\mathbf{x}}}}{\sqrt2}$ and $\mathcal{Q}(\pi^{\mathbf{x}})= -iK^{\mathbf{x}}_{y}\frac{a^y-a^{\dagger y}}{\sqrt2}$.
Notice that the complex exact term of \eqref{eq:symplecticAdaptedSch} is essential in order for the quantized momentum to be self adjoint  and it provides the right prescription for quantization   
 obtained in \cite{corichiSchrodingerFock2004} from an algebraic rather than geometric quantization procedure.

The discussion about the vacuum representation for this quantization is more involved than in the holomorphic case. Notice that, for $A^{\mathbf{x}}_{\mathbf{y}}\neq 0$ the constant function 1 is not annihilated by $a^{x}$. This is because the reference  section $\Psi^S_r$ is not yet the vacuum of the theory. To get the correct vacuum section we profit from the freedom of choice explored in \autoref{sec:Ambiguity} and multiply it by a phase factor $\Psi_0=\Psi^{Sch}_0\Psi^{S}_r$ with  $ \Psi^{Sch}_0\in L^2(\mathcal{N}', D\mu)$.  We will delve further into this issue in \autoref{sssec:SegalBargmann}.

\subsection{Quantization in the  Antiholomorphic and Field-Momenta representations}
\label{sec:momentumQuant}
Another real polarization is $P_M=Re(\mathcal{H}_\Delta)$, in this case we could proceed as we have done so far with the Hilbert space associated to the measure $\mu_c$ and the symplectic potential \eqref{eq:symplecticmomentumnaive} but it soon leads to cumbersome expressions for the quantized operators. This is because the change of coordinates \eqref{eq:changeOfCoordinates} is adapted to deal with the Schrödinger representation in the fields space. Here we will develop antiholomorphic quantization modifying that change of coordinate prescription to obtain a different measure $\mu_c$. Then develop the Schrödinger picture in the field-momentum space. For now on we will denote it simply as \textit{Field-Momentum representation}.  
 In order to quantize this theory in a way akin to the one adapted to the Schrödinger picture we should respect the momentum in the change of coordinates we modify  \eqref{eq:changeOfCoordinates}  to  
 \begin{equation}
 \label{eq:changeAntihol}
     d{\breve{\varphi}^x}=(A^t)^x_yd\pi^y-D^x_yd{\varphi^x} \qquad{d\breve{\pi}^x}=d{\pi^x},
 \end{equation}
 with $\ubreve{\phi}^\mathbf{x}=\frac{1}{\sqrt2}(\breve\pi^\mathbf{x}-i\breve\varphi^\mathbf{x})$ notice that this prescription is conjugate to $\utilde\phi^{\mathbf{x}}$ hence we denote it antiholomorphic coordinate. This treatment, following the arguments below  \eqref{eq:changeOfCoordinates}, modifies \eqref{eq:C} and  the measure considered for this case is $\nu_c$ defined by
\begin{equation}
    \label{eq:gaussianmeasuremomentum}
    \breve{C}(\rho_{\mathbf{x}}, \bar \rho_{\mathbf{x}})=\int_{\mathcal{M}_C}D\nu_c(\phi^\mathbf{x} )e^{i (\overline{\rho_x\phi^x}+\rho_x{\phi}^x) }= e^{\bar\rho_xD^{xy}\rho_y }.
\end{equation}
As in the holomorphic case ${\ubreve{\phi}^{\mathbf{x}}}$ represents a coordinate of $\mathcal{M}_C$ while ${\phi}^{\mathbf{x}}\in \mathcal{N}'_{\mathbb{C}}$ is a placeholder for integration.  This also leads to a antiholomorphic representation in $\mathscr{H}_{\overline{Hol}}=L^2_{\overline{Hol}}(\mathcal{N}'_{\mathbb{C}},D\nu_c)$. The Kähler potential in these coordinates with $D^{-1}_{xy}D^{yz}=\delta^z_x$ is\footnote{To see this result we use \eqref{eq:changeofvar2} and \eqref{eq:magicrelations}. }  $\mathcal{D}=-\bar{\ubreve{\phi}}^xD_{xy}^{-1}{\ubreve{\phi}}^y$  and the symplectic one form
$
\bar\Theta=-{i}\partial_{\bar{\ubreve{\phi}}^x} \mathcal{D}d\bar{\ubreve\phi}^x.$ Thus the quantization mapping becomes
\begin{equation}
\label{eq:antiholquant}
\overline{\mathcal{Q}}(\bar{\ubreve\phi}^{\mathbf{x}}) =\bar\phi^{\mathbf{x}} , \qquad \overline{\mathcal{Q}}(\ubreve\phi^{\mathbf{x}})= -D^{{\mathbf{x}}y}\partial_{\bar\phi^y}.
\end{equation}
These are interpreted as creation $\overline{\mathcal{Q}}(\bar{\ubreve\phi}^{\mathbf{x}})=b^{\dagger,{\mathbf{x}}}$ and annihilation $b^{\mathbf{x}}=\overline{\mathcal{Q}}(\ubreve\phi^{\mathbf{x}})$ operators of the antiholomorphic representation.

Following our steps with the Schrödinger picture we define  the Hilbert space of the momentum representation $\mathscr{H}_M=L^2(\mathcal{N}',D\nu)$ where $\nu$ is the Gaussian measure whose characteristic functional is 

\begin{equation}
    \label{eq:measure_momentum_C}
    \breve{C}_M(\xi_\mathbf{x})= e^{\frac14\xi_xD^{xy}\xi_y}. 
\end{equation}

The field-momentum polarization is therefore adapted to this Cameron Martin Hilbert space, $P_M=Re(\mathcal{H}_{-D})$ and we choose the adapted symplectic potential
\begin{equation}
\label{eq:symplectic}
\breve{\theta}_M=i(\breve\pi^x-i\breve\varphi^x)D^{-1}_{xy}d\breve\pi^y=-\varphi^x\delta_{xy}d\pi^y+i\pi^xD^{-1}_{xy} d\pi^y+\pi^x(AD^{-1})_{xy}d\pi^y.
\end{equation} 
 It follows from this prescription that 
\begin{align}
    \mathcal{Q}(\pi^{\mathbf{y}}) \Phi (\pi^\mathbf{x})&= \pi^{\mathbf{y}} \Phi(\pi^\mathbf{x}),\nonumber\\
     \mathcal{Q}(\varphi^x\delta_{x{\mathbf{y}}})\Phi (\pi^\mathbf{x})&=\left (i\partial_{\pi^{\mathbf{y}}}+i{\pi^xD^{-1}_{x{\mathbf{y}}}}+{\pi^x(AD^{-1})_{x{\mathbf{y}}}}\right ) \Phi(\pi^\mathbf{x}).
    \label{eq:quantizationOfmomentumOperators}
\end{align}
In the momentum field space creation and annihilation operators are 
\begin{equation}
\label{eq:crationannihilationmomentum}
b^{\mathbf{x}}= -\frac{D^{{\mathbf{x}}y}\partial_{\pi^y}-i(A^t)^{\mathbf{x}}_{y}\pi^y}{\sqrt{2}}, \qquad b^{\dagger {\mathbf{x}}}=\sqrt{2}\pi^{\mathbf{x}}+\frac{D^{{\mathbf{x}}y}\partial_{\pi^y}-i(A^t)^{\mathbf{x}}_{y}\pi^y}{\sqrt{2}}.
\end{equation}

These operate dually to creation and annihilation operators in the field space,  satisfying that  $\mathcal{Q}(\pi^{\mathbf{x}})= \frac{b^x+b^{\dagger x}}{\sqrt2}$ and $\mathcal{Q}(\varphi^x)= -i(D^{-1})^{\mathbf{x}}_{y}\frac{b^y-b^{\dagger y}}{\sqrt2}$. The discussion about the vacuum section of this theory is analogous to that in the Schrödinger representation. We will deal with it in \autoref{sssec:SegalBargmann}.

\subsection{Relations between holomorphic, Antiholomorphic, Schrödinger, and Momentum-field pictures.}
Once we have shown different quantizations procedures of a quantum field theory over different spaces $\mathscr{H}_{Hol},\mathscr{H}_{\overline{Hol}},\mathscr{H}_{S}$ and $\mathscr{H}_{M}$ we must study the relations of the quantization procedures under changes of coordinates and representations. We will start by relating $\mathscr{H}_{Hol}$ with $\mathscr{H}_{S}$ and  $\mathscr{H}_{\overline{Hol}}$ with $\mathscr{H}_{M}$ with unitary isomorphism respecting the algebra of creation and annihilation operators in each case. 
 Later, we provide unitary isomorphisms that relate every picture with each other via  quantization preserving integral transforms introducing the Fourier transform.

\subsubsection{Holomorphic and Schrödinger (Antiholomorphic and Momentum-field) pictures: Segal-Bargmann modified transforms}
\label{sssec:SegalBargmann}
Slightly modifying the Segal-Bargmann transform defined in \cite{alonsoGeometricFlavours2023} we can establish a unitary isomorphism  $$\tilde{\mathcal{B}}_{Sch}:L^2(\mathcal{N}',D\mu)\to L^2_{Hol}(\mathcal{N}_{\mathbb{C}}', D\mu_c)$$  
that preserves the algebra spanned by $a^{\mathbf{x}},a^{\dagger \mathbf{x}}$. In order to define this modified Segal-Bargmann transform we should deal with the extra $1/2$ factor appearing in the characteristic functional $C$ \eqref{eq:measure_sch_C}. To do so let us first define $\Psi_{Hol}=\tilde{\mathcal{B}}_{Sch}(\Psi_{Sch})$. Then, we choose
\begin{equation}
    \label{eq:segalBargmanntransform}
    \Psi_{Sch}(\varphi^{\mathbf{x}})=e^{{i}f(\varphi^\mathbf{x})}\int D\mu_S(\pi^{\mathbf{x}})\Psi_{Hol}\big(\sqrt{2}[\varphi^{\mathbf{x}}-i\pi^{\mathbf{x}}]\big),
\end{equation}
where $e^{if(\varphi^\mathbf{x})}$ is a phase factor to be determined. This is similar to the definition in \cite{oecklSchrodingerRepresentation2012},  with the addition of the phase factor. For instance if we take $f=0$, and we denote the corresponding  transformation as $\tilde{\mathcal{B}}$,  it reads:
\begin{equation}
    \label{eq:cration_annTransformed}
\tilde{\mathcal{B}}^{-1}\phi^{\mathbf{x}} \tilde{\mathcal{B}}=\sqrt{2}\varphi^{\mathbf{x}}-\frac{\Delta^{{\mathbf{x}}y}\partial_{\varphi^y}}{\sqrt{2}},\textrm{ and }\tilde{\mathcal{B}}^{-1}\partial_{\phi^{\mathbf{x}}} \tilde{\mathcal{B}}= \frac{1}{\sqrt2}\partial_{\varphi^{\mathbf{x}}}.
\end{equation} 
These relations are derived explicitly in \ref{sse:Segal-Bargmann-SCh}. Thus, departing from the expressions of $a^x$ and $a^{\dagger x}$ shown under \eqref{eq:quantizationholomorphic}  we can not recover the expressions of \eqref{eq:crationannihilation}. To solve this problem we choose, up to a constant 
$$ f(\varphi^\mathbf{x})=\frac{1}2(KA)_{xy}:\varphi^2:\lvert_{\frac{\Delta}2}^{xy}.$$ 
We choose the Wick ordered monomial to get a well defined chaos decomposition even though the pointwise product must be dealt with care \cite{alonsoGeometricFlavours2023}.
 Representing $\tilde{\mathcal{B}}_{Sch}=\tilde{\mathcal{B}}e^{-{i}f(\varphi^\mathbf{x})}$ and $\tilde{\mathcal{B}}_{Sch}^{-1}=e^{{i}f(\varphi^\mathbf{x})}\tilde{\mathcal{B}}^{-1}$ we obtain
\begin{gather}
\tilde{\mathcal{B}}_{Sch}^{-1}\phi^{\mathbf{x}} \tilde{\mathcal{B}}_{Sch}=e^{{i}f}\tilde{\mathcal{B}}^{-1}\phi^{\mathbf{x}} \tilde{\mathcal{B}} e^{-{i}f}=\sqrt2\varphi^{\mathbf{x}}-\frac{\Delta^{{\mathbf{x}}y}\partial_{\varphi^y}-iA^{\mathbf{x}}_{
    y}\varphi^y}{\sqrt 2},\nonumber\\ \tilde{\mathcal{B}}_{Sch}^{-1}\partial_{\phi^{\mathbf{x}}} \tilde{\mathcal{B}}_{Sch}= e^{{i}f}\tilde{\mathcal{B}}^{-1}\partial_{\phi^{\mathbf{x}}}\tilde{\mathcal{B}} e^{-{i}f}=\frac{\partial_{\varphi^x}-i(KA)_{{\mathbf{x}}
y}\varphi^y}{\sqrt 2}.
\label{eq:bargmann_Segal_sch_creat_anhil}
\end{gather} 
With this choice we preserve the form of the creation and annihilation operators for each picture. This may be interpreted as a nontrivial phase in the relation of the Schrödinger and holomorphic vacua as is explained at the end of \autoref{sssec:SchLinearOperators}. Indeed,  $\Psi_0^{Hol}=1$  represents the vacuum of the Fock space in the holomorphic representation because it is annihilated by the operator  $a^{\mathbf{x}}=\Delta^{{\mathbf{x}}y}\partial_{\phi^y}$ as noticed at the end of \autoref{ssec:linearOp}. Computing the Schrödinger vacuum  with \eqref{eq:segalBargmanntransform}  we obtain $\tilde{\mathcal{B}}_{Sch}^{-1}(1)=\Psi^{Sch}_0=\exp\big[\frac{i}2(KA)_{xy}:\varphi^2:\lvert_{\frac{\Delta}2}^{xy}\big]$. This is indeed the true representation of the vacuum because is annihilated by the annihilation operator in \eqref{eq:crationannihilation}.

Similarly for the momentum space we can establish a unitary isomorphism with the antiholomorphic representation
$$\tilde{\overline{\mathcal{B}}}_{Mom}:L^2(\mathcal{N}',D\nu)\to L^2_{\overline{Hol}}(\mathcal{N}_{\mathbb{C}}', D\nu_c).$$

 Let $\hat\Psi_{\overline{Hol}}=\tilde{\overline{\mathcal{B}}}_{Mom}(\hat\Psi_{Mom})$. Then   its inverse is provided by the expression
 
\begin{equation}
    \label{eq:segalBargmanntransformmomentum}
    \hat\Psi_{Mom}(\pi^{\mathbf{x}})=e^{ig}\int D\nu(\varphi^{\mathbf{x}})\hat\Psi_{\overline{Hol}}\big(\sqrt{2}[\pi^{\mathbf{x}}+i\varphi^{\mathbf{x}}]\big).
\end{equation}

Thus,   with $g=\frac{i}2(AD^{-1})_{xy}:\pi^2:\lvert_{-\frac{D}2}^{xy}$ we show in  \ref{sse:Segal-Bargmann-Mom} that \begin{equation}
\label{eq:momcreaanseagal}
    \tilde{\overline{\mathcal{B}}}_{Mom}^{-1}\bar\phi^{\mathbf{x}} \tilde{\overline{\mathcal{B}}}_{Mom}=b^{\dagger {\mathbf{x}}}\textrm{ and }-\tilde{\overline{\mathcal{B}}}_{Mom}^{-1}D^{{\mathbf{x}}y}\partial_{\bar\phi^y} \tilde{\overline{\mathcal{B}}}_{Mom}=b^{\mathbf{x}}.
\end{equation}
This  implies that the creation and annihilation operators of the antiholomorphic representations are transformed into \eqref{eq:crationannihilationmomentum}, the corresponding operators of the momentum-field representation. Also, we see that  the vacuum of the theory, the section annihilated by the annihilation operator $b^{\mathbf{x}}$, is represented by the phase  function $\Psi_0^{Mom}= \exp\big[\frac{i}2(AD^{-1})_{xy}:\pi^2:\lvert_{-\frac{D}2}^{xy}\big].$

\subsubsection{ Preservation of the quantization mapping: Segal-Bargmann and Fourier transforms. }
\label{sssec:FourierTransforms}
The isomorphisms described in the previous section do not respect the different quantization mappings described above. Throughout this section $\mathcal{Q}_s$ and $\overline{\mathcal{Q}}_m$ will represent the Schrödinger \eqref{eq:quantizationOfSchOperators} and momentum-field \eqref{eq:quantizationOfmomentumOperators} quantization mappings  while $\mathcal{Q}$ and $\overline{\mathcal{Q}}$ are the holomorphic \eqref{eq:quantizationholomorphic} and antiholomorphic \eqref{eq:antiholquant} quantization mappings. With these definitions we see that

$$\tilde{\mathcal{B}}_{Sch}\mathcal{Q}_s(\pi^{\mathbf{x}})\tilde{\mathcal{B}}_{Sch}^{-1}=\tilde{\mathcal{B}}_{Sch}iK^{\mathbf{x}}_y\frac{a^{\dagger y}-a^y}{\sqrt{2}}\tilde{\mathcal{B}}_{Sch}^{-1} = iK^{\mathbf{x}}_y \mathcal{Q}\Big(\frac{\utilde\phi^y-\bar{\utilde \phi}^y}{\sqrt{2}}\Big)= \mathcal{Q}(K^{\mathbf{x}}_y\tilde{\pi}^y).  $$
In general,  $K^{\mathbf{x}}_y\tilde{\pi}^y\neq \pi^{\mathbf{x}}$ according to \eqref{eq:changeOfCoordinates}. Thus quantization is not preserved. However, as we prove in \eqref{eq:quantizationpreservation} and \eqref{eq:quantizationpreservation2}, the Bargmann-Seagal transforms defined in \eqref{eq:quantizationholomorphic} and \eqref{eq:antiholquant}.  preserve the   quantization mappings
\begin{align}
    \tilde{\mathcal{B}}^{-1} \mathcal{Q}(\varphi^{\mathbf{x}}) \tilde{\mathcal{B}}=& \mathcal{Q}_s(\varphi^{\mathbf{x}}), 
    &
    \tilde{\mathcal{B}}^{-1} \mathcal{Q}(\pi^{\mathbf{x}}) \tilde{\mathcal{B}} =& \mathcal{Q}_s(\pi^{\mathbf{x}}),
    \nonumber
    \\
    \tilde{\overline{\mathcal{B}}}^{-1} \overline{\mathcal{Q}}(\varphi^{\mathbf{x}}) \tilde{\overline{\mathcal{B}}}=&\overline{\mathcal{Q}}_m(\varphi^{\mathbf{x}}), &
    \tilde{\overline{\mathcal{B}}}^{-1} \overline{\mathcal{Q}}(\pi^{\mathbf{x}}) \tilde{\overline{\mathcal{B}}} =&  \overline{\mathcal{Q}}_m(\pi^{\mathbf{x}}). 
\end{align}
Recall that $ \tilde{\mathcal{B}}^{-1}$ is defined in \eqref{eq:segalBargmanntransform} for $f=0$ and $ {\tilde{\overline{\mathcal{B}}}}^{-1}$ is defined in \eqref{eq:segalBargmanntransformmomentum} for $g=0$.  These are the regular Segal-Bargmann transforms introduced in \cite{alonsoGeometricFlavours2023} with a $\sqrt{2}$ factor multiplying the domain.   
This fact motivates the definition of the Fourier transform as the quantization preserving isometries  that close the following diagram. 

\begin{figure}[htb]
    \centering
        \begin{tikzpicture}
			\def\a{8 em} \def\b{3.2 em}
			\def\p{0.15}
			\def\w{0.65}
			\path
			(-\a,-\b) node[align=center] (D) {
				$\Big(L^2(\mathcal{N}',D\mu)
                ,
                \mathcal{Q}_s
                \Big)
                $   
				\\
				{\scriptsize Schrödinger} 
				}      
			(\a,-\b) node[align=center] (C) {
			$\Big(L^2(\mathcal{N}',D\nu),\,\overline{Q}_m\Big)$
				\\
				{\scriptsize Momentum-field}
				}
			(-\a,\b) node[align=center] (A) {    {\scriptsize holomorphic} \\
				$\Big(L^2_{Hol}(\mathcal{N}_{\mathbb{C}}', D\mu_c), \mathcal{Q}\Big)$     }
            (\a,\b) node[align=center] (B) {
			{\scriptsize Antiholomorphic}
            \\	
            $\Big(L^2_{\overline{Hol}}(\mathcal{N}_{\mathbb{C}}', D\nu_c), \overline{\mathcal{Q}}\Big)$
                };
			\begin{scope}[nodes={midway,scale=.75}]
			\draw[transform canvas={yshift=1.5ex},->] (D) -- (C) node[above]{$\mathcal{F}$};
			\draw[transform canvas={yshift=-1.5ex},->] (C)--(D) node[below]{$\mathcal{F}^{-1}$};
            \draw[transform canvas={xshift=1.5ex},->] (A)--(D) node[right]{$\tilde{\mathcal{B}}^{-1}$};
            \draw[transform canvas={xshift=-1.5ex},->] (D)--(A) node[left]{$\tilde{\mathcal{B}}$};
            \draw[transform canvas={yshift=1.5ex},->] (A) -- (B) node[above]{$\tilde{\mathcal{F}}$};
			\draw[transform canvas={yshift=-1.5ex},->] (B)--(A) node[below]{$\tilde{\mathcal{F}}^{-1}$};
            \draw[transform canvas={xshift=1.5ex},->] (B)--(C) node[right]{$\tilde{\overline{\mathcal{B}}}^{-1}$};
            \draw[transform canvas={xshift=-1.5ex},->] (C)--(B) node[left]{$\tilde{\overline{\mathcal{B}}}$};
			\end{scope}
		\end{tikzpicture}
        \caption{Quantization preserving mappings.}
        \label{fig:quantpreserving}
\end{figure}
(anti)holomorphic coordinates adapted to field \eqref{eq:changeOfCoordinates} or momentum-field \eqref{eq:changeAntihol} representations are related via an antiholomorphic transformation 
\begin{equation}
\label{eq:changeofvar2}
d\bar{\ubreve\phi}^{\mathbf{x}}=i(K+iKA)^{\mathbf{x}}_yd\utilde\phi^y, \qquad -i(D^{-1}+iAD^{-1})^{\mathbf{x}}_yd\ubreve\phi^y=d\bar{\utilde\phi}^{\mathbf{x}}.
\end{equation}
This result can be proved using the relation
 $D=(iA^t+\mathds{1})K(iA-\mathds{1})$ as described in \eqref{eq:magicrelations}. Thus, we define the Fourier transform $\tilde{\mathcal{F}}$ as a relation between  holomorphic and antiholomorphic representations defined by the unitary isomorphism
\begin{equation}
\label{eq:unitary}
\begin{array}{cccc}
\tilde{\mathcal{F}} :&L^2_{Hol}(\mathcal{N}_{\mathbb{C}}', D\mu_c) & \rightarrow & L^2_{\overline{Hol}}(\mathcal{N}_{\mathbb{C}}', D\nu_c) \\ 
 & \Psi_{Hol}(\phi^{\mathbf{x}}) &\mapsto & \hat\Psi_{\overline{Hol}}(\bar\phi^{\mathbf{x}})=\Psi_{Hol}[i(D^{-1}-iAD^{-1})^{\mathbf{x}}_y\bar\phi^y],  
\end{array}
\end{equation}
 with  inverse  
\begin{equation}
    \label{eq:unitaryinv}
    \begin{array}{cccc}
    \tilde{\mathcal{F}}^{-1} :&L^2_{\overline{Hol}}(\mathcal{N}_{\mathbb{C}}', D\nu_c) & \rightarrow & L^2_{Hol}(\mathcal{N}_{\mathbb{C}}', D\mu_c) \\ 
     & \hat\Psi_{\overline{Hol}}(\bar\phi^{\mathbf{x}}) &\mapsto & \Psi_{Hol}(\phi^{\mathbf{x}})=\hat\Psi_{\overline{Hol}}[i(K+iKA)^{\mathbf{x}}_y\phi^y].
    \end{array} 
\end{equation}
For a proof of unitarity see  \ref{ssec:unitaryIsom}. Notice that the transformation acts non-trivially over creation and annihilation operators because 
\begin{equation}
    \label{eq:HolAntiholFourier}
\tilde{\mathcal{F}}\phi^{\mathbf{x}}\tilde{\mathcal{F}}^{-1}=i(D^{-1}-iAD^{-1})^{\mathbf{x}}_y\bar\phi^y,\ \ 
\tilde{\mathcal{F}}\partial_{\phi^{\mathbf{x}}}\tilde{\mathcal{F}}^{-1}=i(K+iKA)^y_{\mathbf{x}}\partial_{\bar\phi^y}
\end{equation}
as we prove in  \ref{ssec:unitaryIsom}. 
However, these are precisely the transformations required to  preserve the quantization  mappings   
 \begin{align*}
     \tilde{\mathcal{F}}\mathcal{Q}(\ubreve\phi^{\mathbf{x}})\tilde{\mathcal{F}}^{-1} &=
 -i(K-iKA)^{\mathbf{x}}_y
 \tilde{\mathcal{F}}\mathcal{Q}(\bar{\utilde\phi}^y)\tilde{\mathcal{F}}^{-1}
 = \overline{\mathcal{Q}}({\ubreve\phi}^{\mathbf{x}}) \\
 \tilde{\mathcal{F}}\mathcal{Q}(\bar{\ubreve\phi}^{\mathbf{x}})\tilde{\mathcal{F}}^{-1} &=
 -i(D^{-1}+iAD^{-1})^{\mathbf{x}}_y
 \tilde{\mathcal{F}}\mathcal{Q}({\utilde\phi}^y)\tilde{\mathcal{F}}^{-1}
 = \overline{\mathcal{Q}}(\bar{\ubreve\phi}^{\mathbf{x}}) 
 \end{align*}
Finally, to close the diagram of \autoref{fig:quantpreserving} the remaining ingredient is the Fourier transform $\mathcal{F}$ defined by 
\begin{equation}
\label{eq:Fourier_momentum}
\begin{array}{cccc}
\mathcal{F} :& L^2(\mathcal{N}',D\mu_S) & \rightarrow & L^2(\mathcal{N}',D\nu_M) \\ 
 & \Psi(\varphi^{\mathbf{x}}) &\mapsto & \hat\Psi(\pi^\mathbf{x})={\tilde{\overline{\mathcal{B}}}}^{-1} \tilde{\mathcal{F}} \tilde{\mathcal{B}}\ \Psi,  
\end{array}
\end{equation}

 Then, as we show explicitly in \ref{ssec:craanhi} the quantization mappings are preserved because   
\begin{align}
    \label{eq:momentumClassical}
    \mathcal{F} \varphi^y\delta_{y{\mathbf{x}}}\mathcal{F}^{-1}&=i\partial_{\pi^{\mathbf{x}}}+i{\pi^yD^{-1}_{y{\mathbf{x}}}}+{\pi^y(AD^{-1})_{y{\mathbf{x}}}} \nonumber \\
    \mathcal{F}^{-1} \pi^{y}\delta_{y{\mathbf{x}}}\mathcal{F}&= -i\partial_{\varphi^{\mathbf{x}}}+i{\varphi^yK_{y{\mathbf{x}}}}-{\varphi^y(K A)_{y{\mathbf{x}}}}.
\end{align}
 Thus the quantization prescription for linear operators in the Schrödinger picture \eqref{eq:quantizationOfSchOperators} is transformed into  the analogous quantization prescription in the field-Momentum picture\eqref{eq:quantizationOfmomentumOperators} and viceversa.  Hence, the quantization mappings $\mathcal{Q}$ are respected by the Fourier transform in these representations. 

The discussion above is carried on over linear functions in each coordinate system. We can  straightforwardly generalize it for arbitrary functions after the choice of one of the ordering prescriptions discussed in \autoref{sec:QuantizHolom}.

In summary, if we choose a function $F(\varphi^{\mathbf{x}},\pi^{\mathbf{x}})$ we can express it in different sets of coordinates, \eqref{eq:change_of_vars} for holomorphic $F(\utilde\phi^{\mathbf{x}},\bar{\utilde\phi}^{\mathbf{x}})$   or \eqref{eq:changeAntihol} for antiholomorphic $F(\ubreve\phi^{\mathbf{x}},\bar{\ubreve\phi}^{\mathbf{x}})$ and the different quantization mappings are related by the diagram in \autoref{fig:quantpreserving} by
\begin{multline}
    \label{eq:quantpreserving_relations}
    \mathcal{Q}_s[F(\varphi^{\mathbf{x}},\pi^{\mathbf{x}})]= \mathcal{F}^{-1}\overline{\mathcal{Q}}_m[F(\varphi^{\mathbf{x}},\pi^{\mathbf{x}})]
    \mathcal{F} = \\
    \tilde{\mathcal{B}}^{-1}{\mathcal{Q}}[F(\utilde{\phi}^{\mathbf{x}},\bar{\utilde\phi}^{\mathbf{x}})]
    \tilde{\mathcal{B}}
    =
    \tilde{\mathcal{B}}^{-1}
    \tilde{\mathcal{F}}^{-1}\overline{\mathcal{Q}}[F(\ubreve\phi^{\mathbf{x}},\bar{\ubreve\phi}^{\mathbf{x}})]
    \tilde{\mathcal{F}}
    \tilde{\mathcal{B}}.
\end{multline}

These transformations do not respect, in general, the algebra of observables. We showed in the previous section that the isomorphisms of algebras are achieved by adding the corresponding phase factors to \eqref{eq:segalBargmanntransform} and \eqref{eq:segalBargmanntransformmomentum} lacking in this section. For this reason the transforms of \autoref{fig:quantpreserving} are not suited for discussions about the vacuum of every representation of the theory. 

For completeness we will show here how this representation fails to preserve the representation of the algebra of observables.  Lets denote $a^{\mathbf{x}},a^{\dagger,{\mathbf{x}}}$ the creation  and annihilation operators of the Schrödinger picture provided by \eqref{eq:crationannihilation}. Let also $b^{\mathbf{x}},b^{\dagger,{\mathbf{x}}}$ represent the creation and annihilation operators of the momentum-field picture given by \eqref{eq:crationannihilationmomentum}. Writing in terms of these operators the expression of the quantized operators $\hat\varphi^{\mathbf{x}}$ and $\hat\pi^{\mathbf{x}}$ the relation above can be rewritten as 
\begin{equation*}
     \mathcal{F}\frac{a^{\dagger {\mathbf{x}}}+a^{\mathbf{x}}}{\sqrt{2}}\mathcal{F}^{-1}= i(D^{-1})_y^{\mathbf{x}}\frac{b^{\dagger y}-b^y}{\sqrt{2}},\qquad
    iK^{\mathbf{x}}_y\mathcal{F}\frac{a^{\dagger y}-a^y}{\sqrt{2}}\mathcal{F}^{-1}= \frac{b^{\dagger {\mathbf{x}}}+b^{\mathbf{x}}}{\sqrt{2}}.
\end{equation*}
Thus we can write 
\begin{align}
    \label{eq:Fourier_creation}
    \mathcal{F}a^{\mathbf{x}}\mathcal{F}^{-1}&= i(\Delta-D^{-1})^{\mathbf{x}}_y\frac{b^y}{2}+i(\Delta+D^{-1})^{\mathbf{x}}_y\frac{b^{\dagger y}}{2} \nonumber \\
    \mathcal{F}a^{\dagger {\mathbf{x}}}\mathcal{F}^{-1}&= -i(\Delta+D^{-1})^{\mathbf{x}}_y\frac{b^y}{2}-i(\Delta-D^{-1})^{\mathbf{x}}_y\frac{b^{\dagger y}}{2} 
    \end{align}
This implies that creation and annihilation operators mix, in general,  in a nontrivial manner under $\mathcal{F}$. 
Notice that, from this expression, it is immediate to conclude that 
the mixing depends on the properties of the complex structure. Indeed, whenever $A^{\mathbf{x}}_\mathbf{y}=0$ we get $D^{-1}=-\Delta$. Hence, in those cases there will be no mixing between creation and annihilation operators under the Fourier transform. Under this assumption $\tilde{\mathcal{B}}_{Sch}$  and $\tilde{\overline{\mathcal{B}}}_{Mom}$ reduce to $\tilde{\mathcal{B}}$  and $\tilde{\overline{\mathcal{B}}}$ respectively and they form a unitary isomorphism of the algebra of observables. The Fourier transforms $\mathcal{F}$  and $\tilde{\mathcal{F}}$ also  produce an isomorphism that respects the canonical commutation relations. To see this result recall that $[a^{\mathbf{x}},a^{\dagger \mathbf{y}}]=\Delta^{\mathbf{x}\mathbf{y}}$ and $[b^{\mathbf{x}},b^{\dagger {\mathbf{y}}}]=-D^{{\mathbf{x}}{\mathbf{y}}}$ and $-D^{{\mathbf{x}}{\mathbf{y}}}=K^{{\mathbf{x}}{\mathbf{y}}}$ for this particular case. In a nutshell, the diagram \autoref{fig:quantpreserving}
only represents unitary isomorphisms of the algebra of observables if $A^{\mathbf{x}}_\mathbf{y}=0$.

\section{Quantization of the algebra of observables of a field theory}
\label{sec:QuantizHolom}

The holomorphic representation of quantum field theory is particularly well behaved to describe a quantum field theory over a Cauchy hypersurface $\Sigma$. In sections above, we showed a detailed account of this representation as well as its relation with several other representations of QFT. Thus, the space of pure states for this theory is  considered to be
$$L^2_{Hol}\big({\mathcal{N}_{\mathbb{C}}'},D\mu_c\big),
$$
for a suited choice of Gaussian measure $D\mu_c$. In this space the Wiener Ito decomposition is particularly simple and Skorokhod integral amounts to simple  multiplication by $\phi^\mathbf{x}$ which is also reflected in the simple form of creation and annihilation operators of \eqref{eq:quantizationholomorphic}. For the discussion of these and further analytical properties see \cite{alonsoGeometricFlavours2023}.  In this section we will present a quick summary of quantization in the holomorphic representation. In this case we will use the new tool of reproducing kernels presented in \cite{alonsoGeometricFlavours2023} to deal with ordering problems in the quantization of nonlinear operators. Since we will not discuss any other quantization mapping, for simplicity, in this and sections below we identify $\phi^{\mathbf{x}}$ with $\utilde\phi^{\mathbf{x}}$ unless otherwise stated.
\subsection{Weyl quantization}

The quantum picture is not complete until we prescribe a mapping from classical functions to quantum operators. This is a quantization that assigns a quantum observable for a suitable class of classical observables. Of course, there is an ambiguity on this prescription usually addressed under the name of ordering problems resulting from the nontrivial commutation relation $[a^{\mathbf{x}},a^{\dagger \mathbf{y}}]=\Delta^{\mathbf{xy}}$. The first choice that we make is called Weyl quantization and denote it $\mathcal{Q}_{Weyl}$. It is defined as follows. Firstly, any holomorphic quantization map must coincide over lineal functions of $\mathcal{M}_{C}$, this is
\begin{equation}
\label{eq:linealquantization}
a^{\mathbf{x}}=\mathcal{Q}_{Weyl}(\bar{\phi}^{\mathbf{x}})= \Delta^{\mathbf{x}y}\partial_{\phi^{y}}, \ \ \ a^{\dagger \mathbf{x}}=\mathcal{Q}_{Weyl}(\phi^{\mathbf{x}})= \phi^{\mathbf{x}}
\end{equation}
Then the  Weyl quantization mapping assigns to each monomial $\phi^n\bar\phi^m$ an averaged assignment of every possible order. We can then define the Weyl quantization to any polynomial classical functional inductively. For linear operators the quantization procedure must coincide with holomorphic geometric quantization  $\mathcal{Q}_{Weyl}[\phi^{\mathbf{x}}]=a^{\dagger \mathbf{x}},   \mathcal{Q}_{Weyl}[\bar\phi^{\mathbf{x}}]=a^{\mathbf{x}}$. For higher order monomials, we proceed  symbolically as
\begin{multline}
\label{eq:weylaverage}
     \mathcal{Q}_{Weyl}[(\phi^n\bar\phi^m)^{\mathbf{u}\vec{\mathbf{x}},\vec{\mathbf{y}}\mathbf{v}}]=\\
   a^{\dagger \mathbf{u}}\mathcal{Q}_{Weyl}[(\phi^{n-1}\bar\phi^m)^{\vec{\mathbf{x}},\vec{\mathbf{y}}\mathbf{v}}]+\mathcal{Q}_{Weyl}[(\phi^{n-1}\bar\phi^m)^{\vec{\mathbf{x}},\vec{\mathbf{y}}\mathbf{v}}]\frac{\Delta^{\mathbf{u}z}}2\frac{\overleftarrow{\delta}}{\delta a^z}\\=\mathcal{Q}_{Weyl}[(\phi^{n}\bar\phi^{m-1})^{\mathbf{u}\vec{\mathbf{x}},\vec{\mathbf{y}}}]a^{\mathbf{v}}+\frac{\Delta^{\mathbf{v} z}}2\frac{{\delta}}{\delta a^{\dagger z}}\mathcal{Q}_{Weyl}[(\phi^{n}\bar\phi^{m-1})^{\mathbf{u}\vec{\mathbf{x}},\vec{\mathbf{y}}}]
\end{multline}
From this symbolic expression it follows that the action of Skorokhod integrals with covariance $\Delta^{\mathbf{\mathbf{xy}}}/2$ is cast into
\begin{align}
\label{eq:WheylSkorokhod}
    \mathcal{Q}_{Weyl}[\partial^{*\phi^x}F]=\mathcal{Q}_{Weyl}[(\phi^x-{\frac{\Delta^{xy}}{2}\partial_{\bar\phi^y}})F]=a^{\dagger x}  \mathcal{Q}_{Weyl}[F]\nonumber \\
    \mathcal{Q}_{Weyl}[\partial^{*\bar\phi^x}F]=\mathcal{Q}_{Weyl}[(\bar\phi^x-{\frac{\Delta^{xy}}{2}\partial_{\phi^y}})F]=\mathcal{Q}_{Weyl}[F]a^{x}  
\end{align}

Thus this  quantization procedure is well suited to quantize classical functions being to a Hilbert space $\mathcal{O}_{cl}$ defined by

\begin{equation}
    \label{eq:obserbablesClasiscossinQuantizar}\mathcal{O}_{cl}=L^2(\mathcal{N}_{\mathbb{C}}',  D\mathcal{W})\textrm{ with }\int  D\mathcal{W}(\phi^{\mathbf{x}})e^{i (\overline{\rho_x\phi^x}+\rho_x\phi^x) }= e^{-\frac{\bar\rho_x\Delta^{xy}\rho_y}{2} }.
\end{equation}

Indeed, for functions in this Hilbert space we can consider a chaos decomposition 

\begin{equation}
\label{eq:f_chaos}
F[\phi, \bar \phi] =\sum F^{(n,\bar{m})}_{\vec{x}\vec{y}}\mathcal{W}_{\frac{\Delta}{2}}(\phi^n\bar\phi^m)^{\vec{x}\vec{y}}.   
\end{equation}
Wick monomials $\mathcal{W}_{\frac{\Delta}{2}}(\phi^n\bar\phi^m)^{\vec{x}\vec{y}}$ were introduced in \cite{alonsoGeometricFlavours2023} and, in this context, may be interpreted as a needed point-splitting regularization procedure in the product of distributions \cite{buchholzLocalityStructure1982}. Notice that, because $\mathcal{M}_F$ is modeled over $\mathcal{N}'_{\mathbb{C}}$, this is needed already at the classical level, this is, for $\mathcal{O}_{cl}$.
Now, quantizing the function $F$ means to construct the operator with the same chaos coefficients. For one such monomial we obtain
\begin{equation}
\label{eq:weyl}
\mathcal{Q}_{Weyl}\big[F^{(n,\bar{m})}_{\vec{x}\vec{y}}\mathcal{W}_{\frac{\Delta}{2}}(\phi^n\bar\phi^m)^{\vec{x}\vec{y}}\big]= F^{(n,\bar m)}_{\vec{x}\vec{y}}(a^{\dagger n})^{\vec{x}}(a^m)^{\vec{y}}.
 \end{equation}
It is straightforward to obtain the action on the total classical function by linearity although we must proceed with care because the resulting operators are, in general, unbounded. 

An important property of Weyl quantization is the transformation complex conjugation into involution 
\begin{equation}
\label{eq:involution}
\mathcal{Q}_{Weyl}\big( F^*\big)= \mathcal{Q}_{Weyl}\big(F\big)^{\dagger}.
\end{equation}
 Therefore, real classical functions  of $\mathcal{O}_{cl}$ are quantized into Hermitian operators. The machinery of reproducing kernels is better suited to deal with the finer details of this quantization we explore it in the next section.

\subsection{Weyl quantization  from reproducing kernels}
\label{sssec:weyl}

We can reexpress Weyl  prescription, acting over the dense subset of Hida test functions $(\mathcal{N}_{\mathbb{C}})\subset L^2_{Hol}\big({\mathcal{N}_{\mathbb{C}}'},D\mu_c\big) $, using reproducing kernels  as explained in the first part of this series. This is, let $\Psi\in (\mathcal{N}_{\mathbb{C}})$ and $f_{\mathbf{x}}\in \mathcal{H}_{\Delta}$ then  we can define the action of a symbol $f$ on the state $\Psi$ as:
\begin{align}
\label{eq:repkernelcreat}
f_x\Delta^{{x}y}\partial_{\sigma^{y}}\Psi(\sigma^{\mathbf{x}})&= \int_{\mathcal{N}'_{\mathbb{C}}}D\mu_c(\phi^{\mathbf{x}})e^{\bar\phi^xK_{xy}\sigma^y}f_x\bar\phi^x\Psi(\phi^{\mathbf{x}})\nonumber \\
f_x\sigma^{x}\Psi(\sigma^{\mathbf{x}})&= \int_{\mathcal{N}'_{\mathbb{C}}}D\mu_c(\phi^{\mathbf{x}})e^{\bar\phi^xK_{xy}\sigma^y}f_x\sigma^x\Psi(\phi^{\mathbf{x}})
\end{align}

This quantization prescription is particularly simple for the algebra trigonometric exponentials $\mathcal{T}(\mathcal{N}_{\mathbb{C}}')$, this is for $\mathcal{E}_{\chi}=\exp\left(i\overline{\chi_x\phi^x}+i\chi_x\phi^x\right)$ we have   
\begin{multline}
\label{eq:WeylQuantization}
\mathcal{Q}_{Weyl}\left[\mathcal{E}_{\chi}\right] \Psi(\sigma^{\mathbf{x}})=\\
\int_{\mathcal{N}'_{\mathbb{C}}}D\mu_c(\phi^{\mathbf{x}})\exp\left({i\chi_x\sigma^x+i\overline{\chi_x \phi^x}-\frac{\bar{\chi}_x\Delta^{xy}\chi_y}{2} + \bar\phi^xK_{xy}\sigma^y}\right)\Psi(\phi^{\mathbf{x}}).
\end{multline}
We know from \cite{alonsoGeometricFlavours2023} that $\mathcal{T}(\mathcal{N}_{\mathbb{C}}')$ is dense in $\mathcal{O}_{cl}$ and therefore we extend Weyl quantization to the whole space by linearity.

\subsubsection{Moyal product}
\label{sec:Moyal}
To complete the picture of Weyl quantization we need to know how composition of operators is translated as a multiplicative operation in the space of classical functions. For that matter we define the Moyal star product, and denote it $\star_m$, as the one that preserves operator composition after quantization, this is for an appropriate subset of classical functions $F,G\in W\subset \mathcal{O}_{cl}$\footnote{We will prove that $W$ is indeed a subset of $\mathcal{O}_{cl}$ in the next section.} 
$$\mathcal{Q}_{Weyl}\big(F\big)\mathcal{Q}_{Weyl}\big(G\big)= \mathcal{Q}_{Weyl}\big(F\star_m G\big).$$

 The space $W$ must be an algebra with the $\star_m$ multiplication and then we denote it the algebra of classical Weyl quantizable  functions. We will characterize this algebra below endowing it with a norm. Using \eqref{eq:WeylQuantization}  together with \eqref{eq:repkernelcreat}  it follows that $\mathcal{E}_{\rho}$ with the Moyal product  is a representation of Weyl relations

\begin{equation}
\label{eq:weyl_relations}
\mathcal{E}_{\rho}\star_{m}\mathcal{E}_{\alpha}=\exp\Big(\frac{\bar\rho_{{x}}\Delta^{xy}\alpha_y-\bar\alpha_{{x}}\Delta^{xy}\rho_y}2\Big)\mathcal{E}_{\rho+\alpha}.
\end{equation}
These Weyl relations will play a crucial role in our discussion about canonical quantization since they represent the canonical commutation relations in an exponential form. Using the identity $e^{\sigma^x\partial_{\phi^x}}F(\phi^{\mathbf{x}},\bar\phi^{\mathbf{x}})= F(\phi^{\mathbf{x}}+\sigma^{\mathbf{x}},\bar\phi^{\mathbf{x}})$ over trigonometric exponentials $\mathcal{E}_{\rho}$, and because of the density of $\mathcal{T}(\mathcal{N}_{\mathbb{C}}')$, we can rewrite Weyl relations and extend  the Moyal star product of any two functions $F,G\in W\subset \mathcal{O}_{cl}$ with an integral representation
\begin{equation}
\label{eq:moyal}
F\star_m G= \int D\mathcal{W}(\sigma^{\mathbf{x}})\int D \mathcal{W}(\zeta^{\mathbf{x}}) F(\phi^{\mathbf{x}}+\sigma^{\mathbf{x}},\bar\phi^{\mathbf{x}}+\bar\zeta^{\mathbf{x}})G(\phi^{\mathbf{x}}+\zeta^{\mathbf{x}},\bar\phi^{\mathbf{x}}-\bar\sigma^{\mathbf{x}}).  
\end{equation}
This expression is tightly related with the one in the covariant formalism of QFT showed in \cite{ditoStarproductApproach1990} . 

\subsubsection{The algebra of Weyl classical quantizable  functions}
Our goal in this section is to study the algebra $W$ and to prove that it is indeed a subset of $\mathcal{O}_{cl}$. To do so we start by noticing that a dense subalgebra should be  $(\mathcal{T}(\mathcal{N}_{\mathbb{C}}'),\star_m)$. Then we must topologize this algebra in such a way that the $\star_m$ product is continuous. The natural way to do this  is,   since $\star_m$ is a reflection of the operator product under the map $\mathcal{Q}_{Weyl}$, the topology is the coarsest one making this map continuous. Thus the topology is the one induced by $\mathcal{Q}_{Weyl}$ with the operator norm in the image.  In more practical terms, we endow $W$ with a norm such that for $F\in W$,

\begin{equation}
\label{eq:weylnorm}
\lVert F \rVert_{Weyl}= \lVert\mathcal{Q}_{Weyl}\big(F \big)\rVert
=\sup_{\lVert\Psi\rVert_{\mu_c}=1} \lVert \mathcal{Q}_{Weyl}\big( F \big)\Psi\rVert_{\mu_c}.
\end{equation}

It is indeed a norm because $Q_{Weyl}$ is linear and respects involution. 
    We start by studying the operator norm of the generators of the algebra, i.e. the trigonometric exponentials. Then \eqref{eq:involution}  allows us  to compute the operator norm  in the following way

\begin{equation}
\label{eq:operatornorm}
\lVert\mathcal{E}_{\chi} \rVert^2_{Weyl}=\sup_{\lVert\Psi\rVert=1}\langle \Psi \vert \mathcal{Q}_{Weyl}\big(\overline{\mathcal{E}_{\chi}}\star_m \mathcal{E}_{\chi} \big)\Psi\rangle= \sup_{\lVert\Psi\rVert=1}\langle \Psi \vert \Psi\rangle=1. 
\end{equation}
Here we have used the relation $\overline{\mathcal{E}_{\chi}}={\mathcal{E}_{-\chi}}$ and  Weyl relations  $\overline{\mathcal{E}_{\chi}}\star_m \mathcal{E}_{\chi}=\mathcal{E}_{0}=1$. This  very important result means that trigonometric exponentials are quantized into unitary  operators of $L^2_{Hol}\big({\mathcal{N}_{\mathbb{C}}'},D\mu_c\big)$.

In order to compute the operator norm of a generic element of $\mathcal{T}(\mathcal{N}_{\mathbb{C}}')$ we must  understand  the action of $\mathcal{Q}_{Weyl}\big(\mathcal{E}_{\chi} \big)$  over a vector $\Psi$. For that matter it is convenient to decompose trigonometric exponentials in terms of holomorphic (and antiholomorphic)  coherent states $\mathcal{K}_{\chi}=\exp\left(\chi_x\phi^x\right)$ in the following way  
\begin{equation}
\label{eq:expfunctions}
\mathcal{E}_{\chi}= \exp\left(\frac{\bar\chi_x\Delta^{xy}\chi_y}{2}\right) \overline{\mathcal{K}_{-i\chi}} \star_m {\mathcal{K}_{i\chi}}.
\end{equation}
Quantized  coherent states act in a straightforward way over the vector $\Psi$  
\begin{align}
    \label{eq:action}
    \mathcal{Q}_{Weyl}\big(\overline{\mathcal{K}_{\chi}}\big)\Psi(\phi^{\mathbf{x}})&=  \Psi(\phi^{\mathbf{x}}+\Delta^{\mathbf{x}y}\bar\chi_y),  \nonumber \\
    \mathcal{Q}_{Weyl}\big({\mathcal{K}_{\chi}}\big)\Psi(\phi^{\mathbf{x}})&=  e^{\chi_x\phi^x}\Psi(\phi^{\mathbf{x}}).  
\end{align}
 Then we can compute for later use  
\begin{multline}
    \label{eq:operatornorm2}
    \langle \Psi \vert \mathcal{Q}_{Weyl}\big( \mathcal{E}_{\chi} \big)\Psi\rangle
    =e^{\frac{\bar\chi_x\Delta^{xy}\chi_y}{2}}\langle \mathcal{Q}_{Weyl}\big( \mathcal{K}_{-i\chi} \big) \Psi \vert \mathcal{Q}_{Weyl}\big( \mathcal{K}_{i\chi} \big)\Psi\rangle\\  = e^{\frac{\bar\chi_x\Delta^{xy}\chi_y}{2}}\int D \mu_c(\phi^{\mathbf{x}})e^{i(\overline{\chi_x\phi^x}+\chi_x\phi^x)}\overline{\Psi(\phi^{\mathbf{x}})} \Psi(\phi^{\mathbf{x}}). 
    \end{multline}
    This expression allows us to compute the Weyl induced norm  of a generic element  $F\in\mathcal{T}(\mathcal{N}_{\mathbb{C}}')$. By definition it can be written, in general in a non unique manner, as $
    F(\phi^{\mathbf{x}},\bar{\phi}^{\mathbf{x}})=\sum_{n=1}^NF_n \mathcal{E}_{\chi^n}.
$ It is easy to see that the norm in $\mathcal{O}_{cl}$ of $F$ is simply computed by means of the characteristic functional \eqref{eq:obserbablesClasiscossinQuantizar}. Then using \eqref{eq:operatornorm2} and the characteristic functional of $D\mu_c$ defined in \eqref{eq:C}, the following equality holds 
\begin{multline}
\label{eq:operatornorm3}
\lVert F \rVert_{cl}^2 =
\sum_{n,m=0}^NF^*_nF_m e^{-\frac{\overline{(\chi^m-\chi^n)}_x\Delta^{xy}{(\chi^m-\chi^n)}_y}{2}} = \langle 1\vert \mathcal{Q}_{Weyl}\big(\bar{F}\star_m F \big)1 \rangle_{\mu_c}.
\end{multline}
Here the state 1 is simply the constant function 1. This expression and \eqref{eq:weylnorm}  provides us with the bound $\lVert F \rVert_{cl}\leq \lVert F \rVert_{Weyl}$ and, by density, this is valid for every $F\in W$. The
equality does not hold in general. 
From this fact we have that  $W\hookrightarrow\mathcal{O}_{cl}$ is a dense subspace  such that the inclusion is continuous. We also have that
$$(W,\star_m,\lVert \cdot \rVert_{Weyl})$$  
is a  $C^*$ algebra with complex conjugation as involution. In this  way $\mathcal{Q}_{Weyl}$ is a $C^*$-isomorphism between $W$ and $\hat W=\mathcal{Q}_{Weyl}(W) $. The latter is the von Neumann algebra of quantum observables which is 
 a representation of Weyl relations \eqref{eq:weyl_relations} as a closed subalgebra of the algebra of bounded operators $\hat W\subset B\left(L^2_{Hol}\big({\mathcal{N}_{\mathbb{C}}'},D\mu_c\big)\right)$. 
 
 An important remark to notice is  that in  previous sections we obtained an expression for  the quantization of linear functions of $\mathcal{O}_{cl}$. The result  is the set of linear combinations of creation and annihilation operators which may be unbounded. This means that, even though we can enlarge $Q_{Weyl}$ to make sense over the whole space of classical functions $\mathcal{O}_{cl}$, the norm $\lVert . \rVert_{cl}$ just provides a lower bound on the operator norm and the Moyal star product $\star_m$ is not well defined outside $W$. In simpler terms, even though we could quantize the whole $\mathcal{O}_{cl}$ we cannot treat it as an algebra.  This fact, even though may seem abstract a priori, is the source of the renormalization program.   In most cases we must deal with Hamiltonians with interaction such that $\hat H\in \mathcal{O}_{cl}$ but not be $W$. In the computation of the propagator of the theory we need to compute products of $\hat H$ with itself.  The naive treatment of this kind of expression leads to divergent integrals that must be treated with care in a renormalization program. We will not deal with this feature in this work and refer elsewhere \cite{nairQuantumField2005,agulloUnitarityUltraviolet2015}  for further information in the standard renormalization program. 
 
Finally let us provide an upper bound to $\lVert F \rVert_{Weyl}$. In general, it is  difficult to obtain an explicit expression of it. Nonetheless, by Hölder inequality we obtain 
$$\lvert \langle \Psi \vert \mathcal{Q}_{Weyl}\big( \mathcal{E}_{\chi} \big)\Psi\rangle \rvert\leq e^{\frac{\bar\chi_x\Delta^{xy}\chi_y}{2}} \lvert \langle \Psi \vert \Psi\rangle \rvert,$$
that for an element written as 
$
    F(\phi^{\mathbf{x}},\bar{\phi}^{\mathbf{x}})=\sum_{n=1}^\infty F_n \mathcal{E}_{\chi^n}
$ leads to 

\begin{multline*}
    \lVert F  \rVert^2_{Weyl}=
    \sup_{\lVert\Psi\rVert=1}\sum_{n,m=1}^\infty  F_n^*F_m \langle \Psi \vert \mathcal{Q}_{Weyl}\big( \mathcal{E}_{\chi^m-\chi^n} \big)\Psi\rangle_{\mu_c} \leq \\ \sum_{n,m=1}^\infty\lvert F_n^*F_m\rvert e^{\frac{\overline{(\chi^n-\chi^m)}_x\Delta^{xy}{(\chi^n-\chi^m)}_y}{2}}.  
\end{multline*}
This is the best that we can do, in turn it leads to a criterion of convergence for a function of $\mathcal{O}_{cl}$ to be in $W$  that may be used in a regularization program.

\subsection{Wick holomorphic quantization}
\label{sec:wick}
The most common ordering in QFT is not Weyl quantization, but Wick quantization, which  assigns to regular monomials the so called \textit{normal order} prescription. This prescription is designed to quantize regular monomials into normal ordered products of creation and annihilation operators that guarantees  null vacuum expectation values.  

In order to find an explicit expression for this operation  we introduce  the Wick operator $\mathcal{W}_{\frac{\Delta}{2}}$ for general functionals. Its action on a monomial must re-order the terms to write the resulting expression in normal ordered. From the properties of Skorokhod integrals \cite{alonsoGeometricFlavours2023}, acting on the constant function 1, it is straightforward to obtain  the expression of this operator as 

\begin{equation}
    \mathcal{W}_{\frac{\Delta}{2}}=\exp\Bigg(-\frac{\Delta^{xy}\partial_{\bar\phi^x}\partial_{\phi^y}}{2}\Bigg). 
\label{eq:wick_iso} 
\end{equation}
Notice that we use the same symbol as for the case of the Wiener Ito chaos decomposition in \cite{alonsoGeometricFlavours2023}, since it is the action of these operator on the suitable pair of fields what produces the suitable polynomials. For the real case this operator is introduced in \cite{cartierFunctionalIntegration2006}.  As an operator, it is invertible and
\begin{equation}
 \label{eq:wick-inverse}
 \mathcal{W}_{\frac{\Delta}{2}}^{-1}=\mathcal{W}_{-\frac{\Delta}2}.
\end{equation}

    Wick quantization is defined from Weyl quantization as  $\mathcal{Q}_{Wick}=\mathcal{Q}_{Weyl}\circ \mathcal{W}_{\frac{\Delta}{2}}$.
    We won't cover this case in full detail, the results of the previous section can be related to this case just by studying the properties of $\mathcal{W}_{\frac{\Delta}{2}}$. Our main interest in this section will be to find a correct star product $\star_{w}$ to find a representation of the Wick relations. Writing $\mathcal{Q}_{Weyl}$ in an integral kernel representation       
    \begin{multline}
        \label{eq:WickQuantization}
        \mathcal{Q}_{Wick}\left[\mathcal{E}_{\chi}\right] \Psi(\sigma^{\mathbf{x}})=\\
        \int_{\mathcal{N}'_{\mathbb{C}}}D\mu_c(\phi^{\mathbf{x}})\exp\left({i\chi_x\sigma^x+i\overline{\chi_x \phi^x}+ \bar\phi^xK_{xy}\sigma^y}\right)\Psi(\phi^{\mathbf{x}}).
        \end{multline}
    Wick star product follows from its definition $\mathcal{Q}_{Wick}\big(F\big)\mathcal{Q}_{Wick}\big(G\big)= \mathcal{Q}_{Wick}\big(F\star_w G\big)$ as 
    \begin{equation}
        \label{eq:wickStar}
        F\star_w G= \int D \mathcal{W}(\zeta^{\mathbf{x}}) F(\phi^{\mathbf{x}},\bar\phi^{\mathbf{x}}+\sqrt{2}\bar\zeta^{\mathbf{x}})G(\phi^{\mathbf{x}}+\sqrt{2}\zeta^{\mathbf{x}},\bar\phi^{\mathbf{x}}).  
        \end{equation}
    Then a representation of the Weyl algebra is obtained by the Wick ordered trigonometric exponentials
    $$:\mathcal{E}_{\chi}:=\exp\left(i\overline{\chi_x\phi^x}+i\chi_x\phi^x-\frac{\bar{\chi}_x\Delta^{xy}\chi_y}{2}\right),$$ 
    \begin{equation}
        \label{eq:weyl_relations_wick}
        :\mathcal{E}_{\rho}:\star_{w}:\mathcal{E}_{\alpha}:=\exp\Big(\frac{\bar\rho_{{x}}\Delta^{xy}\alpha_y-\bar\alpha_{{x}}\Delta^{xy}\rho_y}2\Big):\mathcal{E}_{\rho+\alpha}:.
        \end{equation}

        This procedure is often used in quantum field theory  because classical functions representing Hamiltonians are expressed in terms of regular monomials. Nonetheless, the space of classical functions possesses better analytical functions in Weyl quantization  and the difference between a Hamiltonian written in terms of regular or Wick monomials often amounts to an infinite constant that is physically interpreted as a trivial shift of the ground energy without physical meaning. 

        \subsection{Algebraic quantum field theory: Fock quantization}
        So far we used geometric quantization to find a particular class of representations of a quantum field theory. In those cases we obtain a Hilbert space as representation of the class of pure states and the algebra of observables as the image of a quantization mapping from a class of classical observables. In those cases, though, geometric quantization comes with the ambiguity associated with the choice of polarization then we are forced to study separately holomorphic, Schrödinger and field-momentum representations and find unitary isomorphisms relating each case. The reason why every representation is equivalent has its roots in the abstract study of the $C^*$-algebra of quantum observables, a program called Algebraic Quantum Field Theory (AQFT). We present here a quick summary of  its ingredients in order to introduce the most common QFT representation found in introductory books of the subject, i.e. Fock quantization. We refer to \cite{brunettiAdvancesAlgebraic2015,waldQuantumField1994} for a thorough analysis on the subject.

        \subsubsection{The C* algebra of quantum observables}
        Our goal in this section is to find Weyl relations, that correspond to the  the abstract axioms that characterize the $C^*$-algebra of quantum observables, from the geometric narrative that has been our guideline in this work. Then we find how  AQFT recovers the space of (not necessarily pure) states form this abstract setting.
        
        The first problem that we encountered in order to quantize a classical theory was how to model our phase space $T^*\mathcal{N}$ where we know that there is a canonical symplectic form.  So far, and in the first paper, our approach was to model this space treating both base and fiber as distributions then $\mathcal{M}_F= \mathcal{N}'\times \mathcal{N}'$. This choice, even though hardens the treatment of the classical theory itself, suits the needs of geometric quantization for quantum field theory. In this section we are interested in AQFT thus we start by a different choice of model space for our classical theory. In our model the manifold of fields is $\mathcal{M}_A= \mathcal{N}\times \mathcal{N}$ with coordinates $(\varphi_\mathbf{x},\pi_\mathbf{x})$ and it is endowed with a symplectic form 
\begin{equation}
\label{eq:symplecticalgebraic}
\omega_{\mathcal{M}_A}= \delta^{xy}d\pi_x\wedge d\varphi_y=\int_\Sigma {d^dx}{\sqrt{\lvert h\rvert}}\ [d\pi(x)\otimes d\varphi(x)-d\varphi(x)\otimes d\pi(x)].
\end{equation}

Notice the difference with \eqref{eq:symplecticFormlifted} because here the coordinates are functions and not densities of weight one.
At this stage we can write Weyl relations in a coordinate free manner. Let $\rho,\alpha\in T \mathcal{M}_{A}$ be vector fields, then the Weyl algebra is a $C^*$ algebra with generators $\mathcal{R}(\cdot)$ that fulfil Weyl relations  
\begin{equation}
\label{eq:generatorsweyl}
\mathcal{R}(\rho)^*=\mathcal{R}(-\rho), \ \ \mathcal{R}(\rho)\mathcal{R}(\alpha)=e^{\frac{i}2\omega_{\mathcal{M}_A}(\rho,\alpha)}\mathcal{R}(\rho+\alpha)
\end{equation}

 Notice that particular representations of this algebra are given by \eqref{eq:weyl_relations} and \eqref{eq:weyl_relations_wick}. These are particular examples of a representation of Weyl relations in the algebra of bounded operators\footnote{Here, for simplicity, identify the classical function with the $\star$ product with its image under the corresponding quantization mapping as we know that they are isomorphic } of a particular Hilbert space.

    Once the algebra is known we need to recover the space of states of the theory. To carry on with that study we use the GNS construction. Our starting point is a particular state of the algebra, i.e. a  linear functional $\varpi: \mathscr{A} \rightarrow \mathbb{C}$ which is positive definite (i.e. $\varpi(aa^*)\geq 0$) with $\varpi(1)=1$. With this state we can define the GNS representation of the $C^*$ algebra
     \begin{theorem} {\bf (GNS construction)}
     \label{th:GNS}
     Let $\mathscr{A}$ be a $C^*$-algebra with unit and let $\varpi: \mathscr{A} \rightarrow \mathbb{C}$ be a state. Then there exist a Hilbert space $\mathscr{H}$, a representation $\pi: \mathscr{A} \rightarrow B(\mathscr{H})$ and a vector $\Psi \in \mathscr{H}$ such that,
$$
\varpi(A)=\left\langle\Psi, \pi(A) \Psi\right\rangle_{\mathscr{H}} .
$$
Furthermore, the vector $\Psi$ is cyclic. The triplet $\left(\mathscr{H}, \pi,\Psi\right)$ with these properties is unique (up to unitary equivalence).
     \end{theorem}

In this fashion we recovered the usual prescription of quantum theories where we look for a representation of the algebra of observables as a subalgebra of the space of bounded operators acting on a separable  Hilbert space.  In this  way the GNS construction ensures that is enough to restrict the study to the physically meaningful states $\varpi: \mathscr{A} \rightarrow \mathbb{C}$.  This is, we can study the representation $\pi(\mathscr{A})\subset B(\mathscr{H})$ in our physically relevant setting instead of every possible abstract $C^*-$algebra.

We can study many physical aspects of quantum theories by asking properties over $\varpi$.  For example the quasi free or Gaussian condition ensures that the states are completely determined by two point correlation functions and the  Hadamard condition ensure that causality is preserved \cite{brunettiAdvancesAlgebraic2015}. In our case we are interested in the theory in a Hamiltonian form and therefore such features are part of the particular Hamiltonian that describes the theory. 

Our goal in this section was more modest than the general aim of AQFT as a whole \cite{brunettiAdvancesAlgebraic2015}.  We only pretend to set the groundings to understand the equivalence of the same theory described in different terms. So far we studied  quantization mappings $Q$ to establish particular representations of Weyl relations. This is our $C^*$-algebra of quantum observables always acted over 
$L^2_{Hol}(\mathcal{N'},D\mu_c)$ or similar.  From GNS we now know that this Hilbert space, that seemed the central object in geometric quantization, is in fact irrelevant since it can be reconstructed by means of a particular state $\varpi$. Thus, to recover holomorphic quantization we set $\hat{\mathcal{R}}(\alpha)=Q_{Weyl}(\mathcal{E}_\alpha)$ as our generators of the $C^*$-algebra of observables, then we can reconstruct the Hilbert space knowing that the Hilbert space vector representing the vacuum state is given by the constant function 1. To recover the state of the algebra we look at the GNS construction and by \eqref{eq:operatornorm2} we obtain
\begin{equation}
\label{eq:vacuumHolState}
\varpi_{Hol}(\hat{\mathcal{R}}(\alpha) )=  e^{\frac{\bar\alpha_x\Delta^{xy}\alpha_y}{2}}\int D \mu_c(\phi^{\mathbf{x}})e^{i(\overline{\alpha_x\phi^x}+\alpha_x\phi^x)}=e^{-\frac{\bar\alpha_x\Delta^{xy}\alpha_y}{2}}
\end{equation}
 This construction may seem too involved for our purposes so far but it is crucial to understand the equivalence with Fock quantization.

\subsubsection{Fock space representation}
\label{ssec:FockSpace}

In this section, we will consider how the representation of the $C^*$ algebra can be chosen for the Hilbert space to be the usual Fock space of QFT.  Besides, we will identify the role of the different geometric structures introduced in the manifold of classical fields in the representation.

In a Fock space representation  we start by introducing a  complex structure compatible with \eqref{eq:symplecticalgebraic}, for simplicity we choose the same entries of \eqref{eq:complexStructure} adapted to the new setting
     \begin{equation}
         \label{eq:complexStructurealg}
         -J_{\mathcal{M}_A} =
         (d{\varphi_y},d{\pi_y})\left( \begin{array}{cc}
             A^y_x & \Delta^y_x \\
         D^y_x & -(A^t)^y_x
         \end{array} \right)  \left( \begin{array}{c}
         \partial_{\varphi_{x}} \\
         \partial_{\pi_{x}}
         \end{array} \right).
         \end{equation}
This leads to the positive definite Riemannian metric $\mu_{\mathcal{M}_A}(\cdot,\cdot)= \omega_{\mathcal{M}_A}(J\cdot,\cdot)$. We can use it to define a Hermitian metric on $\mathcal{M}_A$ 

\begin{equation}
\label{eq:Hermitian}
h_{\mathcal{M}_{\mathcal{A}}}=\frac{\mu_{\mathcal{M}_A}-i\omega_{\mathcal{M}_A}}2,
\end{equation}

Moreover, by linearity $(\mathcal{M}_A,h_{\mathcal{M}_{\mathcal{A}}})$ is completed to a complex Hilbert space that we denote $\mathcal{H}_{1ps}$ the Hilbert space of one particle states. We will use this Hilbert space as an starting point to represent our quantum filed theory. The next step is to notice that we must allow states with an arbitrary number of bosonic particles, thus we represent our theory over $\mathcal{H}_{Fock}= \Gamma\mathcal{H}_{1ps}$, the symmetric Fock space.

With this structure at hand we use a change of variables dual to  \eqref{eq:changeOfCoordinates}  
\begin{gather}
    d{\tilde{\varphi}_x}=d{\varphi_x}+(KA)^y_xd\pi_y, \ \ \nonumber 
    d{\tilde{\pi}_x}=K^y_xd{\pi_y},\\ (\psi_{\mathbf{x}},\bar{\psi}_{\mathbf{y}})=\frac{1}{\sqrt{2}}(\tilde{\varphi}_{\mathbf{x}}+i\tilde{\pi}_{\mathbf{x}},\tilde{\varphi}_{\mathbf{y}}-i\tilde{\pi}_{\mathbf{y}}),
    \label{eq:change_of_vars}
\end{gather}
to obtain, in coordinates, $h_{\mathcal{M}_A}=\Delta^{xy}d\bar\psi_x\otimes d\psi_y$. In those coordinates it is clear that $\mathcal{H}_{1ps}=\mathcal{H}_{\Delta}$ where the latter is the completion of $\mathcal{N}_{\mathbb{C}}$ with the Hermitian form $h_{\mathcal{M}_A}$. An element of the Fock space is written in coordinates in a very straightforward manner as  
\begin{equation}
\label{eq:FockCoordinates}
\Psi=(\psi^{(0)},\psi^{(1)}_{x},\psi^{(2)}_{x_1x_2},\cdots,\psi^{(n)}_{\vec{x}_n},\cdots),
\end{equation}
with $\psi^{(n)}_{(\vec{x}_n)}= \psi^{(n)}_{\vec{x}_n}$. Annihilation and creation operators are also easily written as 
\begin{align}
    \bar\chi_xa^x\Psi&=(\bar\chi_x\Delta^{xy}\psi^{(1)}_{y},\cdots,\sqrt{n}\ \bar\chi_x\Delta^{xy}\psi^{(n)}_{y\vec{x}_{n-1}},\cdots), \nonumber \\
    \chi_xa^{\dagger,x}\Psi&=(0,\psi^{(0)}\chi_{x},\cdots,\sqrt{n+1}\ \psi^{(n)}_{(\vec{x}_{n}}\chi^{\phantom{(1)}}_{x_{n+1})},\cdots).    
    \label{eq:FockCreationAnihilation}
\end{align}

Among all the possible choices we pick $1 \in \mathcal{H}_\Delta^0$ as the cyclic vector of the GNS construction because we want to interpret it as the vacuum state of our theory. The last ingredient is the state, a Fock  generator of the Weyl algebra is \cite{corichiSchrodingerFock2004,waldQuantumField1994}, in holomorphic coordinates
 \begin{equation}
 \label{eq:FockGenerator}
 \mathcal{R}_F(\alpha)=\exp(\bar\alpha_x a^x+\alpha_xa^{\dagger,x}).
 \end{equation}

 Using BCH formula with $[a^{\mathbf{x}},a^{\dagger,\mathbf{y}}]=\Delta^{\mathbf{xy}}$ we put annihilation operators to the right of creation ones, then the Fock state acting on the generator, which is the vacuum expectation value of it, is easily computed by

\begin{equation}
\label{eq:Fockstate}
\varpi_{Fock}(\mathcal{R}_F(\alpha))= e^{-\frac{1}2h_{\mathcal{M}_A}(\bar\alpha,\alpha)}
\end{equation}

Thus $\varpi_{Hol}$ of \eqref{eq:vacuumHolState} and $\varpi_{Fock}$ act in the same way over their respective generators of the Weyl algebra and, because of the GNS construction, they are related by a unitary isomorphism and, as such, they represent the same theory. The same is true for the Schrödinger representation \cite{corichiSchrodingerFock2004}. Notice that a unitary isomorphism at the level of the Hilbert space is not enough to establish the equivalence of the theories. 

 Indeed we already knew that Fock space coordinates \eqref{eq:FockCoordinates} are the coefficients of the chaos decomposition of the holomorphic representation under the Segal isomorphism $\mathcal{I}:L_{Hol}^2(\mathcal{M}_C, D\mu_c)\to \mathcal{H}_{Fock}$. But to establish the relation in full we must also preserve the $C^*-$algebra of observables. In this case both theories are  unitary  equivalent representations of the same algebra of observables if we quantize the  $C^*-$algebra of Weyl quantizable functions $(W,\star_m)$ with the mapping $\mathcal{Q}_{Weyl}$ studied on \autoref{sec:QuantizHolom}. 
 
In any case it is important to remark thar Fock quantization is fundamentally different from that of holomorphic quantization   since it relies upon the notion of one particle space structure from which the Fock space is constructed. Kähler structures in this setting are also fundamentally different from those of holomorphic quantization. In holomorphic quantization this structure is densely defined in the domain of the wave function $\Psi(\phi^{\mathbf{x}})$ while  in Fock quantization it is defined dually over the coefficients of the one particle states. 

The relation between both pictures is also more involved than previously expected. In this section we made the design choice of writing Fock operators and vectors in holomorphic coordinates, this is because the explicit isomorphism given by the Segal isomorphism is only obvious  under the change of coordinates \eqref{eq:change_of_vars}. The real power of Fock construction relies in its coordinate free nature, which just depends  on  the Kähler structure of $(\omega,\mu, J)_{\mathcal{M}_A}$. We will keep this section simple at the cost of blurring its real power. Nonetheless, this feature is well known in the literature of QFT in curved spacetimes and we refer elsewhere for a thorough discussion \cite{brunettiAdvancesAlgebraic2015,waldQuantumField1994}.

\section{Geometric quantum field theory}
\label{sec:GQFT}

In this section we present a description of a QFT in purely geometric terms based on Kibble's program of geometrization of quantum mechanics \cite{kibbleGeometrizationQuantum1979,ashtekarGeometricalFormulation1999}. In that program, given a quantum theory defined by a Hilbert space of pure states $\mathcal{H}$ and its algebra of observables $\hat W$, we describe each ingredient in geometrical terms.  The Hilbert space is described as a holomorphic manifold of pure states $\mathscr{P}$  and a Kähler structure  $(\mathcal{G}, \Omega, \mathcal{J})_\mathscr{P}$. The objective of this section is to characterize this new Kähler structure and the algebra of  observables as a subset of quadratic functions $W_2(\mathscr{P})$. 

To model the space as a manifold at this {\it second quantized} level we choose a space for $\mathscr{P}$ that, in our case, will be Hida test functions. Because we do not deal with a particular theory we do not know yet the Kähler structure, thus the discussion is kept in in full generality.

To exploit the geometric insights of this approach we developed a set of tools based on Gaussian and white noise analysis \cite{hidaWhiteNoise1993,kuoWhiteNoise1996,obataWhiteNoise1994,huAnalysisGaussian2016} that will allow to write down some expressions in convenient sets of coordinates.

\subsection{Geometrization of quantum field theory I: The manifold of pure states}
Here we present, and adapt to our proposes, a  summary  Kibble's approach to the geometrization of quantum mechanics \cite{kibbleGeometrizationQuantum1979,ashtekarGeometricalFormulation1999}.  We want to characterize the description of QFT in terms of tensorial objects defined on the manifold of pure states $\mathscr{P}$. It turns out that $\mathscr{P}$ describes a quantum theory by means of a Kähler structure $(\mathcal{G}, \Omega, \mathcal{J})_\mathscr{P}$ that we introduce and study in this section. 

Later we will use this approach to exploit some properties of holomorphic quantization at the level of $\mathscr{P}$. To unleash the full power of the formalism we describe the theory with white noise analysis tools that allows to write tensors in closed forms for particular sets of coordinates.

\subsubsection{Quantum fields in the geometrized setting: The quantum Kähler structure}
 Due to the infinite dimensional nature of the problem the model space used to define the manifold structure is of crucial importance. For a well posed geometric framework we will choose to consider as reference a  particular  Nuclear Fréchet space the set of Hida test functions $\mathscr{N}=(\mathcal{N}_{\mathbb{C}})$.  

As we did above, we will consider a suitable rigged Hilbert space
    \begin{equation}
    \label{eq:riggingQft}
    \mathscr{N}\subset L_{Hol}^2(\mathcal{M}_C, D\mu_c) \subset {\mathscr{N}^{\prime *}}
    \end{equation}
    As a result, the Hilbert space of the states of quantum fields $L_{Hol}^2(\mathcal{M}_C, D\mu_c)$, can be densely modeled as a $C^\infty$-manifold, considering $\mathscr{N}$ as the model space.

    As the model space for our manifold is the space of test functions, we can introduce a system of coordinates associated to them on the space of quantum fields. In this case  we will denote test functions with a superindex  while the subindex stands for distributions.
    $$\Psi^{\phi}\in \mathscr{N},\quad \Upsilon_{\phi}\in \mathscr{N}^{\prime}.$$
    This convention is reverse to that of classical fields, see (1) of \cite{alonsoGeometricFlavours2023}, but turns out to be more convenient in this case because we stick to the geometric convention and test functions will usually be treated as coordinates of the manifold. Of course pointwise meaning holds for test functions while it is lost in the distributional setting. 

    We endow the manifold with a Hermitian form  trivially by using the scalar product of the Hilbert space. 
    Since the topology of $\mathscr{N}$ is finer than that of $L_{Hol}^2(\mathcal{M}_C, D\mu_c)$ we have that its scalar product $\langle,\rangle_{\mu_c}$, inherited by the embedding, is a sesquilinear positive definite and continuous bilinear form on $\mathscr{N}$. Moreover this form is weakly non degenerate. This means that only the zero vector has zero norm also in $\mathscr{N}$ but that the map $\langle\ ,\ \rangle_{\mu_c}:\mathscr{N} \to  {\mathscr{N}^{\prime *}}$ that maps every element $\Psi^{\phi}\in \mathscr{N}$ to $\langle \Psi^{\phi},\cdot\, \rangle_{\mu_c}\in \mathscr{N}^{\prime *}$ is not surjective, although it is clearly injective. 

    In the following we will denote as $\mathscr{P}$ the manifold of pure quantum states. As a linear complex manifold, it can be trivially treated as a real manifold endowed with a Kähler structure $(\mathcal{G}, \Omega, \mathcal{J})_\mathscr{P}$  related to the Hermitian tensor above. To recover the quantum theory we follow \cite{kibbleGeometrizationQuantum1979} and choose a preferred point in $\lvert 0\rangle\in\mathscr{P}$ that represents the vacuum and consider the tangent space of this point $T_{0}\mathscr{P}\simeq \mathscr{N}$. In linear spaces, that is the case that we are treating here, this distinction is blurred by the identification of $\mathscr{P}$ with $T_{0}\mathscr{P}$. In the following we will identify both spaces, when it is possible, to simplify the discussion.    
    
    Notice that in this case the complex structure is completely determined by the definition of the space of quantum states and we do not have freedom to select it as in the case of the manifold of classical fields. This is because we are describing a given theory in geometrical terms.   Nonetheless the picture is incomplete without the representation of the operators. We will deal with this aspect, together with the dynamics, later on. In the next section and below we will develop some important tools to express the geometrization of this section in concrete sets of coordinates.

\subsubsection{Holomorphic quantization in the geometrized setting}  
\label{ssec:QuantizationInGeom}

In this section we address the  geometrization of the space of pure states $\mathscr{P}$ of the  holomorphic quantization procedure. To do so, we look for an efficient way of describing tensors over the manifold $\mathscr{P}$. Thus we follow \cite{hidaWhiteNoise1993} and choose the white noise analysis to describe our tensors. The main idea of the analysis is the following: in the case of regular test functions and distributions the Riemannian metric $h$ over $\Sigma$ provides a canonical way of identifying the duality through the Gel'fand triple      
 \begin{equation}
     \label{eq:riggedOnAManifoldcomp}
     \mathcal{N}_{\mathbb{C}}\subset L^2(\Sigma,\mathrm{dVol}_h) \subset \mathcal{N}_{\mathbb{C}}'.
     \end{equation}
     
To generalize this triple to the second quantized setting we use the white noise  measure $D\beta$, defined in \cite{alonsoGeometricFlavours2023} by the characteristic functional 
 \begin{equation}
     \label{eq:whiteNoise}
     \int_{\mathcal{N}'_{\mathbb{C}}}D\beta(\phi^\mathbf{x}) e^{i(\overline{\rho_x\phi^x}+\rho_x\phi^x)}= e^{-\bar\rho_x\delta^{xy}\rho_y}=\exp\left(-\int_\Sigma \textrm{dVol}(x)\bar\rho(x)\rho(x)\right),
 \end{equation}

 We choose white noise  because it is the immediate generalization of \eqref{eq:riggedOnAManifoldcomp}. The white noise  Hilbert space $L^2(\mathcal{N}'_{\mathbb{C}},D\beta)$ has as  Cameron-Martin Hilbert space $L^2(\Sigma,dVol_h)$. This means that the White noise space is nothing but the bosonic Fock space constructed from it
 as the natural measure for the rigging. With this choice we obtain the natural triple identifying functions with distributions  \begin{equation}
     \label{eq:Triple_hida_second}
     (\mathcal{N}_{\mathbb{C}})\subset L^2_{Hol}(\mathcal{N}'_{\mathbb{C}},  D\beta)\subset (\mathcal{N}_{\mathbb{C}})^{\prime *}.
     \end{equation}

     In this rigging $\beta_{\bar\phi,\sigma}$ is the bilinear form that identifies test functions $\Phi^{\sigma}$ with distributions $\Phi_{\bar\phi}=\beta_{\bar\phi,\sigma}\Phi^{\sigma}$. In this notation distributions are antiholomorphic functionals. It also admits a weak inverse $\beta^{\phi,\bar\sigma}$ such that $\beta^{\gamma,\bar\sigma}\beta_{\bar\sigma,\phi}=\delta_{\phi}^\gamma$ is the evaluation mapping, continuous for Hida test functions as can be shown using the tools of reproducing kernels \cite{alonsoGeometricFlavours2023}.

     In a nutshell, we will use $\beta_{\bar\phi,\sigma}$ and $\beta^{\gamma,\bar\sigma}$ as the auxiliary bilinears to raise and lower indices of our operators and tensors.

Let $\Psi^\phi$ be a test function, representing a state of the manifold $\mathscr{P}$  that is provided by the phase space of holomorphic quantization. 
This notion of canonicity suggest that we must write every bilinear form or operator as the action of an integral kernel with respect to the white noise. In this manner for holomorphic quantization  we have  a complex structure 
\begin{equation}
\label{eq:holo_complex}
    \mathcal{J}_{\mathscr{P}}= i(d\Psi^{\phi} \otimes \partial_{\Psi^{\phi}}- d\overline\Psi^{\bar\sigma} \otimes \partial_{\overline{\Psi}^{\bar\sigma}})
\end{equation} and the Hermitian tensor 
\begin{equation}
\label{eq:hermitian}
h_\mathscr{P}=\frac{\mathcal{G}_{\mathscr{P}}-i\Omega_{\mathscr{P}}}{2}=\varDelta_{ \bar\sigma\phi} 
d\bar\Psi^{\bar\sigma}\otimes d\Psi^{\phi}  .
\end{equation}
Where the bilinear $\varDelta_{\bar\sigma, \phi}$ is easily written as a bivalued kernel with
\begin{multline}
    \label{eq:DeltaBilinear}
    \bar\Upsilon^{\bar\sigma} \varDelta_{\bar{\sigma},\phi}\Psi^{\phi} =\iint D\beta(\sigma) D\beta(\phi)\overline{\Upsilon(\sigma)} e^{\sigma^x\Delta_{xy}\bar{\phi}^y} \Psi(\phi)\\=\int D\mu_c(\phi)\overline{\Upsilon(\phi)}\Psi(\phi)
\end{multline}

Following our discussion about reproducing Kernels in  \cite{alonsoGeometricFlavours2023},  it is simple to understand that this expression, of course, does not have  a pointwise meaning  since the objects are bi-valued Hida distributions.

It is convenient to describe the manifold with real coordinates. Let {$\Psi^{\phi}=\frac{1}{\sqrt{2}}(\tilde\Phi^\phi+i\tilde\Pi^\phi)$} such that the set of coordinates $\left(
    \begin{array}{c}
     \tilde\Phi^\phi \\
     \tilde\Pi^\phi
    \end{array}
   \right)$, are {\it real} in the sense that they are holomorphic functions with real coefficients in its chaos decomposition. This can be written as $\overline{\Phi(\phi^{\mathbf{x}})}=\Phi(\bar{\phi^{\mathbf{x}}})$. In these coordinates 
   \begin{align}
    \mathcal{G}_{\mathscr{P}} =
    (d{\tilde{\Phi}^{\bar{\phi}}} 
    d{\tilde{\Pi}^{\bar{\phi}}})\varDelta_{\bar{\phi}\sigma}   \left( \begin{array}{c}
        d{\tilde{\Phi}^\sigma}\\ d{\tilde{\Pi}_\sigma}
    \end{array} \right), & \ \ 
    \Omega_{\mathscr{P}} =
    (d{\tilde{\Phi}^{\bar{\phi}}} 
    d{\tilde{\Pi}^{\bar{\phi}}})\varDelta_{\bar{\phi}\sigma} \epsilon  \left( \begin{array}{c}
        d{\tilde{\Phi}^\sigma}\\ d{\tilde{\Pi}_\sigma}
    \end{array} \right),\nonumber\\
      \mathcal{J}_{\mathscr{P}} =
    (\partial_{\tilde{\Phi}^{\phi}} 
    \partial_{\tilde{\Pi}^{\phi}})\delta^\phi_\sigma \epsilon  \left( \begin{array}{c}
        d{\tilde{\Phi}^\sigma}\\ d{\tilde{\Pi}_\sigma}
    \end{array} \right)& 
    \hspace{1 cm}\textrm{ for } \epsilon=\left(
        \begin{array}{cc}
        0&-1 \\
        1 & 0
        \end{array} 
    \right).
    \label{eq:KhalerReal}
    \end{align}

    To write down a dynamical system is also interesting to describe the Poisson bracket in this context. To do so we define, associated with these coordinates,  the Wirtinger derivatives

\begin{equation}
\label{eq:wirtingerHolsecond}
\partial_{\Psi^{\phi}}= \frac{1}{\sqrt{2}}(\partial_{\tilde\Phi^\phi}-i\partial_{\tilde\Pi^\phi}), \hspace{1 em} 
\partial_{\bar\Psi^{\bar\phi}}= \frac{1}{\sqrt{2}}(\partial_{\tilde\Phi^{\bar\phi}}+i\partial_{\tilde\Pi^{\bar\phi}}), 
\end{equation}

We then obtain that the inverse bilinear $(\mathcal{K}\varDelta)_\phi^\sigma=\delta_\phi^\sigma$ is   $\mathcal{K}^{\phi,\bar\sigma}=\exp (\bar\sigma^xK_{xy}\phi^y)$. This fact can be proved by, recalling the conventions of \eqref{eq:raiseandlowerconvention},  using the computational rule
\begin{equation}
\label{eq:Multrule}
\int_{\mathcal{N}'_{\mathbb{C}}} D\beta(\sigma)e^{\phi^yA_{yx}\bar\sigma^x+\sigma^xB_{xy}\gamma^y}  = e^{\phi^x(AB)_{xy}\bar\gamma^y }.
\end{equation}
thus the Poisson bracket is 

\begin{equation}
\label{eq:poissonHolo}
    \left\{\cdot,\cdot\right\}_{\mathscr{P}}=-i\mathcal{K}^{\phi,\bar\sigma}\partial_{\Psi^\phi}\wedge\partial_{\bar\Psi^{\bar\sigma}}.
\end{equation}

Let $F(\tilde\Phi^{\phi},\tilde\Pi^\phi)$ and $G(\tilde\Phi^{\phi},\tilde\Pi^\phi)$ be  well behaved real functions over $\mathscr{N}$ then the bracket action is  
 \begin{multline}
         \label{eq:poissonSecondQuantSchAction}
         \left\{F,G\right\}_{\mathscr{P}}=\\
         \iint D\beta(\phi)D\beta(\sigma){\exp (\bar\sigma^xK_{xy}\phi^y)}\left({\frac{{\partial F}}{\partial\tilde\Phi(\sigma)}}\frac{\partial\bar{G
         }}{\partial{\tilde\Pi(\bar{\phi})}}-{\frac{\partial {F}}{\partial {\tilde\Pi({\sigma})}}}\frac{\partial 
         \bar{G}}{\partial\tilde\Phi(\bar\phi)}\right),
         \end{multline}


    The space of Hida test functions is nuclear and Fréchet, among other convenient properties suited for geometry \cite{krieglConvenientSetting1997}, this implies that the notions of Gateaux and Fréchet derivatives coincide and there is no ambiguity in the definition   of this bracket.

\subsubsection{Antiholomorphic quantization in the geometrized setting}  

The tool at hand to express the antiholomorphic picture is the Fourier transform of \eqref{eq:unitary}. For simplicity, we will express this transformation as an internal operator to the holomorphic functions. Thus,  we express the Fourier transform  with an integral kernel as 
    \begin{equation}
    \label{eq:FourierKernel}
    \tilde{\mathcal{F}}^{\phi}_\sigma\left(
        \begin{array}{c}
         \tilde\Phi^\sigma \\
         \tilde\Pi^\sigma
        \end{array}
    \right)=\int D\beta(\sigma) \exp\left[\phi^x
    \left( \begin{array}{cc}
        AD^{-1} & -D^{-1} \\
        D^{-1} & AD^{-1}
    \end{array} \right)_{xy}\bar\sigma^x
    \right]
    \left(
        \begin{array}{c}
         \tilde\Phi(\sigma) \\
         \tilde\Pi(\sigma)
        \end{array}
    \right).
    \end{equation}
    We can also express the inverse as 
    \begin{equation}
        \label{eq:Fourierinverse}
        (\tilde{\mathcal{F}}^{\phi\bar{\sigma}})^{-1}= \exp\left[-\phi^x
        \left( \begin{array}{cc}
            KA & K \\
            -K & KA
        \end{array} \right)_{xy}\bar\sigma^x
        \right].
        \end{equation}

    Expressions with matricial kernels can be easily handled by noticing that the computational rule   \eqref{eq:Multrule} also holds for  those kind of exponents, which leads to an algebra of exponential kernels described in \autoref{tab:multTable} in the appendix,  and its action on a vector is given by
    \begin{equation}
    \label{eq:actionFormula}
    \int D\beta(\sigma) \mathcal{K}(\phi^{\mathbf{x}},\bar\sigma^\mathbf{x})
    \left(
        \begin{array}{c}
         \Phi(\sigma) \\
         \Pi(\sigma)
        \end{array}
    \right)= \mathcal{K}(\phi^{\mathbf{x}},\delta^{\mathbf{x}y}\partial_{\sigma^{y}})  \left.\left(
        \begin{array}{c}
         \Phi(\sigma) \\
         \Pi(\sigma)
        \end{array}
    \right)\right\rvert_{\sigma=0}
    \end{equation}
    Following these rules it is easy to see that
    \begin{equation}
    \label{eq:fourierActionDarboux}
    {\varDelta}^\phi_\sigma=(\tilde{\mathcal{F}}^t\mathcal{D}\tilde{\mathcal{F}})^\phi_\sigma\ \  \textrm{ for }\mathcal{D}^{\phi,\bar\sigma}=\exp (-\phi^xD_{xy}\bar\sigma^y)
    \end{equation}
    Where we define we define the transpose of an operator as  
    
    \begin{equation}
        \label{eq:1psWN}
        {\left( \begin{array}{cc}
            \mathcal{B} & \mathcal{C}\\
        \mathcal{E}  & \mathcal{F}
        \end{array} \right)^t}_\phi^\sigma= 
        {\left( \begin{array}{cc}
            \mathcal{B}^t & \mathcal{E}^t\\
        \mathcal{C}^t  & \mathcal{F}^t
        \end{array} \right)}_\phi^\sigma, \textrm{ with }(\mathcal{B}^t)_\phi^\sigma= \beta^{\sigma \bar\gamma}\beta_{\bar\delta\phi}\mathcal{B}_{\bar\gamma}^{\bar\delta}.
        \end{equation} 
        
        Using $\tilde{\mathcal{F}}$ as a change of coordinates we see that this is enough to turn \eqref{eq:KhalerReal} into the appropriate expressions for the antiholomorphic representation even if they are represented by holomorphic functions. We will denote the coordinates of the antiholomorphic representation with  $\Bigg(
            \begin{array}{c}
             {\hat\Phi}^\sigma \\
             {\hat\Pi}^\sigma
            \end{array}
        \Bigg)= \tilde{\mathcal{F}}^{\phi}_\sigma\Bigg(
            \begin{array}{c}
             \tilde\Phi^\sigma \\
             \tilde\Pi^\sigma
            \end{array}
        \Bigg)$.

\subsubsection{Geometrization in Darboux second quantized (s.q.) coordinates }
Our approach in this section is to craft a notion of Darboux second quantized (s.q) coordinates that provides some sense of canonical coordinates in this more involved  framework. The choice, consistent with the discussion of the previous sections is to choose as a canonical form for the symplectic structure the one corresponding to the white noise Hilbert space which is  a direct generalization of Darboux coordinates in \eqref{eq:symplecticalgebraic}. 
Thus the canonical form is 
 \begin{equation}
     \label{eq:symplecticDarbouxSecond}
     \Omega_{\mathscr{P}}= \beta_{\bar\phi,\sigma}d\Pi^{\bar\phi}\wedge d\Phi^{\sigma}= \int D\beta(\phi)\left[d \Pi(\bar\phi)\otimes d \Phi(\phi)- d \Phi(\bar\phi)\otimes d \Pi(\phi)\right].
     \end{equation}
     
    We can simply find a system of Darboux s.q.  coordinates with a change of variable $
  \left( 
      \begin{array}{c}
       \tilde\Phi^\phi \\
       \tilde\Pi^\phi
      \end{array}
  \right)
  = \
  \mathcal{C}\left(
  \begin{array}{c}
   \Phi^\phi \\
   \Pi^\phi
  \end{array}
\right)$ that provides a complex structure

\begin{equation}
    \label{eq:darbouxJ}
    \mathcal{C}=
        \left( \begin{array}{cc}
            \mathcal{K}  &  0\\
            0  &  \mathds{1}
        \end{array} \right)
        , \ \ \ \mathcal{C}^{-1}=
          \left( \begin{array}{cc}
            \varDelta &  0\\
            0  & \mathds{1}
        \end{array} \right), \ \ \ \mathcal{J}= 
        \left( \begin{array}{cc}
            0 &  \varDelta\\
            -\mathcal{K}  & 0
        \end{array} \right).
    \end{equation}

    Finally, this symplectic structure induces a Poisson bracket over functions  in Darboux s.q. coordinates    
    \begin{equation}
    \label{eq:PoissonBracketSecondQuantizedDarboux}
    \left\{\cdot,\cdot\right\}_{\mathscr{P}}=\beta^{\phi,\bar\sigma}\partial_{\Phi^\phi}\wedge\partial_{\Pi^{\bar\sigma}}.
    \end{equation}
    that over smooth real functions acts like 
     \begin{equation}
     \label{eq:poissonSecondQuantDarbouxAction}
     \left\{F,G\right\}_{\mathscr{P}}=\int D\beta(\phi)\left({\frac{{\partial F}}{\partial\Phi(\phi)}}\frac{\partial \overline{G}}{\partial\Pi(\bar\phi)}-{\frac{\partial F}{\partial\Pi(\phi)}}\frac{\partial \overline{G}}{\partial\Phi(\bar\phi)}\right).
     \end{equation}

     This set of coordinates will be important later to discuss the particular expression for the Hermitian of an operator.

\subsection{Geometrization of quantum field theory II: The algebra of observables}

 Once we described the space of pure states of the theory in the geometrized setting and we explained how to describe concrete sets of coordinates using white noise analysis, the remaining task is to describe an algebra of observables that fulfills Weyl relations and is unitary equivalent to the ones described so far. As before, we start in full generality and unveil several interesting properties in particular sets of coordinates using Fock quantization.

\subsubsection{Quantum observables in the geometrized setting: quadratic forms}
We must address first the issue of how to cast into this formalism a previously existing algebra. The description of the space of pure states $\mathscr{P}$ that we want to achieve is geometrical in nature then it follows that  a description of the algebra should follow the next three conditions.

\begin{itemize}
    \item The algebra of observables is a subset of the space of $C^{\infty}(\mathscr{P})$.
    \item It contains a dense representation of the Weyl algebra  \eqref{eq:generatorsweyl}. This implies that the product of observables cannot be the pointwise product of functions.
    \item  The Hamiltonian vector field generated from an element of the algebra $F$ given by $X_{F}=\{\cdot,F\}_{\mathscr{P}}$ is such that its flow preserves the Kähler structure. This   means that $\mathcal{L}_{X_F} h_{\mathscr{P}}=0$. This implies that quantum observables generates symmetries that cannot transform the structure $(\mathcal{G}, \Omega, \mathcal{J})_\mathscr{P}$ that defines the quantum phase space. If we want to change this structure with respect to a given evolution it must be introduced as a change in the Kähler manifold from an external source such as an external gravitational field \cite{alonsobujHybridGeometrodynamics2024}.
\end{itemize}

To achieve this,we will be considering the sesquilinear forms defined in holomorphic s.q. coordinates as 
\begin{equation}
\label{eq:function_A}
f_{\hat{G}}(\bar \Psi, \Psi)=\int D\mu_c(\phi) \overline{\Psi (\phi)} \hat G \Psi (\phi),
\end{equation}
for $G\in \mathcal{O}_{cl}$, $\hat G= \mathcal{Q}_{Weyl}(G)$. 
In the following, we will be using a notation similar to the one used above to represent these objects. In formal terms, we will write
\begin{equation}
\label{eq:function_A_coord}
f_{\hat G}(\bar \Psi, \Psi)=  \langle  \Psi, \hat  G \Psi\rangle_{\mathscr{P}}=  \bar\Psi^{\bar\phi}\varDelta_{\bar\phi, \sigma}  \hat G^\sigma_{\gamma} \Psi^\gamma, 
\end{equation}
where $\hat G^\sigma_{\gamma}$ represents the integral kernel representation of the operator $\mathcal{Q}_{Weyl}(G)$. The product of sesquilinear forms follows directly from an extension of the Moyal product  
\begin{equation}
\label{eq:moyalseconsQuantized}
f_{\hat G}\ast_m f_{\hat H}:=f_{\mathcal{Q}_{Weyl}({G\star_m H})}= f_{\hat G \hat H}.
\end{equation}

With these quadratic functions  we can  find a  $C^*$-isomorphism from the algebra  $(W,\star_m,\lVert\cdot\rVert_{Weyl})$, described in \autoref{sec:Moyal}, to the set

\begin{equation}
    \label{eq:algebra_set}
    W_2(\mathscr{P})= \left \{ f_{\hat G}(\bar \Psi, \Psi) \vert \  \hat G\in \hat{W} \right \},
\end{equation}
endowed with the product already described and a norm 
\begin{equation}
\label{eq:nomKibble}
\lVert f_{\hat G} \rVert:= \sup\left\{[{f_{\hat G^\dagger}}\ast_m f_{\hat{G}}(\Psi)]^{\frac12}, \lVert \Psi\rVert_{h_{\mathscr{P}}}=1\right\}=  \lVert G \rVert_{Weyl}.
\end{equation}
Then it is clear that $W_2(\mathscr{P})$ is a representation of Weyl relations. 

The most interesting property of this product is that trivializes the Lie algebra representation of the commutator of quantum operators. It can be readily checked that  
\begin{equation}
\label{eq:lieAlgebraIsomorphism}
\{f_{\hat G},f_{\hat H}\}_{\mathscr{P}}= -i(f_{\hat G}\ast_m f_{\hat H}- f_{\hat H}\ast_m f_{\hat G})= f_{-i[\hat G,\hat H]}.
\end{equation}
Where the l.h.s. of this expression only depends on geometrical properties of the manifold, i.e. on its Kähler structure. In summary, a dense representation of Weyl relations $W_2(\mathscr{P})$ is contained in the algebra of sesquilinear forms endowed with the product $\ast_m$.

 Due to the fact that quadratic functions in \eqref{eq:function_A_coord} do not mix holomorphic an antiholomorphic coordinates this representation also fulfills the last condition. To see this notice that $\mathcal{L}_{X_{f_A}}\Omega_{\mathscr{P}}=0$ because the symplectic structure is always preserved by the Hamiltonian flow of  $f_A$.  We can explicitly write down the   Hamiltonian vector field as using  \eqref{eq:poissonHolo} as 
$X_{f_A}=-i\Big(\Psi^{\sigma}A^{\phi}_{\sigma}\partial_{\Psi^\phi }- \bar{\Psi}^{\bar{\sigma}}(A^t)^{\bar\phi}_{\bar\sigma}\partial_{\bar\Psi^{\bar\phi} }\Big)$. For the complex structure \eqref{eq:holo_complex} we thus obtain 

\begin{multline*}
        \mathcal{L}_{X_{f_A}}\big(\mathcal{J}
        _{\mathscr{P}}\big)^{\sigma}_\phi=
        \big(\mathcal{J}
        _{\mathscr{P}}\big)^\alpha_\phi\partial_{\Psi^\alpha}X_{f_A}^\sigma 
        -
        \partial_{\Psi^\sigma}X_{f_A}^\alpha
        \big(\mathcal{J}
        _{\mathscr{P}}\big)_\alpha^\phi+ 
        \\
        \big(\mathcal{J}
        _{\mathscr{P}}\big)^\alpha_\phi\partial_{\bar\Psi^\alpha}X_{f_A}^\sigma 
        -
    \partial_{\bar\Psi^\sigma}X_{f_A}^\alpha
        \big(\mathcal{J}
        _{\mathscr{P}}\big)_\alpha^\phi
        =0
\end{multline*}

Finally, due to the Kähler compatibility condition $\mathcal{G}_\mathscr{P}=\Omega_\mathscr{P}(\cdot,\mathcal{J}_\mathscr{P}\cdot)$ we get for \eqref{eq:hermitian} $\mathcal{L}_{X_{f_A}}h_\mathscr{P}=0$.

In particular we want the mean value of an observable \eqref{eq:function_A_coord} to be a real quantity. For that matter restrict observables to be associated with self-adjoint operators. To simplify the presentation we skip the discussion on domain problems and we restrict the definition to bounded operators, this is to operators  $A\in W_2(\mathscr{P})\subset \mathcal{O}_{cl}$. Then self-adjoint operators correspond to the  set of quadratic polynomials on $\mathscr{P}$ provided by :
\begin{equation}
\label{eq:functions_set}
\mathcal{F}_2(\mathscr{P})= \left \{ f_{\hat A}(\bar \Psi, \Psi) \vert \  \hat A^{\dagger}=\hat A  \right \}.
\end{equation}
In our notation this is 
$\varDelta_{\bar\phi, \sigma}  \hat A^\sigma_{\rho} \mathcal{K}^{\rho, \bar \gamma} , 
=  \overline{\hat A
_{\phi}^{\gamma} }
$, which is equivalent, at the level of classical functions, to $ A^*=A\in W_2(\mathscr{P})\subset \mathcal{O}_{cl}$. Thus the algebra of quantum observables might be seen as the quantization of real classical functions. In addition, those were the legit observables of a classical theory in the first place.

In summary $\mathcal{F}_2(\mathscr{P})$ is the algebra of observables of the quantum theory. We can endow it with a product $\ast$ that leaves   $\mathcal{F}_2(\mathscr{P})$ invariant, introducing the Jordan product $f_{\hat A}\circ f_{\hat B}=(f_{\hat A}\ast_m f_{\hat B}+f_{\hat B}\ast_m f_{\hat A})/2=f_{(\hat A\hat B+\hat B\hat A)/2}$, we have that 
\begin{equation}
\label{eq:liejordan}
f_{\hat A}\ast_m f_{\hat B}= f_{\hat A}\circ f_{\hat B}+\frac{i}2 \{f_{\hat A},f_{\hat B}\}_{\mathscr{P}}
\end{equation}
is the internal product to the algebra of observables.

\subsubsection{Hermitian operators in real coordinates}

The $\dagger$ operator in different systems of coordinates has different expressions. In general, the Hermitian of an operator is given by the symplectic form, such that $\Omega(\cdot,A^\dagger)=\Omega(A,\cdot)$.   It is for this reason that the simplest expression of the $\dagger$ operation is in Darboux s.q. coordinates. Here we will distinguish Hermitian operators $A^\dagger$ of adjoint operators. We call $A^\dagger$ the adjoint of $A$ only if $[A,\mathcal{J}]=0$ otherwise it will mix the holomorphic and antiholomorphic subspaces not respecting the definition of the whole scalar product. 
Thus in real coordinates

\begin{equation}
    \label{eq:g_h}
    f_{\mathcal{O}}(\Phi,\Pi)= \frac12\Omega_{\mathscr{P}}\left(\left(
        \begin{array}{c}
         \Phi^\phi \\
         \Pi^\phi
        \end{array}
    \right) ,(\mathcal{J}\mathcal{O})_\phi^\sigma\left(
        \begin{array}{c}
         \Phi^\phi \\
         \Pi^\phi
        \end{array}
    \right) \right).
    \end{equation} 
Then \eqref{eq:functions_set} is 
\begin{equation}
    \label{eq:functions_set_darb}
    \mathcal{F}_2(\mathscr{P})= \left \{ f_{\hat A}(\Phi,\Pi) \vert \  \hat A^{\dagger}=\hat A \textrm{ and } [\hat{A},\mathcal{J}]=0 \right \}.
    \end{equation}

Recall that we assume $\hat{A}$ bounded to avoid domain problems in this definition.
    Our goal in this section  it then to describe the $\dagger$ operator.  To distinguish the canonical case of Darboux s.q. coordinates we denote the adjoint operator with respect to the white noise canonical symplectic form  by $\dardag$
\begin{equation}
\label{eq:daggerDarboux}
\mathcal{O}^\dardag= \epsilon^t \mathcal{O}^t \epsilon.
\end{equation}
In this operation the $\epsilon$ matrix of \eqref{eq:KhalerReal} implements the anti-symmetry of the symplectic form.
We can represent, using this definition, the adjoint operator in other systems of coordinates. With a change of variables like \eqref{eq:darbouxJ} we get an expression for the holomorphic $\dagger_{H}$ and antiholomorphic $\dagger_{\bar{H}}$ cases 

\begin{equation}
    \label{eq:daggerHol}
    \mathcal{O}^{\dagger_{H}}=\mathcal{K}\mathcal{O}^\dardag{\varDelta}, \ \ \mathcal{O}^{\dagger_{\bar{H}}}= \mathcal{D}^{-1}\mathcal{O}^\dardag\mathcal{D}.
    \end{equation}

where $(\mathcal{D}^{-1})^{\phi\bar\sigma}=\exp (-\bar\sigma^xD_{xy}^{-1}\phi^y)$. 
In particular, it is interesting to notice that the Fourier transform is not internal to any of the two systems of coordinates and therefore we obtain $\tilde{\mathcal{F}}^{\dagger}= \mathcal{K}\tilde{\mathcal{F}}^\dardag\mathcal{D}.$
and it is immediate to see that 
\begin{equation}
    \label{eq:fourierMirror}
    \tilde{\mathcal{F}}^{\dagger}=\tilde{\mathcal{F}}^{-1}
\end{equation}

\subsubsection{The algebra of observables of Fock quantization in the geometrized setting}
\label{sssec:Fock_in_coords}

Our aim now is to describe a quantization mapping $\mathcal{Q}$ in this setting. In Fock quantization, once we choose the Hilbert space of pure states that we can describe as the holomorphic representation in coordinates \eqref{eq:FockCoordinates}, the representation of Weyl  relations stems from the definition of creation and annihilation operators fulfilling the canonical commutation relations. Thus we will limit our discussion here to the linear case.  It turns out that creation and annihilation operators are represented by the Skorokhod integral and its adjoint \eqref{eq:quantizationholomorphic}, based on the Malliavin derivative. In particular, for a generic system of coordinates  we can define  a pair of adjoint operators $\mathfrak{D}^{\dagger,x}$ and $\mathfrak{D}^x$ that in holomorphic coordinates are just

\begin{equation}
\label{eq:holcoords}
\tilde{\mathfrak{D}}^x=
\Delta^{xy}\partial_{\phi^y} \textrm{ and } \tilde{\mathfrak{D}}^{\dagger,x}=\phi^x
\end{equation}
they are indeed adjoint and not only Hermitian because the complex structure in those coordinates is equivalent to multiply by the constant matrix $\epsilon$. 

Another desirable property for our quantization mapping is to map the imaginary unity to the complex structure $\mathcal{Q}(i)=\mathcal{J}$. To see this condition in place we must contract the operators  with a direction of $\chi\in T_0\mathcal{M}_A\simeq \mathcal{M}_A$, the tangent space of the manifold on which the Hermitian structure \eqref{eq:Hermitian} is defined. In holomorphic coordinates for $\mathcal{M}_A$, according to this quantization rule, it must be  expressed as  $\tilde\chi_{\mathbf{x}}=\frac{1}{\sqrt2}(\tilde\lambda_{\mathbf{x}}+\mathcal{J}\tilde\eta_{\mathbf{x}})$, therefore we have
\begin{equation}
    [\mathfrak{D}(\bar\chi_1),\mathfrak{D}^{\dagger}(\chi_2)]=\frac{1}{2}\left[\mu(\bar\chi_1,\chi_2)-\mathcal{J}\omega(\bar\chi_1,\chi_2)\right]
\end{equation}
where the bilinears are the ones of the Hermitian structure \eqref{eq:Hermitian}. Then we can treat them as the creation and annihilation operators even though they act on the antiholomorphic part as well.

It is also interesting to express the direction $\chi$ in canonical coordinates $(\lambda_x,\eta_x)$ related with the holomorphic coordinates by the change of variable \eqref{eq:change_of_vars}.
Lets particularize our example to the case of the holomorphic representation. Operators $\tilde{\mathfrak{D}},\tilde{\mathfrak{D}}^{\dagger_H}$ contracted with a direction $(\lambda_x,\eta_x)$ (and its conjugate respectively) are given by 
\begin{align}
\tilde{\mathfrak{D}}(\bar\chi)=\tilde{\mathfrak{D}}(\lambda_x,\eta_x)=&
\frac{\lambda_x}{\sqrt2}\Delta^{xy}\partial_{\phi^y}-
\frac{\eta_x}{\sqrt2}\left( \begin{array}{cc}
    A^t & -\delta \\
    \delta & A^t
\end{array} \right)^{xy}\partial_{\phi^y}\nonumber \\
\tilde{\mathfrak{D}}^{\dagger_H}(\chi)=\tilde{\mathfrak{D}}(\lambda_x,\eta_x)^{\dagger_H}=&
\frac{\lambda_x{\phi^x}}{\sqrt2}-
\frac{(\eta K)_x}{\sqrt2}\left( \begin{array}{cc}
    A & \delta \\
    -\delta & A
\end{array} \right)_y^{x}{\phi^y}
\label{eq:holMalliavin}
\end{align}

Looking at  \eqref{eq:changeOfCoordinates} we can see that the first coordinate selects directions of the $\varphi^\mathbf{x}$ canonical coordinate while the  second selects the direction $\pi^\mathbf{x}$. Thus, in this system of coordinates we get  
\begin{equation}
\label{eq:position_momenta_in_sq_coords}
    \lambda_x\hat{\varphi}^x={\tilde{\mathfrak{D}}(\lambda_x,0)+\tilde{\mathfrak{D}}(\lambda_x,0)^{\dagger_H}}, \ \ \eta_x\hat{\pi}^x= {\tilde{\mathfrak{D}}(0,\eta_x)+\tilde{\mathfrak{D}}(0,\eta_x)^{\dagger_H}}.
\end{equation}
In this way we get $[\lambda_x\hat{\varphi}^x,\eta_x\hat{\pi}^x]=\epsilon \lambda_x\delta^{xy}\eta_y$ which are the canonical commutation relations.

It is interesting to compute the Fourier transform for these operators. Let $\hat{\mathfrak{D}}(\lambda_x,\eta_x)=\tilde{\mathcal{F}}\tilde{\mathfrak{D}}(\lambda_x,\eta_x)\tilde{\mathcal{F}}^{-1}$ and  $\hat{\mathfrak{D}}(\lambda_x,\eta_x)^{\dagger_{\bar{H}}}=\tilde{\mathcal{F}}\tilde{\mathfrak{D}}(\lambda_x,\eta_x)^{\dagger_H}\tilde{\mathcal{F}}^{-1}$, then their explicit expression is 
\begin{align}
\hat{\mathfrak{D}}(\lambda_x,\eta_x)&= -\frac{\eta_xD^{xy}}{\sqrt2}\partial_{\phi^y}-
\frac{\lambda_x}{\sqrt2}\left( \begin{array}{cc}
    A & \delta \\
    -\delta & A
\end{array} \right)^{xy}\partial_{\phi^y},\nonumber \\
\hat{\mathfrak{D}}(\lambda_x,\eta_x)^{\dagger_{\bar{H}}}&= \frac{\eta_x{\phi^x}}{\sqrt2}+ 
\frac{(\lambda D^{-1})_x}{\sqrt2}\left( \begin{array}{cc}
    A^t & -\delta \\
    \delta & A^t
\end{array} \right)^{x}_y{\phi^y}.
\label{eq:FourierTransformedSkorokhodMalliavin}
\end{align}

This is, if we switch from the holomorphic to the antiholomorphic picture with the Fourier transform we see how the roles of $\lambda$ and $\eta$ exchange. The duality is  also adapted to the covariance of each picture as described in Section \ref{sec:momentumQuant}. This is a mere reflection of the preservation of quantization by $\tilde{\mathcal{F}}$ showed in the diagram of   \autoref{fig:quantpreserving}.

To complete the quantization procedure the only remaining ingredient is the choice of ordering. The discussion is completely equivalent to that of \autoref{sec:QuantizHolom} and therefore we will skip it in this section. Once this is considered, the algebra of quadratic observables \eqref{eq:algebra_set} is just given by \eqref{eq:g_h}.

\section{Time Evolution: The (modified) Schrödinger equation}
\label{sec:timeevolution}

At this stage we are interested  in postulating a dynamical evolution for the system. In regular quantum mechanics this evolution is just the Schrödinger equation, but in dynamical spacetimes the situation is more complicated. Our task at this point is to derive a dynamical equation from first principles, consistent  with our geometrical viewpoint, and constrained to reproduce the Schrödinger equation in the situations in which there is no ambiguity in its derivation.
More precisely, we must match the unmodified Schrödinger equation in stationary spacetime.

 Our starting point is the \textit{a priori} assumption  that we can tell kinematics from dynamics. This is just the requirement of a consistent definition of the manifold of pure states $\mathscr{P}$, its Kähler structure and its algebra of observables in a foliation $\Sigma_t$ independent way.  This distinction is crucial when we want to describe the system in a dynamical spacetime with backreaction of the quantum degrees of freedom on the classical spacetime \cite{alonsoGeometricFlavours2023}. 

 In this section our point of view is that of a parametric theory, in that case the parameter $t$ is treated as a label. This label describes a foliation of spacetime into leaves generated by a different the embedding of the Cauchy hypersurface $\Sigma$ at each value of $t$. This is expressed as a parametric dependence of the objects defining the geometric structure at each leaf, the Riemannian metric $h^{ij}(t)$, its associated momenta $\pi_{ij}(t)$ and lapse and shift functions $N(t), N^i(t)$. In our case the Kähler structure depend on those objects and therefore it will acquire a parametric dependence  $(\Omega(t),\mathcal{G}(t),\mathcal{J}(t))_{\mathscr{P}}$. The particular dependence is irrelevant for our discussion in this section, we will show a particular example in \autoref{sec:KleinGordonExample}. However, results obtained  in this section are somehow limited. To understand the source and nature of the dependence on the  time parameter we should go beyond the parametric theory and study the structure of admissible $h^{ij}$ and $\pi_{ij}$ providing the Hamiltonian structure of General Relativity. This study is the subject of  \cite{alonsobujHybridGeometrodynamics2024} and will not be covered here.

\subsection{Covariance of the time derivative}
\label{ssec:CovariantDerivative}
 
The first aspect to take into account is that, because the full Kähler structure  $(\Omega(t),\mathcal{G}(t),\mathcal{J}(t))_{\mathscr{P}}$ depends on time, as well as the quantization procedure itself $\mathcal{Q}(t)$,  we can not assume explicit time independence for every set of s.q. coordinates and as such we must define a covariant derivative for the time parameter $\nabla_t$ that preserves the structure and behaves well under changes of s.q. coordinates.

Recall that we know from geometric quantization that the points of the manifold of pure states $\Psi$ represent different sections $\Psi_s$ of an Hermitian line bundle $\pi_{\mathcal{M}_C,B}:B\rightarrow \mathcal{M}_C$, associated with a $U(1)$-principal bundle on the manifold of classical  fields $\phi^{\mathbf{x}}\in\mathcal{M}_C$.  

    In this scenario the manifold of classical fields must be enlarged to accommodate time in the base of the bundle. In concordance with the picture in which  $\mathcal{M}_C$ accounts for the Cauchy data of a theory in a globally hyperbolic spacetime, the bundle incorporating time in this theory is $\pi_{\mathcal{M}_C,B}:B_t\rightarrow \mathcal{M}_C\times \mathbb{R}$. The covariant derivative in the tangent directions of $\mathcal{M}_C$ has been discussed in regards to geometric quantization. In this section we must study the covariant derivative in the tangent directions to the time parameter that we denote $\nabla_t$.

    Recall also that again we know from geometric quantization that we can describe general sections as functions, This is because we factor out an special section $\Psi_r$ such that $\Psi_s=\Psi_r\Psi$ and the nontrivial structure of the bundle is rephrased in terms of the quantization procedure $\mathcal{Q}$. 
    In that  construction we use the twofold nature of line bundles. On one hand they are vector bundles and, as such they are represented as vectors but, on the other, they are also principal bundles using the multiplication of the fiber\footnote{Excluding the zero section, irrelevant in our construction.} $\mathbb{C}\backslash \{0\}$. 

When we treat the line bundle as a real manifold the twofold nature of the line bundle is lost. In that context we substitute the fiber of the complex line bundle $B_t$ by a two dimensional real space keeping only the vector bundle structure. We denote this bundle $\pi_{\mathcal{M}_C,\tilde B_t}:\tilde B_t\rightarrow \mathcal{M}_C\times  \mathbb{R}$.  This has implications in the decomposition of the section $\Psi_s=\Psi_r\Psi$. We choose to describe functions  $\Psi$  with  vectors $\left(
    \begin{array}{c}
     \Phi^\phi \\
     \Pi^\phi
    \end{array}
\right)$.  As is implied by \eqref{eq:alphacondition}, and discussed further in \autoref{sec:holom}, we must introduce a vacuum  section $\Psi_0$, that is regarded as the multiplication by a phase with respect to the reference section $\Psi_0$. This  should be recovered from a principal bundle structure.  Then $\Psi_0$ is part of the  associated $O(2)$-principal bundle.

The quantization procedures introduced in the previous sections are crafted to be blind to this subtlety and treat the points of the manifold of pure states as simple functions. This is nonetheless different in the case of the time derivative.  The quantization procedure depends itself on time thorough the dependence of the Kähler structure. 
We can study its interplay with the covariant time derivative for linear operators using the prequantization procedure \eqref{eq:actionPrecuantumOperator}.  Consider thus the connection defined on the product $\mathcal{M}_C\times \mathbb{R}$ by addition of the pullbacks of the connection one-forms with respect to the canonical projections. We will use $\nabla$ to represent the covariant derivative with respect to this new connection. By introducing the curvature tensor $R(\partial_t,X)=\nabla_t\nabla_X-\nabla_X\nabla_t$  we get that 

\begin{equation}
\label{eq:falireInComm}
\left[\nabla_t,\mathcal{Q}(F)\right]=\partial_t{F}-\mathcal{J} R(\partial_t,X_F)
\end{equation}

In order to get a quantization that assigns the same operators independently of the time parameter that we choose it would be desirable to get 
\begin{equation}
\label{eq:condition}
\left[\nabla_t,\mathcal{Q}(F)\right]=\mathcal{Q}(\partial_tF)
\end{equation}
but this condition does not hold in general. This condition is further explored in \cite{alonsobujHybridGeometrodynamics2024}  where we provide an example of  connection  satisfying this condition under mean values. In that work we also obtain the connection from a different bundle structure.   As a consistency criterion, we ask for the vacuum section to be parallel transported by this connection

\begin{equation}
\label{eq:nablaovervacuum}
\nabla_t \tilde{\Psi}_0= \partial_t\tilde{\Psi}_0+[\Gamma,\tilde{\Psi}_0]=0
\end{equation}
Sections $\Psi_s$ without dynamical evolution, that we will discuss later on, must be also parallely transported and therefore 
 \begin{equation}
 \label{eq:covtime}
 \nabla_t  \left(
    \begin{array}{c}
     \Phi^\phi \\
     \Pi^\phi
    \end{array}
\right)=\frac{\partial}{\partial t}\left(
    \begin{array}{c}
     \Phi^\phi \\
     \Pi^\phi
    \end{array}
\right)+\Gamma_\sigma^\phi 
\left(
    \begin{array}{c}
     \Phi^\sigma \\
     \Pi^\sigma
    \end{array}
\right)
=0 
 \end{equation}

\subsubsection{The connection in holomorphic coordinates}
\label{sssec:Holcon}

In realified holomorphic coordinates  $\left(
    \begin{array}{c}
     \tilde\Phi^\phi \\
     \tilde\Pi^\phi
    \end{array}
   \right)$ the complex structure is constant and as such the regular time derivative preserves it. In order to preserve also the symplectic structure, and therefore the whole Kähler structure is enough to  respect the adjoint operator.
   
    Nonetheless, this condition does not select a unique covariant derivative. As we mentioned before, we would like to meet the requirement \eqref{eq:condition} but this connection is cumbersome to deal with as we show in \cite{alonsobujHybridGeometrodynamics2024}. To provide a simple example, we pick the connection that preserves the Fourier transform $\tilde{\mathcal{F}}$. This, as we explained in Section \ref{sssec:FourierTransforms}, is motivated by the fact that the choice of the holomorphic and antiholomorphic representation is, in last instance arbitrary.  All things considered, let $\mathcal{O}^\phi_\sigma$ be an operator and $\hat{\mathcal{O}}=\mathcal{F}\mathcal{O}\mathcal{F}^{-1}$,  thus we postulate that the covariant derivative acts over this operator in a way such that, in these holomorphic coordinates, it is written as
    
   \begin{equation}
   \label{eq:covderHolHol}
   \tilde\nabla_t{\mathcal{O}}= \frac14\left[\frac{\partial {\mathcal{O}}}{\partial t}+\left(\frac{\partial {\mathcal{O}}^{\dagger_H}}{\partial t}\right)^{\dagger_H}\right]+
   \frac14\tilde{\mathcal{F}}^{-1}\left[\frac{\partial \hat{\mathcal{O}}}{\partial t}+
   \left(\frac{\partial {\hat{\mathcal{O}}}^{\dagger_{\bar{H}}}}{\partial t}\right)^{\dagger_{\bar{H}}}\right]\tilde{\mathcal{F}}
   \end{equation}
   We can also define $\hat\nabla_t$ by exchanging $\tilde{\mathcal{F}} \leftrightarrow \tilde{\mathcal{F}}^{-1}$ and $\dagger_H\leftrightarrow \dagger_{\bar{H}}$. Thus it follows  \begin{equation}
        (\tilde\nabla_t{\mathcal{O}})^{\dagger_H}=\tilde\nabla_t({\mathcal{O}}^{\dagger_H})\textrm{ and }\hat\nabla_t{\hat{\mathcal{O}}}
        = \tilde{\mathcal{F}}\tilde\nabla_t{\mathcal{O}}\tilde{\mathcal{F}}^{-1}.
    \end{equation}
    
    We can write this expression with a connection such that  $\tilde\nabla_t{\mathcal{O}}=\partial_t\mathcal{O}+[\tilde\Gamma,\mathcal{O}]$.  As we show in \ref{app:Connection}, we can write in these coordinates the connection 
    
    \begin{equation}
        \label{eq:ConnectionTerm}
        \tilde\Gamma^\phi_\sigma=\frac14\left[{\mathcal{K}}^{\phi\bar{\gamma}}\dot{\varDelta}_{\bar\gamma\sigma}+\big(\tilde{\mathcal{F}}^{-1}\dot{\tilde{\mathcal{F}}}\big)^\phi_\sigma+(\mathcal{D}\tilde{\mathcal{F}})^{-1}_{\sigma\bar\gamma}\frac{d(\mathcal{D}{\tilde{\mathcal{F}}})^{\bar{\gamma}\phi}}{dt}\right],
    \end{equation}
    where  $\cdot$ represents  the time derivative. In particular if the complex structure $J_{\mathcal{M}_F}$ of \eqref{eq:complexStructure} is time independent this connection is null. 
 We derive the explicit expression and compute the derivative of creation and annihilation operators in the \ref{app:Connection}. We also show that an  \textit{easier-to-handle} expression for this connection is 
 \begin{equation}
\tilde\Gamma=\tilde\nabla_t (\mathfrak{D}^{\dagger x})K_{xy}\mathfrak{D}^{y}=-\mathfrak{D}^{\dagger,x}\tilde\nabla_t (K_{xy}\mathfrak{D}^{y}).
 \end{equation}

To show that this object indeed behaves as a connection consider a  change of s.q. coordinates $\Psi^\phi \rightarrow \mathcal{C}^\phi_\sigma\Psi^\sigma$ such as \eqref{eq:darbouxJ}. Then using \eqref{eq:covderHolHol} the covariant derivative gives  $\mathcal{C}\tilde\nabla_t\big({\mathcal{O}}\big)\mathcal{C}^{-1}=\tilde\nabla_t\big(\mathcal{C}{\mathcal{O}}\mathcal{C}^{-1}\big)+\big[\mathcal{C}\dot{\mathcal{C}}^{-1}, \mathcal{C}{\mathcal{O}}\mathcal{C}^{-1}\big]$. Thus it fulfills the  inhomogeneous transformation law of a connection.  
    $$\Gamma\rightarrow \mathcal{C}\Gamma\mathcal{C}^{-1}+\mathcal{C}\frac{d \mathcal{C}^{-1}}{dt}$$

Notice however that this connection is of the form $\Gamma=a^{\dagger,x}\mathcal{K}_{xy}a^y$ for some kernel $\mathcal{K}$. This is, it does not possess quadratic terms with repeated creation or annihilation operators. For this type of connections it is impossible to obtain mixing between creation and annihilation operators with the parallel transport equation. In other terms, this prescription of the connection does not 
allow for Bogoliubov transformations to be generated by parallel transport as is the case in \cite{kozhikkalBogoliubovTransformation2023}.   For instance particle production in an expanding universe is not recovered with this choice of connection. Ultimately this is a consequence of the violation of condition \eqref{eq:condition}. In \cite{alonsobujHybridGeometrodynamics2024}  we show that a connection rooted in this condition indeed leads to a kind of connection that allow  Bogoliubov transformations to occur. However, the treatment of that connection is cumbersome and only fulfil \eqref{eq:condition} under mean values.  The example showed in this work is just a simple model that provides an interesting case of study with the formalism developed so far.

\subsection{Dynamics: The Modified Schrödinger equation }

In this section we will modify the evolution, i.e.  the Schrödinger equation, to respect the  Kähler structure at the \textit{second quantized}level  $(\mathcal{G}, \Omega, \mathcal{J})_\mathscr{P}$. In particular we want it to preserve the symplectic structure.   
For that matter we opt for a Hamiltonian flow generated by an Hermitian operator $\hat{H}$ to postulate evolution. Then, for any function $F(t,\Psi,\bar\Psi)$, its evolution is generated by some $f_{\hat H}$
\begin{equation}
\label{eq:schrodinger}
\frac{d}{dt}F=\partial_tF+\{F,f_{\hat H}\}_{\mathscr{P}}
\end{equation}
where $\hat H$ is a self adjoint operator. 
The form of $f_{\hat H}$ is \eqref{eq:g_h}. With the Poisson Bracket $\{\cdot,\cdot\}_{\mathscr{P}}$,  we can readily write down the modified Schrödinger in any set of real coordinates 
\begin{equation}
\label{eq:modifSchrodinger}
\nabla_t\left(
    \begin{array}{c}
     \Phi^\phi \\
     \Pi^{\phi} 
    \end{array}
\right)= -\mathcal{J}\hat H_\sigma^\phi \left(
    \begin{array}{c}
     \Phi^\sigma \\
     \Pi^\sigma
    \end{array}
\right)
.
\end{equation}

In static and stationary spacetimes we may find a time independent expression of complex structure $J_{\mathcal{M}_F}$ of \eqref{eq:complexStructure}. In this limiting case  is  the  connection term is null. In this case $\nabla_t$ is simply $\partial_t$ and this equation becomes the regular Schrödinger equation. 

It is clear that this equation does not have a solution on $(\mathcal{N})\times (\mathcal{N})$. In Quantum theory it is desirable to solve this equation over a Hilbert space. Unfortunately for Quantum Field theory this is not possible either, to solve this equation we must look for solutions on $(\mathcal{N})'\times (\mathcal{N})'$. 
This is our way out to deal with different instantaneous Hilbert spaces described in \cite{agulloUnitarityUltraviolet2015,alonsobujHybridGeometrodynamics2024}. A similar equation is postulated in \cite{kozhikkalBogoliubovTransformation2023} to deal with this problem.

\section{Application: Klein Gordon  theory on curved spacetimes}
\label{sec:KleinGordonExample}

We have seen above how the choice of a complex structure on the phase space of classical fields is a crucial ingredient of the quantization procedure. Nonetheless, from a physical point of view, given a relativistic classical field theory, it is an open question how to identify the correct complex structure to define the quantization procedure. There exists a successful answer for stationary space times \cite{ashtekarQuantumFields1975,corichiSchrodingerFock2004,muchComplexStructures2021}, but the solution for arbitrary models is, to the best of our knowledge, still missing. In this section, we review this treatment form the geometrical perspective developed above, in the simple case of a Klein-Gordon model.

\subsection{The classical theory and the choice of the complex structure on the classical phase space}

To start with,  we consider a $d+1$ globally hyperbolic spacetime $(\mathcal{M},\mathbf{g}, \nabla_g)$
 endowed with a pseudo-Riemannian structure $\mathbf{g}$ of Lorentzian 
signature, with sign convention  $(-,+,\cdots,+)$, and the Levi-Civita connection $\nabla_g$. $\mathcal{M}$ is diffeomorphic to $\mathbb{R}\times\Sigma$ where $\Sigma$ is a $d-$dimensional compact manifold diffeomorphic
to every space-like Cauchy hypersurface of $\mathcal{M}$. A one parameter family of embeddings   $\Sigma\to \mathcal{M}$ is determined by a scalar function $\hat t:\mathcal{M}\to \mathbb{R}$ such that $\Sigma_t=\{\hat t(x)=t \textrm{ with }x\in \mathcal{M}\}$ i.e. $\Sigma_t$ are the level sets of $\hat{t}$. The structure of the whole spacetime project into $(\Sigma_t,\mathbf{h}_t,\mathbf{D}:t)$
where $\mathbf{h}$ is a Riemannian structure and $\mathbf{D}$  the corresponding Levi-Cività connection $\mathbf{D}\mathbf{h}=0$. 

We can also define the dynamics directly on $(\Sigma,\mathbf{h},\mathbf{D})$. To this end we take a coordinate system $\mathbf{x}=(x^i)$ $i=1,\cdots,d$ over $\Sigma$ that induces a coordinate system over $\mathcal{M}$ {as} of $(t,\mathbf{x})$. In this coordinate system we obtain the relation
\begin{equation}
\label{eq:shiftlapse}
\partial_t= N\mathbf{n}+N^i\frac{\partial}{\partial{x^i}},
\end{equation}
where $N \mathbf{n}=\vec\nabla\hat{t}$ and $N=[-\mathbf{g}(\vec\nabla\hat{t},\vec\nabla\hat{t})]^{-\frac12}$ is the lapse function while $\vec{N}=N^i\frac{\partial}{\partial{x^i}}$ is the shift vector. Thus, if we provide as data $(N,\vec{N})$ over $\Sigma$, the embedding is fully determined.

In the Klein-Gordon classical theory the dynamics of a scalar field can be defined as a parametric curve $\mathbb{R}\to \mathcal{N}$, where $\mathcal{N}$ is a nuclear space of test functions. We denote this solution by $\varphi_{\mathbf{x}}(t)$. Associated to the space of solutions, we build a Gel'fand triple with respect to the Hilbert space $L^2(\Sigma, d\mathrm{Vol})$, where $d\mathrm{Vol}$ represents the measure associated with the 3-metric tensor $h$. 

With these ingredients we can introduce a Lagrangian formalism, as it is usually done in QFT. From the usual Lagrangian density in four dimensions, we can write the following bilinear form:

\begin{multline}
    \mathcal{L}(t)=\int_{\Sigma}d\mathrm{Vol}\frac{1}{2}\Bigl[\frac{(\dot{\varphi})^2}{N} -
  \frac{N^i}{N}2\dot{\varphi}D_i\varphi+
  \\ 
  \Bigl(\frac{N^iN^j}{N}-Nh^{ij}\Bigr)(D_i\varphi) (D_j\varphi)
  -m^2(\varphi)^2\Bigr]
\end{multline}
Where $N,N^i$ and $h^{ij}$ are functions on $\Sigma$.

In order to define the canonical momentum we must choose between different options. One possible path is to define $\pi^{\mathbf{x}}$ as a density and therefore to build it as an element of  $\mathcal{N}'$, which is the choice made in \cite{corichiSchrodingerFock2004}. Another option is to consider $\varphi_{\mathbf{x}}$ as an element of a Rigged Hilbert space and profit from this structure to let $\pi_{\mathbf{x}}$ be an element of the same space as the field $\varphi_{\mathbf{x}}$. This option is the one we choose to describe Fock quantization in \autoref{ssec:FockSpace} thus, our model manifold is $\mathcal{M}_A$ introduced in that section.   With respect to the Gel'fand triple and the injection $\mathcal{N}\subset L^2(\Sigma,dVol)$  we obtain as canonical momentum 
\begin{equation}
\label{eq:momentum}
\pi_{\mathbf{x}} = \delta_{{\mathbf{x}}y}\frac{\partial{ \mathcal{L}}}{\partial \dot{\varphi}_y} =
  \frac{1}{N}\Big[\dot{\varphi}_x-N^iD_i\varphi_x\Big]
\end{equation}

The Hamiltonian is defined using the same rigging structure  $\mathcal{H}= \delta^{xy}\pi_x\dot\varphi_y-\mathcal{L}$ and can be written as 
\begin{equation}
\label{eq:HmiltonianKG}
H=\int_\Sigma d^dx\sqrt{h}[N \mathcal{H}+N^i\mathcal{H}_i]
\end{equation}
Where $\mathcal{H}$ is the so called superHamiltonian and  $\mathcal{H}_i$ the supermomenta given by
\begin{align}
\label{eq:superham}
\mathcal{H}&=\frac12[\pi^2+h^{ij}D_i\varphi D_j\varphi+m^2\varphi^2]\\
\mathcal{H}_i&= \pi D_i\varphi
\end{align}

As we saw in \eqref{eq:symplecticalgebraic}, with this definition we can endow $\mathcal{M}_A$ with a Poisson bracket, which in our notation reads
\begin{equation}
\label{eq:classicpoisson}
\{P,Q\}_{\mathcal{M}_A}=\delta_{xy}\Big(\frac{\partial P}{\partial\varphi_x}\frac{\partial Q}{\partial\pi_y}-\frac{\partial P}{\partial\pi_x}\frac{\partial Q}{\partial\varphi_y}\Big).
\end{equation}

Notice that the definition of the Hilbert space, and hence of the rigging, depends on the complex structure and thus on time. Therefore, if
we write Liouville equation with the Hamiltonian and the Poisson bracket,
it is also necessary to introduce a covariant derivative for time already at the classical level. The usual prescription for classical theory is to consider $(\varphi_{\mathbf{x}},\pi^{\mathbf{x}})$ as time independent. In such a case,  we get a correction term for $\pi_x$  and the evolution of the classical theory is written in our coordinates as:

\begin{equation}
\label{eq:evoutionclas}
\left(\frac{d}{dt}+ \Gamma_c\right) 
\left(
    \begin{array}{c}
        \varphi_{\mathbf{x}} \\
        \pi_{\mathbf{x}}
    \end{array}
    \right)
    =\Big\{
        \left(
    \begin{array}{c}
        \varphi_{\mathbf{x}} \\
        \pi_{\mathbf{x}}
    \end{array}
    \right), H
    \Big\}
    =
    F
    \left(
    \begin{array}{c}
        \varphi_{\mathbf{x}} \\
        \pi_{\mathbf{x}}
    \end{array}
    \right)
\end{equation}
 with $$\Gamma_c=\left(
        \begin{array}{cc}
         0 & 0 \\
         0 & \dot{\delta}_{xz} \delta^{zy}
        \end{array}
\right) \textrm{ and }
    F=
    \left(
        \begin{array}{cc}
         N^iD_i & N \\
         -\Theta & \Lambda
        \end{array}
    \right)$$ where 
 $\dot{\delta}_{xz} \delta^{zy}\varphi_y= -\frac{\dot{h}}{2h}\varphi(x)$,  $-\Theta=ND^iD_i+(D^iN)D_i-Nm^2$ and $\Lambda=N^iD_i+D_iN^i$.  This implies that $\Theta^{x}_{y}$ is a symmetric operator.



We can connect this classical dynamics with the quantum one considering one particle states, which are those that can be written in holomorphic second-quantized coordinates for the holomorphic quantization as $\Psi^\phi_{1PS}=\tilde\Phi_x\phi^x+i\tilde\Pi_x\phi^x$. The main problem with this representation is that it relies on the rigging provided by $L^2_{Hol}(\mathcal{N}'_{\mathbb{C}},D\mu_c)$  that, in turn, depends on the complex structure \eqref{eq:complexStructure}  which is, for now, unknown.  Notice that, as the complex structure is built on the phase space of fields defined on the submanifold $\Sigma$, which in general evolves in time, it  will be, again in general, time dependent. In \cite{ashtekarQuantumFields1975} it is proved that, for static and stationary spacetimes, there exists a unique complex structure $J_{\mathcal{M}_A}$, introduced in \eqref{eq:complexStructurealg}, once the submanifold $\Sigma$, the 3-metric $h_{jk}$ (and hence the covariant derivative $D$) and operators $N_0$ and $\vec N$ have been fixed.   The expression of such a complex structure in our notation reads 
\begin{equation}
\label{eq:complex_KG}
J_{\mathcal{M}_A}=\lvert F \rvert^{-1}F,
\end{equation}
for $F$ the classical linear evolution defined by Equation \eqref{eq:evoutionclas}.

For simplicity we will take this prescription as well  in the general case.  Nonetheless, this prescription requires further investigation \cite{muchComplexStructures2021}.
We can write down an explicit expression for $J_{\mathcal{M}_A}$ using the prescription \eqref{eq:complex_KG}  studying first two limiting cases where $J_{\mathcal{M}_A}$ can be written down straightforwardly.

\paragraph*{Null Shift}
In this case we set $N^i=0$ and we get 

\begin{equation}
\label{eq:jotanullshif}
J_0=\left(
    \begin{array}{cc}
     0 & \Theta^{-\frac12}N^{\frac12} \\
     -N^{-\frac12}\Theta^{\frac12} & 0
    \end{array}
\right)
\end{equation}

\paragraph*{Huge shift}
In this case we consider $N^iN_i\gg N^2$ and therefore we set $N\simeq 0$. This is, of course, a limiting approximation because the lapse function can never vanish. Then we get

\begin{equation}
    \label{eq:jotahugeshif}
    J_{\infty}=\left(
        \begin{array}{cc}
         (-\alpha^2)^{-\frac12} N^iD_i  & 0  \\
         0 & (-\Lambda^2)^{-\frac12}\Lambda
        \end{array}
    \right)
    \end{equation}
with  $\alpha=N^iD_i$.

\paragraph*{Interpolating cases}

The other cases can be obtained as 

\begin{equation}
\label{eq:generalJ}
J_{\mathcal{M}_A}=A_0J_0+A_\infty J_\infty
\end{equation}
Where $A_0=\lvert F\rvert^{-1}\lvert F_0\rvert$ 
and $A_\infty=\lvert F\rvert^{-1}\lvert F_\infty\rvert$ are operators  being

\begin{gather}
\label{eq:fexpressions}
\lvert F_0\rvert= \left(
    \begin{array}{cc}
     (-N\Theta)^{\frac12}  & 0  \\
     0 & (-\Theta N)^{\frac12}
    \end{array}
\right),\hspace{ 1 em}
\lvert F_\infty\rvert= \left(
    \begin{array}{cc}
     (-\alpha^2)^{\frac12}  & 0  \\
     0 & (-\Lambda^2)^{\frac12}
    \end{array}
\right), \\
\lvert F\rvert^2=  \lvert F_0\rvert^2+ \lvert F_\infty\rvert^2- \big(\lvert F_0\rvert J_0J_\infty \lvert F_\infty\rvert+\lvert F_\infty\rvert J_\infty J_0 \lvert F_0\rvert\big)
\end{gather}

In general this expression is difficult to work out explicitly because of the need to compute $\lvert F\rvert^{-1}$. The derivation of the general complex structure can be found in \cite{corichiSchrodingerFock2004} with the corresponding changes due to $\pi$ being considered a density instead of a scalar.

Remember that what we refer to as the time-dependence of these structures reflects a dependence on  the geometric Cauchy data ($h_{ij},\pi^{ij}$) and on the lapse and shift functions ($N,N^i$). This dependence is necessary in order to preserve the physical interpretation of our QFT throughout the foliation. 

Indeed, for a given background spacetime the geometric content on each hypersurface is the one given by the choice of   the foliation and thus, we have a description of the spacetime in terms of a curve $\gamma:\mathbb{R}\rightarrow\mathcal{M}_G\times\mathcal{M}_N$ such that $\gamma(t)=(h_{ij}(t),\pi^{ij}(t),N(t),N^i(t))$ is the geometrical content at the hypersurface $\Sigma_t$, which is considered known for $ t\geq 0$. Thus, the dependence on these geometric variables of the Kähler structure described above manifest itself as an additional dependence on the Hamiltonian time. 

 On the other hand, in a theory  where, instead of being a background, the metric evolves coupled to the quantum matter and therefore exhibits  backreaction from the matter evolution, this geometric curve is unknown beforehand and one must make the K\"ahler structure depend on the  gravitational degrees of freedom, instead of time. This is the subject of study of  \cite{alonsobujHybridGeometrodynamics2024}.

\subsection{The quantum  model}
Having chosen a complex structure on $\mathcal{M}_A$, we can define $\mathcal{M}_C$ and proceed with the quantization of the theory.

 Lets first discuss the limiting case  $N\simeq 0$. This case may be physically interpreted as the appearance of a horizon in our spacetime. This limiting case breaks the formalism developed above in the following sense. The  
 covariance of the Gaussian measure defined in \eqref{eq:C} is   $\Delta^{\mathbf{xy}}=0$. This means that the measure represents a Dirac delta with infinite dimensional domain   around $\phi^{\mathbf{x}}=0$. This fact might actually be reconciled with a physical interpretation. However, the momentum operator of the Schrödinger picture \eqref{eq:quantizationOfSchOperators} needs the definition of the inverse $K_{\mathbf{xy}}$. In this case the momentum operator diverges with unclear consequences. This, in turn, may be interpreted as the splitting of the  underlining Cauchy hypersurface $\Sigma$ in two regions, inside an outside the horizon. In each of these regions the formalism exposed so far can be developed without further complications. However, once the spacetime develops one of these horizons the purity of the quantum state in each of these regions may be compromised. We may, for example, trace out the degrees of freedom  corresponding to wavefunctions completely supported inside the horizon. See e.g. \cite{waldQuantumField1994} for this kind of partial trace in spacetimes with bifurcating Killing Horizons.    Our goal in this section is to study the limit $N^i=0$.  

\subsubsection{Obtaining the quantum evolution for flat FLRW  spacetimes. }

Consider a situation in which  the shift functions are null $N^i=0$. In that situation it follows that

\begin{equation}
\label{eq:energyinverse}    
\varphi_x\Delta^{xy}\xi_y=\int_\Sigma d^dx\sqrt{h} \varphi(x)\Theta^{-\frac12}N^{\frac12}\xi(x)
\end{equation}
Also, the Hamiltonian in holomorphic coordinates is written $
{H}=\bar\phi^x\Theta_{xy}\phi^y$ and the quantum Hamiltonian acting over this space is obtained with Wick quantization as  
\begin{equation}
\label{eq:quantumH}
\hat{H}=  \Theta_{xy}(a^{\dagger})^xa^y
\end{equation}
In holomorphic s.q. coordinates this expression is just  $ \phi^y(\sqrt{\Theta N})_y^x\partial_{\phi^x}$ where $N^x_y\varphi_x= N \varphi_y$

The diagonal elements of the complex structure \eqref{eq:complexStructurealg} are  $A=0$, then $D^{-1}=-\Delta$ and the Fourier transform \eqref{eq:FourierKernel} is particularly simple to compute. In this scenario, we derive in \ref{app:Connection}  that, denoting $(\dot{\delta^\circ}\delta_\circ)^x_z=\dot{\delta}^{xz}\delta_{zy}$, the connection \eqref{eq:ConnectionTerm} reduces to  
\begin{equation*}
    \tilde{\Gamma}=\frac12\phi^xK_{xy}\dot{\Delta}^{yz}\partial_{\phi^z}-\frac14\phi^xK_{xy} (\dot{\delta^\circ}\delta_\circ)^y_u{\Delta}^{uz}\partial_{\phi^z}+
    \frac14\phi^x (\dot{\delta^\circ}\delta_\circ)^z_x\partial_{\phi^z}.
\end{equation*}

 It is important to notice that \eqref{eq:condition} does not hold with this prescription of complex structure. 
In the holomorphic setting the Schrödinger equation is particularly simple 
 
\begin{multline}
\label{eq:SchrodingerEq}
i \left[\frac{\partial}{\partial t} +\frac12\phi^y K_{yz}\dot{\Delta}^{zx} \partial_{\phi^x}-\frac14\phi^xK_{xy} (\dot{\delta^\circ}\delta_\circ)^y_u{\Delta}^{uz}\partial_{\phi^z}+
    \frac14\phi^x (\dot{\delta^\circ}\delta_\circ)^z_x\partial_{\phi^z} \right] \Psi=\\    \phi^y(\sqrt{\Theta N})_y^x\partial_{\phi^x}\Psi
\end{multline}
There are new terms in the equation that involves the time derivative $\dot{\Delta}$ or $\dot{\delta}_{\mathbf{xy}}$. The left hand side  term is a  non self adjoint correction to the time derivative needed in the evolution to preserve the probabilistic nature of quantum mechanics. In the absence of this term the time derivative breaks the Kähler structure $(\mathcal{G}, \Omega, \mathcal{J})_\mathscr{P}$ and norm conservation is not guaranteed. This is an already known fact for the Schrödinger equation in curved spacetimes \cite{hofmannClassicalQuantum2015,hofmannNonGaussianGroundstate2017,hofmannQuantumComplete2019,eglseerQuantumPopulations2021}. 

Let us particularize our example further for the case of a Klein-Gordon Theory on a  flat Friedman-Lemaitre Robertson-Walker (FLRW) spacetime see e.g. \cite{corichiSchrodingerFock2004}. The Riemannian metric in this case is 
\begin{equation}
    g=-dt \otimes dt + a^2(t) (dx\otimes dx+dy\otimes dy+dz\otimes dz).
\end{equation}
Thus the induced metric in coordinates is $h_{ij}=a^2(t)\delta_{ij}$. In this case the lapse $N=1$ while the shift $N^i=0$ is null.  Also $\sqrt{h}= a^3$ and 

\begin{equation*}
    \xi_x\delta^{xy}\chi_y= \iint_{\mathbb{R}^3\times \mathbb{R}^3} dx^3dy^3\  a^3(t)\xi(x) \delta^3(x-y)\chi(y)
\end{equation*}
Consider the  classical Klein-Gordon  Hamiltonian \eqref{eq:HmiltonianKG} in this background 
\begin{equation}
    {H}_{KG}= \int_{\mathbb{R}^3} \frac{a^3(t)}{2}\Big(\pi^2(x)-\frac{\varphi(x)\nabla^2\varphi(x)}{a^2(t)}+ m^2\varphi(x)^2\Big)dx^3.
\end{equation}
Defining $M(t)= a(t)m$, the parameters of the quantum theory are
\begin{align}
    \Theta^{x}_{y}= \frac{\delta^x_y}{a^2}(-\nabla^2+ M^2),& \hspace{ 1cm}
    \Delta^{x}_y= {a}{\delta^x_y}\Big({\sqrt{-\nabla^2+M^2}}\Big)^{-1}, \nonumber\\
        \hat H= \frac{1}{a}\phi^x&\left(\sqrt{-\nabla^2+M^2}\right)_{x}^y\partial_{\phi^y}.
\end{align}
With these expressions we obtain a modified  Schrödinger equation given by 
\begin{equation}
\label{eq:schrodinger_flrw}
i\left[\partial_t+ \frac12\frac{\dot{a}}{a}\phi^y \left(1-\frac{M^2}{M^2-\nabla^2}\right)^x_y\partial_{\phi^x}\right]\Psi= \frac{1}{a}\phi^x\left(\sqrt{-\nabla^2+M^2}\right)_{x}^y\partial_{\phi^y}\Psi
\end{equation}
Hence, in  this scenario we obtain a modification of the propagator of the theory proportional to the Hubble parameter $\frac{\dot{a}}{a}$. 

As we discussed at the end of \autoref{sssec:Holcon}, the physical significance of this particular choice of connection is unclear. Moreover, the violation of \eqref{eq:condition} is  hard to reconcile with  a coupling with gravity.
For these reasons we skip this discussion and refer to  \cite{alonsobujHybridGeometrodynamics2024}.  There the authors discussed the physical requirements needed to couple the formalism developed in this work with a dynamical theory of gravity and proposed a different candidate of connection in agreement to those requirements.

\subsubsection{Static spacetimes}
 
In the particular case of a static spacetime there is a time-like Killing vector field and a foliation that, in a local region, has constant lapse  function $N$ and null shift.  Therefore the non self-adjoint term disappears recovering the usual Schrödinger equation. In this particular framework we can solve Schrödinger equation with the integral kernel expression

\begin{equation}
\label{eq:solutionsch}
\Psi(\phi,t)=
\int D\beta(\sigma) \exp[\phi^x(\delta_{xy}-it E )_{xy}\bar\sigma^y]\Psi(\sigma,0)
\end{equation}

Where $E=\Theta^{-\frac12}N^{\frac12}\Theta$. In the particular case Minkowski space-time we can split space and time with a foliation in which $N=1$, in this case $E=\sqrt{-\nabla^2+m^2}$  which is the energy operator. 

\section{Conclusions}
\label{sec:conclusions}
In this paper we have made extensive use of the mathematical tools presented in the first part of this series \cite{alonsoGeometricFlavours2023} to study the mathematical formalism underlying different approaches to quantization of a system describing a scalar field in  quantum field theory. The focus of this work is the geometric interpretation of the tools of Gaussian analysis to describe the Hamiltonian picture of a quantum field theory of a scalar field over the space of Cauchy data. 

Following this idea, we started our analysis with geometric quantization. The first step of this program is prequantization. In this context we used Bochner-Minlos theorem to define the measure of the prequantum Hilbert space and the Malliavin derivative as  the central tool to define a connection on such space. 

The next step is the choice of a polarization. Each choice of polarization amounts to a particular picture of the QFT to be described. We discussed the (anti)holomorphic, Schrödinger and Field-Momentum pictures. In this case the tools of Gaussian analysis suited for the discussion are the Wiener-Ito decomposition theorem that provides the particle interpretation of the theory through the Segal isomorphism with the Bosonic Fock space. But this setting is also suited for the discussion of more general relations among these spaces. 

In particular we studied the conservation of the quantization mappings obtained with the aforementioned geometric quantization procedure. 
We showed how the Segal-Bargmann transform studied in \cite{alonsoGeometricFlavours2023} preserves quantization between the Schrödinger and holomorphic (Field-momentum and Antiholormorphic) prescriptions. We also defined a new notion of Fourier transform among infinite dimensional spaces that preserve quantization between the Schrödinger and momentum-field representations. 

General operators other than constant or linear operators  fail to be consistently quantized by the procedure of geometric quantization. For this reason we had to study problems of ordering in this picture. We found that trigonometric exponentials and coherent states are algebras of functions specially well suited for the discussion of  Weyl and Wick quantization and the study of the star products that arise in these quantization procedures.

In accordance with our geometric approach to this problem, we abandoned the description of the spaces of pure states in terms of  Hilbert spaces and treated them as Kähler manifolds. The space of Hida test functions is specially suited for this matter. This point of view led us to a deeper understanding of the different tools described so far, in particular we described the ingredients of Fock quantization in the holomorphic and antiholomorphic representation and its relation through the Fourier transform. 

From this geometric perspective we have been able to propose a Hamiltonian evolution that stems from the Kähler structure of the problem. In this case we introduce a quantum connection to preserve it. In turn, this modifies the Schrödinger equation in such a way that losses of norm, found in other approaches to quantization, are not allowed in our prescription. 

However, there is an infinite family of connections that achieves this purpose. In our case we found a unique choice based on the indistinguishability of the holomorphic and Antiholomorphic representations. We illustrated the choice of this connection with an example of Klein Gordon theory in a generic curved spacetime and a flat FLRW spacetime. Nonetheless,  we discussed why this choice does not led to physically desirable results. For instance it forbids the appearance of Bogouliuvob transformation in the parallel transport of creation and annihilation operators.  

In conclusion, we introduced from geometrical arguments a modification in the Schödinger equation of the QFT of a scalar field. This modification is provided by a connection term in the time derivative. This connection must be introduced in order to preserve the geometric structure of geometric quantum field theory discussed above.  There is, however, some ambiguity left in the choice of connection with physical implications. This ambiguity is further explored in \cite{alonsobujHybridGeometrodynamics2024} and its phenomenology will be the subject of future investigation. Other subject pending of study is the generalization to other QFTs. In this regard the formalism may be enlarged, studying its connection with stochastic quantization, to include  Fermions \cite{albeverioGrassmannianStochastic2020} and gauge fields \cite{masujimaPathIntegral2008}.

\ 

\section{Declarations}

The authors  declare there is no conflict of interest. 

\section{Acknowledgments}
 The authors would like to thank  Prof. Carlos Escudero Liébana for pointing out to very useful bibliography on the connection of Malliavin calculus with QFT. Furthermore, we would like to thank Prof M. Schneider for stimulating conversations that sparked our curiosity in the phenomenology of quantum completeness in relation with our framework.
 
The authors acknowledge partial financial support of Grant PID2021-
123251NB-I00 funded by MCIN/AEI/10.13039/501100011033 and by the
European Union, and of Grant E48-23R funded by Government of Aragón.
C.B-M and D.M-C acknowledge financial support by Gobierno de Aragón
through the grants defined in ORDEN IIU/1408/2018 and ORDEN
CUS/581/2020 respectively.

\newpage

\appendix

\section{Segal-Bargmann and Fourier transforms}

In this appendix we will prove that the Segal-Bargmann transforms and the Fourier  transform conserve the quantization mappings of \autoref{sssec:FourierTransforms}. We also provide proof for some relations exposed in  \autoref{sssec:SegalBargmann}. We will use techniques developed in \cite{alonsoGeometricFlavours2023}.

\subsection{Schrödinger picture}
\label{sse:Segal-Bargmann-SCh}
\noindent Let $\tilde{\mathcal{B}}(\Psi_S)=\Psi_H$ then we get the relation between both pictures  with 
\begin{align*}
    &\tilde{\mathcal{B}}^{-1}(\Psi_H)(\varphi^{\mathbf{x}})=\int D\mu_S(\pi^{\mathbf{x}})\Psi_{H}(\sqrt{2}\phi^\mathbf{x})=
    \\ 
    &
    \hspace{ 2 em}
    \sum_{n=0}^\infty \int D\mu_S(\pi^{\mathbf{x}}) 2^{\frac{n}2}\psi^{(n)}_{\vec{x}_n}{[(\varphi-
    i\pi)^n]^{\vec{x}_n}}=
    \\ 
    &
    \hspace{ 4 em}
    \sum_{n=0}^\infty  \psi^{(n)}_{\vec{x}_n}{\frac{\partial^{n}}{\partial \xi^{n}_{\vec{x}_n}}}\int D\mu_S(\pi^{\mathbf{x}}) e^{\sqrt{2}\xi_z(\varphi-i\pi)^z}\bigg\lvert_{\xi_{\mathbf{x}}=0}= 
    \\ 
    &
    \hspace{ 6 em}
    \sum_{n=0}^\infty  \psi^{(n)}_{\vec{x}_n}
    {\frac{\partial^n}{\partial \xi^n_{\vec{x}_n}}}\Bigg[   e^{\sqrt{2}{\xi_z\varphi^z} } \int D\mu_S(\pi^{\mathbf{x}}) e^{-i{\sqrt{2}\xi_z\pi^z}}\Bigg]_{\xi_{\mathbf{x}}=0}= 
    \\ 
    &
    \hspace{ 8 em}\sum_{n=0}^\infty \psi^{(n)}_{\vec{x}_n}
    \frac{\partial^n}{\partial {\xi}^n_{\vec{x}_n}}
    e^{\sqrt{2}\xi_x\varphi^x-\frac{\xi_u\Delta^{uv}\xi_v}{2}}\Big\lvert_{\xi_\mathbf{x}=0}=
    \\ 
    &
    \hspace{ 10 em}
    \sum_{n=0}^\infty 2^{\frac{n}2}
    \psi^{(n)}_{\vec{x}_n}
    \frac{\partial^n}
    {\partial \tilde{\xi}^n_{\vec{x}_n}}
    e^{\xi_x\varphi^x-\frac{\tilde{\xi}_u\Delta^{uv}\tilde{\xi}_v}{4}}\Big\lvert_{\tilde{\xi}_\mathbf{x}=0}=
    \\
    &\hspace{ 12 em}
    \sum_{n=0}^\infty2^{\frac{n}2} \psi^{(n)}_{\vec{x}_n}:\varphi^n:\lvert_{\frac{\Delta}2}^{\vec{x}_n}=
     \Psi_{S}(\varphi^{\mathbf{x}}).
\end{align*}

\noindent To compute $\tilde{\mathcal{B}}^{-1}\phi^x\tilde{\mathcal{B}}$ lets compute the transform of $\chi_x\phi^x\Psi_H(\phi^{\mathbf{x}})$. We get 

\begin{align*}
    &\tilde{\mathcal{B}}^{-1}(\chi_x\phi^x\Psi_H)(\varphi^{\mathbf{x}})=\int D\mu_S(\pi^{\mathbf{x}})\chi_y\sqrt{2}\phi^y\Psi_{H}(\sqrt{2}\phi^\mathbf{x})=
    \\ 
    &\hspace{3 em}
    \sum_{n=0}^\infty \sqrt{2}\chi_y\int D\mu_S(\pi^{\mathbf{x}}) 2^{\frac{n}2}(\varphi-
    i\pi)^y\psi^{(n)}_{\vec{x}_n}{[(\varphi-
    i\pi)^n]^{\vec{x}_n}}=
    \\
    &\hspace{6 em}\sum_{n=0}^\infty  \chi_{(y}\psi^{(n)}_{\vec{x}_n)}{\frac{\partial^{n+1}}{\partial \xi^{n+1}_{\vec{x}_n,y}}}\int D\mu_S(\pi^{\mathbf{x}}) e^{\sqrt{2}\xi_z(\varphi-i\pi)^z}\bigg\lvert_{\xi_{\mathbf{x}}=0}= 
    \\
    &
    \hspace{ 9 em}\sum_{n=0}^\infty \sqrt{2}\chi_y\frac{\partial}{\partial \xi_{y}}
    2^{\frac{n}2}
    \psi^{(n)}_{\vec{x}_n}
    \frac{\partial^n}{\partial \xi^n_{\vec{x}_n}}e^{\xi_x\varphi^x-\frac{\xi_u\Delta^{uv}\xi_v}{4}}\Big\lvert_{\xi_\mathbf{x}=0}\\
    & \hspace{12 em}=
     \chi_x\Big(\sqrt{2}\varphi^x-\frac{\Delta^{xy}\partial_{\varphi^y}}{\sqrt2}\Big)\Psi_{S}(\varphi^{\mathbf{x}}).
\end{align*}
Here we used $\frac{\partial}{\partial \xi_{y}}\exp\Big({\xi_x\varphi^x-\frac{\xi_u\Delta^{uv}\xi_v}{4}}\Big)={\varphi^y}-\frac{\Delta^{yx}}{2}\partial_{\varphi^x}$. Then we proved

\begin{equation}
    \label{eq:a0}
    \tilde{\mathcal{B}}^{-1}\phi^{\mathbf{x}}\tilde{\mathcal{B}}= \sqrt{2}\varphi^{\mathbf{x}}-\frac{\Delta^{{\mathbf{x}}y}\partial_y}{\sqrt2}.
\end{equation}

Finally for $\tilde{\mathcal{B}}^{-1}\partial_{\phi^{\mathbf{x}}}\tilde{\mathcal{B}}$ we compute the transform of $\chi_x\Delta^{xy}\partial_{\phi^y}\Psi_H(\phi^{\mathbf{x}})$. Notice that 

\begin{align*}
    \tilde{\mathcal{B}}^{-1}(\chi_x\Delta^{xy}\partial_{\phi^y}\Psi_H)(\varphi^{\mathbf{x}})= &\tilde{\mathcal{B}}^{-1}\Bigg(\sum_{n=0}^\infty \chi_y\Delta^{yz}\partial_{\phi^z}\psi^{(n)}_{\vec{x}_n}{(\phi^n)^{\vec{x}_n}}
    \Bigg)=
    \\ 
    \chi_{y}\Delta^{yz}\sum_{n=0}^\infty \tilde{\mathcal{B}}^{-1}\Bigg(\partial_{\phi^z}\psi^{(n)}_{\vec{x}_n}{(\phi^n)^{\vec{x}_n}}\Bigg)=&\chi_{y}\Delta^{yz}  \sum_{n=0}^\infty
    \tilde{\mathcal{B}}^{-1}\Big(\psi^{(n)}_{z,\vec{x}_{n-1}}{n(\phi^{n-1})^{\vec{x}_{n-1}}}\Big) =
    \\
    \chi_{y}\Delta^{yz}\sum_{n=0}^\infty2^{\frac{n-1}2} n\psi^{(n)}_{z,\vec{x}_n}:\varphi^{n-1}:\lvert_{\frac{\Delta}2}^{\vec{x}_{n-1}}=&
     \frac{\chi_{y}\Delta^{yz}\partial_{\varphi^z}}{\sqrt{2}}\Psi_{S}(\varphi^{\mathbf{x}}).
\end{align*}
Thus we proved 
\begin{equation}
    \label{eq:a+0}
    \tilde{\mathcal{B}}^{-1}\partial_{\phi^{\mathbf{x}}}\tilde{\mathcal{B}}= \frac{\partial_{\varphi^{\mathbf{x}}}}{\sqrt{2}}
\end{equation}
Adding the corresponding phase factor we also derive  \eqref{eq:cration_annTransformed} from these expressions.

Now we will prove that $\tilde{\mathcal{B}}$ preserves the quantization mappings. According to \eqref{eq:changeOfCoordinates} we can write $\varphi^x=\frac{\utilde\phi^{\mathbf{x}}+\bar{\utilde\phi}^{\mathbf{x}}}{\sqrt{2}}$ and $\pi^x=iK^{\mathbf{x}}_y\frac{\utilde\phi^y-\bar{\utilde\phi}^y}{\sqrt{2}}-(KA)^{\mathbf{x}}_y\frac{\utilde\phi^y+\bar{\utilde\phi}^y}{\sqrt{2}}$. Using the definition \eqref{eq:quantizationholomorphic} an the relations proven above we see that 

\begin{align}
    \tilde{\mathcal{B}}^{-1} \mathcal{Q}(\varphi^{\mathbf{x}}) \tilde{\mathcal{B}}&= \varphi^x= \mathcal{Q}_s(\varphi^{\mathbf{x}}), \nonumber \\
    \tilde{\mathcal{B}}^{-1} \mathcal{Q}(\pi^y\delta_{y{\mathbf{x}}}) \tilde{\mathcal{B}} &= -i\partial_{\varphi^{\mathbf{x}}}+i{\varphi^yK_{y{\mathbf{x}}}}-{\varphi^y(K A)_{y{\mathbf{x}}}} = \mathcal{Q}_s(\pi^y\delta_{y{\mathbf{x}}}) \label{eq:quantizationpreservation}
\end{align}

\subsection{Momentum field picture}
\label{sse:Segal-Bargmann-Mom}
Similarly we define an isomorphism 

$$\tilde{\overline{\mathcal{B}}}:L^2(\mathcal{N}',D\nu_M)\to L^2_{\overline{Hol}}(\mathcal{N}_{\mathbb{C}}', D\nu_c).$$
such that if  $\hat\Psi_{\overline{H}}=\tilde{\overline{\mathcal{B}}}(\hat\Psi_{M})$ then  the inverse is provided by \eqref{eq:segalBargmanntransformmomentum} with phase factor $g=0$. In terms of  the chaos decomposition this is

\begin{align*}
    &\tilde{\overline{\mathcal{B}}}^{-1}(\hat\Psi_{\bar{H}})(\pi^{\mathbf{x}})=\int D\nu_M(\varphi^{\mathbf{x}})\Psi_{\bar{H}}(\sqrt{2}\bar\phi^\mathbf{x})=
    \\ 
    &
    \hspace{ 2 em}
    \sum_{n=0}^\infty \int D\nu_M(\varphi^{\mathbf{x}}) 2^{\frac{n}2}{\hat{\psi}}^{(n)}_{\vec{x}_n}{[(\pi+i\varphi
    )^n]^{\vec{x}_n}}=
    \\ 
    &
    \hspace{ 4 em}
    \sum_{n=0}^\infty  {\hat{\psi}}^{(n)}_{\vec{x}_n}{\frac{\partial^{n}}{\partial \xi^{n}_{\vec{x}_n}}}\int D\nu_M(\varphi^{\mathbf{x}}) e^{\sqrt{2}\xi_z(\pi+i\varphi)^z}\bigg\lvert_{\xi_{\mathbf{x}}=0}= 
    \\ 
    &
    \hspace{ 6 em}
    \sum_{n=0}^\infty  {\hat{\psi}}^{(n)}_{\vec{x}_n}
    {\frac{\partial^n}{\partial \xi^n_{\vec{x}_n}}}\Bigg[   e^{\sqrt{2}{\xi_z\pi^z} } \int D\nu_M(\varphi^{\mathbf{x}}) e^{i{\sqrt{2}\xi_z\varphi^z}}\Bigg]_{\xi_{\mathbf{x}}=0}= 
    \\ 
    &
    \hspace{ 8 em}\sum_{n=0}^\infty {\hat{\psi}}^{(n)}_{\vec{x}_n}
    \frac{\partial^n}{\partial {\xi}^n_{\vec{x}_n}}
    e^{\sqrt{2}\xi_x\pi^x+\frac{\xi_uD^{uv}\xi_v}{2}}\Big\lvert_{\xi_\mathbf{x}=0}=
    \\ 
    &
    \hspace{ 10 em}
    \sum_{n=0}^\infty 2^{\frac{n}2}
    {\hat{\psi}}^{(n)}_{\vec{x}_n}
    \frac{\partial^n}
    {\partial \tilde{\xi}^n_{\vec{x}_n}}
    e^{\xi_x\pi^x+\frac{\tilde{\xi}_uD^{uv}\tilde{\xi}_v}{4}}\Big\lvert_{\tilde{\xi}_\mathbf{x}=0}=
    \\
    &\hspace{ 12 em}
    \sum_{n=0}^\infty2^{\frac{n}2} {\hat{\psi}}^{(n)}_{\vec{x}_n}:\pi^n:\lvert_{-\frac{D}2}^{\vec{x}_n}=
     \Psi_{M}(\pi^{\mathbf{x}}).
\end{align*}

To compute $\tilde{\overline{\mathcal{B}}}^{-1}\bar\phi^{\mathbf{x}}\tilde{\overline{\mathcal{B}}}$ lets compute the transform of $\chi_x\bar\phi^x\hat\Psi_H(\bar\phi^{\mathbf{x}})$. We get 

\begin{align*}
    &\tilde{\overline{\mathcal{B}}}^{-1}(\chi_x\bar\phi^x\hat\Psi_H)(\varphi^{\mathbf{x}})=\int D\mu_S(\pi^{\mathbf{x}})\chi_y\sqrt{2}\bar\phi^y\hat\Psi_{H}(i\sqrt{2}\bar\phi^\mathbf{x})=
    \\ 
    &\hspace{3 em}
    \sum_{n=0}^\infty \sqrt{2}\chi_y\int D\nu_M(\varphi^{\mathbf{x}}) 2^{\frac{n}2}(
    \pi+i\varphi)^y{\hat{\psi}}^{(n)}_{\vec{x}_n}{[(\pi+i\varphi)^n]^{\vec{x}_n}}=
    \\
    &\hspace{6 em}\sum_{n=0}^\infty  \chi_{(y}{\hat{\psi}}^{(n)}_{\vec{x}_n)}{\frac{\partial^{n+1}}{\partial \xi^{n+1}_{\vec{x}_n,y}}}\int D\nu_M(\varphi^{\mathbf{x}}) e^{\sqrt{2}\xi_z(\pi+i\varphi)^z}\bigg\lvert_{\xi_{\mathbf{x}}=0}= 
    \\
    &
    \hspace{ 9 em}\sum_{n=0}^\infty \sqrt{2}\chi_y\frac{\partial}{\partial \xi_{y}}
    2^{\frac{n}2}
    {\hat{\psi}}^{(n)}_{\vec{x}_n}
    \frac{\partial^n}{\partial \xi^n_{\vec{x}_n}}e^{\xi_x\pi^x+\frac{\xi_uD^{uv}\xi_v}{4}}\Big\lvert_{\xi_\mathbf{x}=0}\\
    & \hspace{12 em}=
     \chi_x\Big(\sqrt{2}\pi^x+\frac{D^{xy}\partial_{\pi^y}}{\sqrt2}\Big)\hat\Psi_{M}(\pi^{\mathbf{x}}).
\end{align*}
Where we used $\frac{\partial}{\partial \xi_{y}}\exp\Big({\xi_x\pi^x+\frac{\xi_uD^{uv}\xi_v}{4}}\Big)={\pi^y}+\frac{D^{yx}}{2}\partial_{\pi^x}$. Then 

\begin{equation}
    \label{eq:creamom}
    \tilde{\overline{\mathcal{B}}}^{-1}\bar\phi^{\mathbf{x}}\tilde{\overline{\mathcal{B}}}= \sqrt{2}\pi^{\mathbf{x}}+\frac{D^{{\mathbf{x}}y}\partial_{\pi^y}}{\sqrt2}
\end{equation}

Finally for $\tilde{\overline{\mathcal{B}}}^{-1}\partial_{\bar\phi^{\mathbf{x}}}\tilde{\overline{\mathcal{B}}}$ we compute the transform of $-\chi_xD^{xy}\partial_{\bar\phi^y}\hat\Psi_H(\bar\phi^{\mathbf{x}})$.
\begin{align*}
    \tilde{\overline{\mathcal{B}}}^{-1}(-\chi_xD^{xy}\partial_{\bar\phi^y}\hat\Psi_H)(\pi^{\mathbf{x}})= &\tilde{\overline{\mathcal{B}}}^{-1}\Bigg(\sum_{n=0}^\infty -\chi_yD^{yz}\partial_{\bar\phi^z}{\hat{\psi}}^{(n)}_{\vec{x}_n}{(\bar\phi^n)^{\vec{x}_n}}
    \Bigg)=
    \\ 
    -\chi_{y}D^{yz}\sum_{n=0}^\infty \tilde{\overline{\mathcal{B}}}^{-1}\Bigg(\partial_{\bar\phi^z}{\hat{\psi}}^{(n)}_{\vec{x}_n}{(\bar\phi^n)^{\vec{x}_n}}\Bigg)=&-\chi_{y}D^{yz}  \sum_{n=0}^\infty
    \tilde{\overline{\mathcal{B}}}^{-1}\Big({\hat{\psi}}^{(n)}_{z,\vec{x}_{n-1}}{n(\bar\phi^{n-1})^{\vec{x}_{n-1}}}\Big)=
    \\
    -\chi_{y}D^{yz}\sum_{n=0}^\infty2^{\frac{n-1}2} n{\hat{\psi}}^{(n)}_{z,\vec{x}_n}:\pi^{n-1}:\lvert_{-\frac{D}2}^{\vec{x}_{n-1}}=&
     -\frac{\chi_{y}D^{yz}\partial_{\pi^z}}{\sqrt2}\hat\Psi_{M}(\pi^{\mathbf{x}}).
\end{align*}
Thus proving 
\begin{equation}
    \label{eq:anihimon}
    \tilde{\overline{\mathcal{B}}}^{-1}\partial_{\bar\phi^{\mathbf{x}}}\tilde{\overline{\mathcal{B}}}= \frac{1}{\sqrt2}\partial_{\pi^{\mathbf{x}}}
\end{equation}
Adding the phase factor of \eqref{eq:segalBargmanntransformmomentum} we also  prove the relations \eqref{eq:momcreaanseagal}.

Now we will prove that $\tilde{\overline{\mathcal{B}}}$ preserves the quantization mappings. According to \eqref{eq:changeAntihol} we can write $\varphi^{\mathbf{x}}=-i(D^{-1})^{\mathbf{x}}_y\frac{\ubreve\phi^y- \bar{\ubreve\phi}^y}{\sqrt{2}}+(AD^{-1})^{\mathbf{x}}_y\frac{\ubreve\phi^y+\bar{\ubreve\phi}^y}{\sqrt{2}}$ and $\pi^{\mathbf{x}}=\frac{\ubreve{\phi}^{\mathbf{x}}+\bar{\ubreve{\phi}}^{\mathbf{x}}}{\sqrt{2}}$. Using the definition \eqref{eq:antiholquant} an the relations proven above we see that 

\begin{align}
    \tilde{\overline{\mathcal{B}}}^{-1} \overline{\mathcal{Q}}(\varphi^y\delta_{y{\mathbf{x}}}) \tilde{\overline{\mathcal{B}}}&=i\partial_{\pi^{\mathbf{x}}}+i{\pi^yD^{-1}_{y{\mathbf{x}}}}+{\pi^y(AD^{-1})_{y{\mathbf{x}}}}= \overline{\mathcal{Q}}_m(\varphi^y\delta_{y{\mathbf{x}}}), \nonumber \\
    \tilde{\overline{\mathcal{B}}}^{-1} \overline{\mathcal{Q}}(\pi^{\mathbf{x}}) \tilde{\overline{\mathcal{B}}} &= \pi^{\mathbf{x}} = \overline{\mathcal{Q}}_m(\pi^{\mathbf{x}}) \label{eq:quantizationpreservation2}
\end{align}

\subsection{Holomorphic and Antiholomorphic Fourier transform}
\label{ssec:unitaryIsom}
Let $\hat\Psi(\bar\phi^\mathbf{x})=\tilde{\mathcal{F}}[ \Psi(\phi^\mathbf{x})]$ and $\hat\Phi(\bar\phi^\mathbf{x})=\tilde{\mathcal{F}}[ \Phi(\phi^\mathbf{x})]$ we denote the chaos  decomposition of each function 

\begin{align}
\label{eq:aligning}
    \Psi(\phi^{\mathbf{x}})= & \sum_{n=0}^\infty \psi^{(n,0)}_{\vec{x}_n} (\phi^n)^{\vec{x}_n}, &
    \Phi(\phi^{\mathbf{x}})= & \sum_{n=0}^\infty \varphi^{(n,0)}_{\vec{x}_n} (\phi^n)^{\vec{x}_n}, \\
    \hat\Psi(\bar\phi^{\mathbf{x}})= & \sum_{n=0}^\infty \hat\psi^{(0,\bar n)}_{\vec{x}_n} (\bar\phi^n)^{\vec{x}_n}, &
    \hat\Phi(\phi^{\mathbf{x}})= & \sum_{n=0}^\infty \hat\varphi^{(0,\bar n)}_{\vec{x}_n} (\bar\phi^n)^{\vec{x}_n}.
\end{align}
By the definition of $\tilde{\mathcal{F}}$ in \eqref{eq:unitary} in terms of the chaos decomposition we get

\begin{equation}
    \label{eq:deomopositionFourier}
\hat \psi^{(0,\bar n)}_{\vec{x}_n}=  (i)^n\psi^{(n,0)}_{\vec{y}_n}{\big[(D^{-1}-iAD^{-1})^n\Big]^{\vec{y}_n}}_{\vec{x}_n}.
\end{equation}
To prove that  $\tilde{\mathcal{F}}$  is an isometric isomorphism we derive from the relations displayed under \eqref{eq:complexStructure}    that 

\begin{align}
\Delta^{\mathbf{xy}}=& -(D^{-1}+iAD^{-1})^{{\mathbf{x}}}_u D^{uv} (D^{-1}-iAD^{-1})^{{\mathbf{y}}}_v,\nonumber\\
D^{{\mathbf{xy}}}=&-(K+iKA)^{\mathbf{x}}_u \Delta^{uv} (K-iKA^{-1})^{{\mathbf{y}}}_v,\nonumber\\
\delta^{\mathbf{x}}_{\mathbf{y}}=  & -(K-iKA)^{{\mathbf{x}}}_u(D^{-1}+iAD^{-1})^{u}_{\mathbf{y}}, \nonumber\\
\delta^{\mathbf{x}}_{\mathbf{y}}=  & -(D^{-1}-iAD^{-1})^{\mathbf{x}}_u(K+iKA )^{u}_{\mathbf{y}}.
\label{eq:magicrelations}
\end{align}
Recall from the Wiener-Ito decomposition theorem, \cite{alonsoGeometricFlavours2023} with definitions of the measures given by \eqref{eq:C} and \eqref{eq:gaussianmeasuremomentum},  that  

\begin{align}
    \label{eq:complexchaosDecomposition}
    \int D\nu_c(\phi^\mathbf{x}) \overline{\hat\Phi(\bar\phi^\mathbf{x})}\hat\Psi(\bar\phi^\mathbf{x})
    = \nonumber \sum_{n=0}^\infty
     & n!\ (-1)^n
     \overline{\hat\varphi^{(0,\bar{n})}_{\vec{x}_n}}
      \big[D^n\big]^{\vec{x}_n\vec{y}_n} 
      \hat\psi^{(0,\bar{n})}_{\vec{y}_n}
    =\nonumber \\ 
    \sum_{n=0}^\infty
    n!\ (i^2 )^n
    \overline{\varphi^{(n,0)}_{\vec{u}_n}}
    {[(D^{-1}+iAD^{-1})^n\Big]^{\vec{u}_n}}_{\vec{x}_n}
     & \big[D^n\big]^{\vec{x}_n\vec{y}_n} 
     {[(D^{-1}-iAD^{-1})^n\Big]^{\vec{v}_n}}_{\vec{y}_n}
     \psi^{(n,0)}_{\vec{v}_n} = \nonumber\\
     \sum_{n=0}^\infty
    n!\
    \overline{\varphi^{(n,0)}_{\vec{u}_n}}
    \big[\Delta^n\big]^{\vec{u}_n\vec{v}_n} 
     \psi^{(n,0)}_{\vec{v}_n} = & \int D\mu_c(\phi^\mathbf{x}) \overline{\Phi(\phi^\mathbf{x})}\Psi(\phi^\mathbf{x})
    \end{align}
This proves the  isometry. The isomorphism is shown inverting \eqref{eq:deomopositionFourier} with

\begin{equation}
    \label{eq:deomopositionFourierinverse}
(i)^n\hat \psi^{(0,\bar n)}_{\vec{x}_n}{[(K+iKA)^n]^{\vec{x}_n}}_{\vec{{\mathbf{y}}}_n}=  \psi^{(n,0)}_{\vec{{\mathbf{y}}}_n}.
\end{equation}
In turn, using Wiener-Ito decompositions and \eqref{eq:deomopositionFourierinverse}
, \eqref{eq:deomopositionFourier}
 it follows that

 \begin{equation}
 \label{eq:transCreaDestru}
     \tilde{\mathcal{F}}(\phi^{\mathbf{x}}\Psi)= i(D^{-1}-iAD^{-1})^{\mathbf{x}}_y\bar\phi^y\hat\Psi\,  \textrm{ and } \,\tilde{\mathcal{F}}(\partial_{\phi^{\mathbf{x}}}\Psi)= i(K+iKA)^y_{\mathbf{x}}\partial_{\bar\phi^y}\hat\Psi.
 \end{equation}
 proving \eqref{eq:HolAntiholFourier}.
\subsection{Fourier transform $\mathcal{F}$ of linear operators}
\label{ssec:craanhi}

Lets use the short hand notation  
\begin{align}
     a^{\mathbf{x}}_0=&\Delta^{{\mathbf{x}}y}\frac{\partial_{\phi^y}}{\sqrt 2}, &  a^{\dagger {\mathbf{x}}}_0=&\sqrt{2}\varphi^{\mathbf{x}}-\Delta^{{\mathbf{x}}y}\frac{\partial_{\phi^y}}{\sqrt 2}, \nonumber\\
     b^{\mathbf{x}}_0=& - D^{{\mathbf{x}}y}\frac{\partial_{\pi^y}}{\sqrt 2}, & b^{\dagger {\mathbf{x}}}_0=&  \sqrt{2}\pi^{\mathbf{x}}+\frac{D^{{\mathbf{x}}y}\partial_{\pi^y}}{\sqrt2}.
     \label{eq:a0b0}
\end{align}
We are interested in the action of $\mathcal{F}={\tilde{\overline{\mathcal{B}}}}^{-1} \tilde{\mathcal{F}} \tilde{\mathcal{B}}$ on linear operators. 
By virtue of \eqref{eq:a0},\eqref{eq:a+0}, \eqref{eq:creamom}, \eqref{eq:anihimon} and the action of the $\tilde{\mathcal{F}}$ operator \eqref{eq:transCreaDestru} is immediate to see that\footnote{Recalling \eqref{eq:magicrelations} the relations displayed under \eqref{eq:complexStructure}.} we get 

\begin{align}
    \mathcal{F}a^{\mathbf{x}}_0\mathcal{F}^{-1}&= -i(D^{-1}+iAD^{-1})^{{\mathbf{x}}}_y{b^y_0}, \nonumber \\
    \mathcal{F}a^{\dagger {\mathbf{x}}}_0\mathcal{F}^{-1}&= i(D^{-1}-iAD^{-1})^{\mathbf{x}}_y{b^{\dagger y}_0}, \nonumber
    \\
    \mathcal{F}^{-1}b^{\mathbf{x}}_0\mathcal{F} & =  -i(K-iKA)^{{\mathbf{x}}}_y{a^y_0},\nonumber
    \\
    \mathcal{F}^{-1}b^{\dagger {\mathbf{x}}}_0\mathcal{F} & =  i(K+iKA)^{{\mathbf{x}}}_y{a^{\dagger y}_0},
    \end{align}
Notice that  $\varphi^{\mathbf{x}}= \frac{a^{\mathbf{x}}_0+a^{\dagger {\mathbf{x}}}_0}{\sqrt2}$ and   $\pi^{\mathbf{x}}= \frac{b^{\mathbf{x}}_0+b^{\dagger {\mathbf{x}}}_0}{\sqrt2}$ thus we can compute 
\begin{align}
    \mathcal{F}\varphi^{\mathbf{x}}\mathcal{F}^{-1} &= \mathcal{F} \Big(\frac{a^{\mathbf{x}}_0+a^{\dagger {\mathbf{x}} }_0}{\sqrt2}\Big)\mathcal{F}^{-1}=i(D^{-1})^{\mathbf{x}}_{y}\frac{b^{\dagger y}_0-b^y_0}{\sqrt2}+ (AD^{-1})^{\mathbf{x}}_y\frac{b^{\dagger y}_0+b^y_0}{\sqrt2} \\
    \mathcal{F}^{-1}\pi^{\mathbf{x}}\mathcal{F} &=
     \mathcal{F}^{-1} \Big(\frac{b^{\mathbf{x}}_0+b^{\dagger {\mathbf{x}} }_0}{\sqrt2}\Big)\mathcal{F}= iK^{\mathbf{x}}_{y}\frac{a^{\dagger y}_0-a^y_0}{\sqrt2}+ (KA)^{\mathbf{x}}_y\frac{a^{\dagger y}_0+a^y_0}{\sqrt2}
\end{align}
Plugging in the definitions \eqref{eq:a0b0} and we obtain
\begin{align}
    \mathcal{F} \varphi^y\delta_{y{\mathbf{x}}}\mathcal{F}^{-1}&=i\partial_{\pi^{\mathbf{x}}}+i{\pi^yD^{-1}_{y{\mathbf{x}}}}+{\pi^y(AD^{-1})_{y{\mathbf{x}}}} \nonumber \\
    \mathcal{F}^{-1} \pi^{y}\delta_{y{\mathbf{x}}}\mathcal{F}&= -i\partial_{\varphi^{\mathbf{x}}}+i{\varphi^yK_{y{\mathbf{x}}}}-{\varphi^y(K A)_{y{\mathbf{x}}}},
\end{align}
proving \eqref{eq:momentumClassical} and as such the preservation of the quantization mapping.

\section{Explicit expressions for the connection and the quantized operators in s.q. coordinates.}
\label{app:Connection}
In this appendix we will compute explicitly the expressions of the connection in  \eqref{eq:covderHolHol} in holomorphic s.q. coordinates. 
Let $\mathcal{O}^\phi_\sigma$ be an operator then the expression of the covariant derivative is 
   \begin{align}
\tilde\nabla_t{\mathcal{O}}= \frac14\left[\frac{\partial {\mathcal{O}}}{\partial t}+\left(\frac{\partial {\mathcal{O}}^{\dagger_H}}{\partial t}\right)^{\dagger_H}\right]+
   \frac14\tilde{\mathcal{F}}^{-1}\left[\frac{\partial \hat{\mathcal{O}}}{\partial t}+
   \left(\frac{\partial {\hat{\mathcal{O}}}^{\dagger_{\bar{H}}}}{\partial t}\right)^{\dagger_{\bar{H}}}\right]\tilde{\mathcal{F}}
   \end{align}

To compute this term recall the kernels with respect to white noise of the Fourier transform  \eqref{eq:FourierKernel} and its inverse.
 We will start by the linear operators \eqref{eq:holcoords}. For simplicity consider a real direction $\xi_x$ then we see that 
\begin{align}
     \tilde{\mathcal{F}}\xi_x\tilde{\mathfrak{D}}^x\tilde{\mathcal{F}}^{-1}&=
- \xi_x\left( \begin{array}{cc}
    A & \delta \\
    -\delta & A
\end{array} \right)^{xy}\partial_{\phi^y},\\
     \tilde{\mathcal{F}}\ \xi_x\tilde{\mathfrak{D}}^{\dagger_H,x}\tilde{\mathcal{F}}^{-1}&=\xi_x
        \left( \begin{array}{cc}
            AD^{-1} & D^{-1} \\
            -D^{-1} & AD^{-1}
        \end{array} \right)_{y}^x\phi^y.
\end{align}

Let the time derivative be represented by a $\cdot$, consider also $\xi_x$  time independent. Following the discussion after \eqref{eq:quantizationholomorphic} we will consider that the coordinates $(\phi^x,\bar\phi^x)$ are just placeholders for integration and, as such do not depend on the time parameter.

Let us consider the derivative of the operator $O=\xi_x\tilde{\mathfrak{D}}^x$. We can compute every term of the derivative using the rule \eqref{eq:actionFormula} and the definitions of the $\dagger$ operators given by \eqref{eq:daggerHol}. We get  $\dot{O}=\xi_x{\dot{\Delta}^{xy}\partial_{\phi^y}}$, $\dot{O}^{\dagger_H}=0$. For the other piece of the derivative we get
   \begin{align*}
       \tilde{\mathcal{F}}^{-1}\frac{d \tilde{\mathcal{F}}\mathcal{O}\tilde{\mathcal{F}}^{-1}}{d t}\tilde{\mathcal{F}} =& -{\xi_x} \left( \begin{array}{cc}
    \dot{A} & \dot{\delta} \\
    -\dot{\delta} & \dot{A}
\end{array} \right)^{xy} 
\left( \begin{array}{cc}
    {A}D^{-1} & -D^{-1} \\
    D^{-1} & {A}D^{-1}
\end{array} \right)^{z}_y 
\partial_{\phi^z}, \\
\tilde{\mathcal{F}}^{-1}\left(\frac{d {\tilde{\mathcal{F}}\mathcal{O}^{\dagger_{{H}}}\tilde{\mathcal{F}}^{-1}}}{d t}\right)^{\dagger_{\bar{H}}}\tilde{\mathcal{F}} =& -{\xi_x} \left[\frac{d}{dt}
\left( \begin{array}{cc}
    {A}D^{-1} & D^{-1} \\
   -D^{-1} & {A}D^{-1}
\end{array} \right)^{x}_y\right] 
\left( \begin{array}{cc}
    A^t & -\delta\\
    \delta & A^t
\end{array} \right)^{yz} 
\partial_{\phi^z}.
   \end{align*}
The hardest term to compute is the second one. First notice that 

$$\frac{d {\tilde{\mathcal{F}}\mathcal{O}^{\dagger_{{H}}}\tilde{\mathcal{F}}^{-1}}}{d t}= \xi_x
        \frac{d}{dt}\left( \begin{array}{cc}
            AD^{-1} & -D^{-1} \\
            D^{-1} & AD^{-1}
        \end{array} \right)_{y}^x\phi^y $$
And the $\dagger_{\bar{H}}$ operator is given by substituting $\phi^x$ by $-D^{xy}\partial_{\phi^y}$ and transposing the matrix (in this particular case in which $\epsilon$ commutes with the operator). Applying the transformations leads to the desired result.  
   
Once this is computed, simply by using the definitions of \eqref{eq:complexStructure} it follows that 

$$\Delta^{xy}=-\left( \begin{array}{cc}
    {A} & \delta \\
    -\delta & {A}
\end{array} \right)^{xu}D^{-1}_{uv}\left( \begin{array}{cc}
    A^t & -\delta\\
    \delta & A^t
\end{array} \right)^{vy}    $$
Then  
\begin{align}
    \nabla_t{\mathfrak{D}}^x=\frac12{\dot{\Delta}^{xy}\partial_{\phi^y}}&-\frac{1}4 \left( \begin{array}{cc}
    \dot{A} & \dot{\delta} \\
    -\dot{\delta} & \dot{A}
\end{array} \right)^{xy} 
\left( \begin{array}{cc}
    {A}D^{-1} & -D^{-1} \\
    D^{-1} & {A}D^{-1}
\end{array} \right)^{z}_y 
\partial_{\phi^z} 
\nonumber
\\& 
+\frac{1}4  
\left( \begin{array}{cc}
    {A}D^{-1} & -D^{-1} \\
    D^{-1} & {A}D^{-1}
\end{array} \right)^{x}_y 
\left( \begin{array}{cc}
    \dot{A}^t & \dot{\delta} \\
    -\dot{\delta} & \dot{A}^t
\end{array} \right)^{yz}
\partial_{\phi^z}
\label{eq:covdercreation}
\end{align}
For the creation operator we use the fact $(\nabla_tO)^{\dagger_H}=\nabla_t(O^{\dagger_H})$ and simply  replace $\partial_{\phi^x}$ by $K_{xy}\phi^y$ and transpose the matrices.

Now we will express the connection $\Gamma$. In this fashion see that because of \eqref{eq:daggerHol} we can express  

\begin{equation}
    \big(O^{\dagger_H}\big)_{\bar\phi}^{\bar\sigma}=\epsilon^{\textrm{T}} \Delta_{\bar\phi\gamma}(O^{\textrm{T}})^{\gamma}_{\tau}\mathcal{K}^{\tau\bar\sigma}\epsilon,
    \label{eq:transpose}
\end{equation}
where $\textrm{T}$ indicates the transpose as a matrix. We use this definition, different from \eqref{eq:1psWN}, to make sure that it is a time dependent operation.   

\begin{equation}
    \left(\frac{d {\mathcal{O}}^{\dagger_H}}{d t}\right)^{\dagger_H}=\dot{\mathcal{O}}+[{\mathcal{K}}^{\phi\bar{\gamma}}\dot{\varDelta}_{\bar\gamma\sigma},\mathcal{O}] 
\end{equation}
We use here the identity  $\dot{\mathcal{K}}^{\phi\bar{\gamma}}\varDelta_{\bar\gamma\sigma}=-{\mathcal{K}}^{\phi\bar{\gamma}}\dot{\varDelta}_{\bar\gamma\sigma}$ that is a consequence of being one the inverse of the other and thus its contraction is $\delta^\phi_\sigma$ which is time independent. Similarly we get $\tilde{\mathcal{F}}^{-1}\dot{\tilde{\mathcal{F}}}=-\dot{\tilde{\mathcal{F}}}^{-1}{\tilde{\mathcal{F}}} $ and 

\begin{equation}
     \tilde{\mathcal{F}}^{-1}\frac{d \tilde{\mathcal{F}}\mathcal{O}\tilde{\mathcal{F}}^{-1}}{d t}\tilde{\mathcal{F}}=\dot{\mathcal{O}}+[\tilde{\mathcal{F}}^{-1}\dot{\tilde{\mathcal{F}}},\mathcal{O}] 
\end{equation}

The last term is a combination of the two above.

\begin{equation}
    (\tilde{\mathcal{F}}^{-1})\left(\frac{\partial {\tilde{\mathcal{F}}\mathcal{O}^{\dagger_{H}}\tilde{\mathcal{F}}^{-1}}}{\partial t}\right)^{\dagger_{\bar{H}}}\tilde{\mathcal{F}}=\dot{\mathcal{O}}+\Big[(\mathcal{D}\tilde{\mathcal{F}}^{\phi\bar\gamma})^{-1}\frac{d(\mathcal{D}{\tilde{\mathcal{F}}})_{\bar{\gamma}\sigma}}{dt},\mathcal{O}\Big] 
\end{equation}

Then, explicitly we get that in holomorphic second quantized coordinates 
\begin{equation}
    \tilde\Gamma^\phi_\sigma=\frac14\left[{\mathcal{K}}^{\phi\bar{\gamma}}\dot{\varDelta}_{\bar\gamma\sigma}+\big(\tilde{\mathcal{F}}^{-1}\dot{\tilde{\mathcal{F}}}\big)^\phi_\sigma+(\mathcal{D}\tilde{\mathcal{F}}^{\phi\bar\gamma})^{-1}\frac{d(\mathcal{D}{\tilde{\mathcal{F}}})_{\bar{\gamma}\sigma}}{dt}\right]
    \label{eq:gammaDerivation}
\end{equation}

If we particularize to the case in which the diagonal elements of the complex structure are  $A=0$, then $D^{-1}=-\Delta$ and the Fourier transform \eqref{eq:FourierKernel} is particularly simple to compute:
\begin{align*}
        \varDelta^{\phi\bar{\sigma}}&= \exp\left[\phi^x
         \Delta_{xy}\bar\sigma^x
        \right],& 
        {\mathcal{K}}^{\phi\bar{\sigma}}&= \exp\left[\phi^x
        K_{xy}\bar\sigma^x
        \right],
    \\
    \tilde{\mathcal{F}}^{\phi\bar{\sigma}}&= \exp\left[-\phi^x
        \epsilon  \Delta_{xy}\bar\sigma^x
        \right],& (\tilde{\mathcal{F}}^{\phi\bar{\sigma}})^{-1}&= \exp\left[\phi^x
        \epsilon K_{xy}\bar\sigma^x
        \right].
\end{align*}
Therefore we compute 

\begin{align*}
    {\mathcal{K}}^{\phi\bar{\gamma}}\dot{\varDelta}_{\bar\gamma\sigma}&= \delta^\phi_\sigma\phi^xK_{xy}\dot{\Delta}^{yz}\partial_{\sigma^z}, \\
    \tilde{\mathcal{F}}^{-1}\dot{\tilde{\mathcal{F}}}&= \delta^\phi_\sigma\phi^xK_{x}^y\dot{\Delta}_{y}^z\partial_{\sigma^z},
    \\
    (\mathcal{D}\tilde{\mathcal{F}}^{\phi\bar\gamma})^{-1}\frac{d(\mathcal{D}{\tilde{\mathcal{F}}})_{\bar{\gamma}\sigma}}{dt}&=\delta^\phi_\sigma\phi^x\delta_{xy}\dot{\delta}^{yz}\partial_{\sigma^z}.
\end{align*}
and denoting $(\dot{\delta^\circ}\delta_\circ)^x_z=\dot{\delta}^{xz}\delta_{zy}$
\begin{equation*}
    \tilde{\Gamma}=\frac12\phi^xK_{xy}\dot{\Delta}^{yz}\partial_{\phi^z}-\frac14\phi^xK_{xy} (\dot{\delta^\circ}\delta_\circ)^y_u{\Delta}^{uz}\partial_{\phi^z}+
    \frac14\phi^x (\dot{\delta^\circ}\delta_\circ)^z_x\partial_{\phi^z}.
\end{equation*}
With this expression is easy to check that \eqref{eq:covdercreation} holds. In fact we can derive this expression from a different perspective. Notice that  \eqref{eq:gammaDerivation} leads always to a connection $\tilde\Gamma=\phi^x\Gamma_{x}^y\partial_{\phi^y}$. In this way we get 

\begin{equation}
    \tilde\nabla_t (\phi^x)\partial_{\phi^x}=\tilde\nabla_t (\mathfrak{D}^{\dagger x})K_{xy}\mathfrak{D}^{y}=\Gamma
    \label{eq:unGammaMAs}
\end{equation}

Also  as a direct consequence of  $(\tilde\nabla_t{\mathcal{O}})^{\dagger_H}=\tilde\nabla_t({\mathcal{O}}^{\dagger_H})$ we obtain  
\begin{multline}
    (\tilde\nabla_t{\mathcal{O}})^{\dagger_H}=
    \Big(
    \partial_t{\mathcal{O}}+
    \Big[
    \tilde\Gamma
    ,
    \mathcal{O}
    \Big]
    \Big)^{\dagger_H}= 
    \partial_t{\mathcal{O}^{\dagger_H}}-
    \Big[
    \tilde\Gamma^{\dagger_H}-
    {\mathcal{K}}^{\phi\bar{\gamma}}\dot{\varDelta}_{\bar\gamma\sigma}
    ,
    \mathcal{O}^{\dagger_H}
    \Big]
\end{multline}
Then $\tilde\Gamma= -\tilde\Gamma^{\dagger_H}+\phi^xK_{xy}\dot{\Delta}^{yz}\partial_{\phi^z}=-\tilde\Gamma^{\dagger_H}-\mathfrak{D}^{\dagger,x}\dot{K}_{xy}\mathfrak{D}^{y} $. Using this we obtain from \eqref{eq:unGammaMAs} that

\begin{equation}
    \Gamma=-\mathfrak{D}^{\dagger,x}K_{xy}\tilde\nabla_t (\mathfrak{D}^{y})+\mathfrak{D}^{\dagger,x}\dot{K}_{xy}\mathfrak{D}^{y} = -\mathfrak{D}^{\dagger,x}\tilde\nabla_t (K_{xy}\mathfrak{D}^{y}).
\end{equation}

We can check this equation directly from the induced form of the connection $\tilde\Gamma=\phi^x\Gamma_{x}^y\partial_{\phi^y}$.

        \begin{landscape} 
            \begin{center}
            \vspace{3cm}

            $$\int D\beta(\sigma) A(\phi,\bar\sigma)B(\sigma,\bar\gamma):$$

            \begin{table}[h]  
              \centering 
              \renewcommand{\arraystretch}{3}
              \begin{tabular}{|c||c|c|c|c|}\hline
                \diagbox{$A$}{$B$} & $\sqrt{2}\cosh(N)\cos(V)$ &
                 $\sqrt{2}\cosh(N)\sin(V)$ &
                 $\sqrt{2}\sinh(N)\cos(V)$ &
                 $\sqrt{2}\sinh(N)\sin(V)$
                 \\
                \hline \hline

                $\sqrt{2}\cosh(M)\cos(U)$  & $\cosh\boxplus\cos\circleddash+\cosh\boxminus\cos\oplus$ & 0 & 0 & $-\sinh\boxplus\sin\circleddash+\sinh\boxminus\sin\oplus$\\  \hline
                $\sqrt{2}\cosh(M)\sin(U)$  & 0 &$\sinh\boxplus\cos\circleddash-\sinh\boxminus\cos\oplus$ & $\cosh\boxplus\sin\circleddash+\cosh\boxminus\sin\oplus$ &0  \\  \hline
                $\sqrt{2}\sinh(M)\cos(U)$ & 0 & $-\cosh\boxplus\sin\circleddash+\cosh\boxminus\sin\oplus$ & $\sinh\boxplus\cos\circleddash+\sinh\boxminus\cos\oplus$ &0 \\ \hline
                $\sqrt{2}\sinh(M)\sin(U)$  & $\sinh\boxplus\sin\circleddash+\sinh\boxminus\sin\oplus$ & 0 & 0 & $\cosh\boxplus\cos\circleddash-\cosh\boxminus\cos\oplus$ \\
                \hline
              \end{tabular}
              \caption{Result of the integral $\int D\beta(\sigma) A(\phi,\bar\sigma)B(\sigma,\bar\gamma)$ where the kernels $A$ and $B$ are specified in the first column and row respectively}
              \label{tab:multTable}
            \end{table}

            Where  $M=\phi^xM_{xy}\bar\sigma^y, U=\phi^xU_{xy}\bar\sigma^y , N= \sigma^xN_{xy}\bar\gamma^y$ and $V= \sigma^xV_{xy}\bar\gamma^y$.
            
            Also  $ \boxplus=\phi^x(MN+UV)_{xy}\bar\gamma^y,$  $\boxminus=\phi^x(MN-UV)_{xy}\bar\gamma^y,$ $\oplus=\phi^x(UN+MV)_{xy}\bar\gamma^y$  and $\circleddash=\phi^x(UN-MV)_{xy}\bar\gamma^y$ 
        \end{center}
            \end{landscape}


\begin{thebibliography}{10}
                \expandafter\ifx\csname url\endcsname\relax
                  \def\url#1{\texttt{#1}}\fi
                \expandafter\ifx\csname urlprefix\endcsname\relax\def\urlprefix{URL }\fi
                \expandafter\ifx\csname href\endcsname\relax
                  \def\href#1#2{#2} \def\path#1{#1}\fi
                
                \bibitem{alonsoGeometricFlavours2023}
                J.~L. Alonso, C.~{Bouthelier-Madre}, J.~{Clemente-Gallardo}, D.~{Mart{\'i}nez-Crespo}, Geometric flavours of {{Quantum Field}} theory on a {{Cauchy}} hypersurface. {{Part I}}: {{Geometric}} quantization (Jun. 2023).
                \newblock \href {http://arxiv.org/abs/2306.14844} {\path{arXiv:2306.14844}}, \href {https://doi.org/10.48550/arXiv.2306.14844} {\path{doi:10.48550/arXiv.2306.14844}}.
                
                \bibitem{alonsobujHybridGeometrodynamics2024}
                J.~L. Alonso, C.~Bouthelier~Madre, J.~{Clemente-Gallardo}, D.~{Mart{\'i}nez-Crespo}, Hybrid geometrodynamics: {{A Hamiltonian}} description of classical gravity coupled to quantum matter, Classical and Quantum Gravity (2024).
                \newblock \href {https://doi.org/10.1088/1361-6382/ad3459} {\path{doi:10.1088/1361-6382/ad3459}}.
                
                \bibitem{gelfandGeneralizedFunctions1964}
                I.~M. Gel'fand, N.~J. Vilenkin, Generalized {{Functions}}: {{Applications}} of Harmonic Analysis, no. Volume 4 in Generalized Functions / {{I}}. {{M}}. {{Gel}}'fand, {{G}}. {{E}}. {{Shilov}}, AMS Chelsea Publishing, Providence, Rhode Island, 2016.
                
                \bibitem{hidaBrownianMotion1980}
                T.~Hida, Brownian {{Motion}}, Springer US, New York, NY, 1980.
                \newblock \href {https://doi.org/10.1007/978-1-4612-6030-1} {\path{doi:10.1007/978-1-4612-6030-1}}.
                
                \bibitem{hidaWhiteNoise1993}
                T.~Hida, H.-H. Kuo, J.~Potthoff, L.~Streit, White {{Noise}}, Vol.~72, Springer Netherlands, Dordrecht, 1993.
                \newblock \href {https://doi.org/10.1007/978-94-017-3680-0} {\path{doi:10.1007/978-94-017-3680-0}}.
                
                \bibitem{kuoWhiteNoise1996}
                H.-H. Kuo, White Noise Distribution Theory, Probability and Stochastics Series, CRC Press, Boca Raton, Fla., 1996.
                
                \bibitem{obataWhiteNoise1994}
                N.~Obata, White {{Noise Calculus}} and {{Fock Space}}, Vol. 1577 of Lecture {{Notes}} in {{Mathematics}}, Springer Berlin Heidelberg, Berlin, Heidelberg, 1994.
                \newblock \href {https://doi.org/10.1007/BFb0073952} {\path{doi:10.1007/BFb0073952}}.
                
                \bibitem{huAnalysisGaussian2016}
                Y.~Hu, Analysis on {{Gaussian Spaces}}, World Scientific, 2016.
                \newblock \href {https://doi.org/10.1142/10094} {\path{doi:10.1142/10094}}.
                
                \bibitem{kondratievGeneralizedFunctionals1996}
                Y.~G. Kondratiev, P.~Leukert, J.~Potthoff, L.~Streit, W.~Westerkamp, Generalized {{Functionals}} in {{Gaussian Spaces}}: {{The Characterization Theorem Revisited}}, Journal of Functional Analysis 141~(2) (1996) 301--318.
                \newblock \href {http://arxiv.org/abs/math/0303054} {\path{arXiv:math/0303054}}, \href {https://doi.org/10.1006/jfan.1996.0130} {\path{doi:10.1006/jfan.1996.0130}}.
                
                \bibitem{sampedroSpaceInfinite2020}
                J.~C. Sampedro, On the space of infinite dimensional integrable functions, Journal of Mathematical Analysis and Applications 488~(1) (2020) 124043.
                \newblock \href {https://doi.org/10.1016/j.jmaa.2020.124043} {\path{doi:10.1016/j.jmaa.2020.124043}}.
                
                \bibitem{nunnoMalliavinCalculus2009}
                G.~D. Nunno, B.~{\O}ksendal, F.~Proske, Malliavin {{Calculus}} for {{L{\'e}vy Processes}} with {{Applications}} to {{Finance}}, Vol.~44, Springer Berlin Heidelberg, Berlin, Heidelberg, 2009.
                \newblock \href {https://doi.org/10.1007/978-3-540-78572-9} {\path{doi:10.1007/978-3-540-78572-9}}.
                
                \bibitem{glimmQuantumPhysics1987}
                J.~Glimm, A.~Jaffe, Quantum Physics: A Functional Integral Point of View, 2nd Edition, Springer-Verlag, New York, 1987.
                
                \bibitem{cartierFunctionalIntegration2006}
                P.~Cartier, C.~{DeWitt-Morette}, Functional Integration: Action and Symmetries, Cambridge University Press, 2006.
                
                \bibitem{cartierRigorousMathematical1997}
                P.~Cartier, C.~{DeWitt-Morette}, A.~Wurm, D.~Collins, A {{Rigorous Mathematical Foundation}} of {{Functional Integration}}, in: Functional {{Integration}}, Vol. 361 of {{NATO ASI Series}}, Springer US, Boston, MA, 1997, pp. 1--66.
                \newblock \href {https://doi.org/10.1007/978-1-4899-0319-8} {\path{doi:10.1007/978-1-4899-0319-8}}.
                
                \bibitem{westerkampRecentResults2003}
                W.~Westerkamp, Recent {{Results}} in {{Infinite Dimensional Analysis}} and {{Applications}} to {{Feynman Integrals}}, Ph.D. thesis (Feb. 2003).
                \newblock \href {http://arxiv.org/abs/math-ph/0302066} {\path{arXiv:math-ph/0302066}}.
                
                \bibitem{brunettiAdvancesAlgebraic2015}
                R.~Brunetti, C.~Dappiaggi, K.~Fredenhagen, J.~Yngvason, Advances in {{Algebraic Quantum Field Theory}}, Springer International Publishing, Cham, 2015.
                \newblock \href {https://doi.org/10.1007/978-3-319-21353-8} {\path{doi:10.1007/978-3-319-21353-8}}.
                
                \bibitem{ashtekarQuantumFields1975}
                A.~Ashtekar, A.~Magnon, P.~R. S.~L. A, Quantum fields in curved space-times, Proceedings of the Royal Society of London. A. Mathematical and Physical Sciences 346~(1646) (1975) 375--394.
                \newblock \href {https://doi.org/10.1098/rspa.1975.0181} {\path{doi:10.1098/rspa.1975.0181}}.
                
                \bibitem{corichiSchrodingerFock2004}
                A.~Corichi, J.~Cortez, H.~Quevedo, Schr{\"o}dinger and {{Fock}} representation for a field theory on curved spacetime, Annals of Physics 313~(2) (2004) 446--478.
                \newblock \href {https://doi.org/10.1016/j.aop.2004.05.004} {\path{doi:10.1016/j.aop.2004.05.004}}.
                
                \bibitem{muchComplexStructures2021}
                A.~Much, R.~Oeckl, Complex structures for {{Klein}}--{{Gordon}} theory on globally hyperbolic spacetimes, Classical and Quantum Gravity 39~(2) (2021) 025015.
                \newblock \href {http://arxiv.org/abs/1812.00926} {\path{arXiv:1812.00926}}, \href {https://doi.org/10.1088/1361-6382/ac3fbd} {\path{doi:10.1088/1361-6382/ac3fbd}}.
                
                \bibitem{woodhouseGeometricQuantization1997}
                N.~M.~J. Woodhouse, Geometric {{Quantization}}, second edition Edition, Oxford {{Mathematical Monographs}}, Oxford University Press, Oxford, 1997.
                
                \bibitem{hallQuantumTheory2013}
                B.~C. Hall, Quantum Theory for Mathematicians, no. 267 in Graduate Texts in Mathematics, Springer, New York, 2013.
                
                \bibitem{tuynmanMetaplecticCorrection2016}
                G.~M. Tuynman, The metaplectic correction in geometric quantization, Journal of Geometry and Physics 106 (2016) 401--426.
                \newblock \href {https://doi.org/10.1016/j.geomphys.2016.04.007} {\path{doi:10.1016/j.geomphys.2016.04.007}}.
                
                \bibitem{oecklSchrodingerRepresentation2012}
                R.~Oeckl, The {{Schr{\"o}dinger}} representation and its relation to the holomorphic representation in linear and affine field theory, Journal of Mathematical Physics 53~(7) (2012) 072301.
                \newblock \href {http://arxiv.org/abs/1109.5215} {\path{arXiv:1109.5215}}, \href {https://doi.org/10.1063/1.4731770} {\path{doi:10.1063/1.4731770}}.
                
                \bibitem{oecklAffineHolomorphic2012}
                R.~Oeckl, Affine holomorphic quantization, Journal of Geometry and Physics 62~(6) (2012) 1373--1396.
                \newblock \href {http://arxiv.org/abs/1104.5527} {\path{arXiv:1104.5527}}, \href {https://doi.org/10.1016/j.geomphys.2012.02.001} {\path{doi:10.1016/j.geomphys.2012.02.001}}.
                
                \bibitem{oecklHolomorphicQuantization2012}
                R.~Oeckl, Holomorphic quantization of linear field theory in the general boundary formulation, Symmetry, Integrability and Geometry: Methods and Applications (SIGMA) 8 (2012) 31.
                \newblock \href {http://arxiv.org/abs/1009.5615} {\path{arXiv:1009.5615}}, \href {https://doi.org/10.3842/SIGMA.2012.050} {\path{doi:10.3842/SIGMA.2012.050}}.
                
                \bibitem{ditoStarproductApproach1990}
                J.~Dito, Star-product approach to quantum field theory: {{The}} free scalar field, Letters in Mathematical Physics 20~(2) (1990) 125--134.
                \newblock \href {https://doi.org/10.1007/BF00398277} {\path{doi:10.1007/BF00398277}}.
                
                \bibitem{albeverioGrassmannianStochastic2020}
                S.~Albeverio, L.~Borasi, F.~C. De~Vecchi, M.~Gubinelli, Grassmannian stochastic analysis and the stochastic quantization of {{Euclidean Fermions}}, arXiv (2020) 1--84\href {http://arxiv.org/abs/2004.09637} {\path{arXiv:2004.09637}}.
                
                \bibitem{masujimaPathIntegral2008}
                M.~Masujima, Path Integral Quantization and Stochastic Quantization, 2nd Edition, Springer, Berlin, 2008.
                
                \bibitem{kibbleGeometrizationQuantum1979}
                T.~W.~B. Kibble, Geometrization of quantum mechanics, Communications in Mathematical Physics 65~(2) (1979) 189--201.
                \newblock \href {https://doi.org/10.1007/BF01225149} {\path{doi:10.1007/BF01225149}}.
                
                \bibitem{ashtekarGeometricalFormulation1999}
                A.~Ashtekar, T.~A. Schilling, Geometrical {{Formulation}} of {{Quantum Mechanics}}, in: A.~Harvey (Ed.), On {{Einstein}}'s {{Path}}: {{Essays}} in {{Honor}} of {{Engelbert Schucking}}, Springer, New York, NY, 1999, pp. 23--65.
                \newblock \href {https://doi.org/10.1007/978-1-4612-1422-9_3} {\path{doi:10.1007/978-1-4612-1422-9_3}}.
                
                \bibitem{agulloUnitarityUltraviolet2015}
                I.~Agullo, A.~Ashtekar, Unitarity and ultraviolet regularity in cosmology, Physical Review D 91~(12) (2015) 124010.
                \newblock \href {https://doi.org/10.1103/PhysRevD.91.124010} {\path{doi:10.1103/PhysRevD.91.124010}}.
                
                \bibitem{kozhikkalBogoliubovTransformation2023}
                M.~M. Kozhikkal, A.~Mohd, Bogoliubov transformation and {{Schr{\"o}dinger}} representation on curved space, Physical Review D 108~(8) (2023) 085028.
                \newblock \href {https://doi.org/10.1103/PhysRevD.108.085028} {\path{doi:10.1103/PhysRevD.108.085028}}.
                
                \bibitem{hofmannClassicalQuantum2015}
                S.~Hofmann, M.~Schneider, Classical versus quantum completeness, Physical Review D 91~(12) (2015) 125028.
                \newblock \href {https://doi.org/10.1103/PhysRevD.91.125028} {\path{doi:10.1103/PhysRevD.91.125028}}.
                
                \bibitem{hofmannNonGaussianGroundstate2017}
                S.~Hofmann, M.~Schneider, Non-{{Gaussian}} ground-state deformations near a black-hole singularity, Physical Review D 95~(6) (2017) 065033.
                \newblock \href {http://arxiv.org/abs/1611.07981} {\path{arXiv:1611.07981}}, \href {https://doi.org/10.1103/PhysRevD.95.065033} {\path{doi:10.1103/PhysRevD.95.065033}}.
                
                \bibitem{hofmannQuantumComplete2019}
                S.~Hofmann, M.~Schneider, M.~Urban, Quantum complete prelude to inflation, Physical Review D 99~(6) (2019) 065012.
                \newblock \href {https://doi.org/10.1103/PhysRevD.99.065012} {\path{doi:10.1103/PhysRevD.99.065012}}.
                
                \bibitem{eglseerQuantumPopulations2021}
                L.~Eglseer, S.~Hofmann, M.~Schneider, Quantum populations near black-hole singularities, Physical Review D 104~(10) (2021) 105010.
                \newblock \href {https://doi.org/10.1103/PhysRevD.104.105010} {\path{doi:10.1103/PhysRevD.104.105010}}.
                
                \bibitem{krieglConvenientSetting1997}
                A.~Kriegl, P.~Michor, The {{Convenient Setting}} of {{Global Analysis}}, Vol.~53, American Mathematical Society, Providence, Rhode Island, 1997.
                \newblock \href {https://doi.org/10.1090/surv/053} {\path{doi:10.1090/surv/053}}.
                
                \bibitem{dodsonGeometryFrechet2016}
                C.~T.~J. Dodson, G.~Galanis, E.~Vassiliou, Geometry in a {{Fr{\'e}chet Context}}, Cambridge University Press, Cambridge, 2016.
                \newblock \href {https://doi.org/10.1017/CBO9781316556092} {\path{doi:10.1017/CBO9781316556092}}.
                
                \bibitem{longSchrodingerWave1998}
                D.~V. Long, G.~M. Shore, The {{Schr{\"o}dinger}} wave functional and vacuum states in curved spacetime, Nuclear Physics B 530~(1-2) (1998) 247--278.
                \newblock \href {http://arxiv.org/abs/hep-th/9605004} {\path{arXiv:hep-th/9605004}}, \href {https://doi.org/10.1016/S0550-3213(98)00408-8} {\path{doi:10.1016/S0550-3213(98)00408-8}}.
                
                \bibitem{longSchrodingerWave1996}
                D.~V. Long, G.~M. Shore, The {{Schrodinger Wave Functional}} and {{Vacuum State}} in {{Curved Spacetime II}}. {{Boundaries}} and {{Foliations}}, Nuclear Physics B 530~(1-2) (1996) 279--303.
                \newblock \href {http://arxiv.org/abs/gr-qc/9607032} {\path{arXiv:gr-qc/9607032}}, \href {https://doi.org/10.1016/S0550-3213(98)00409-X} {\path{doi:10.1016/S0550-3213(98)00409-X}}.
                
                \bibitem{waldQuantumField1994}
                R.~M. Wald, Quantum {{Field Theory}} in {{Curved Spacetime}} and {{Black Hole Thermodynamics}}, 1st Edition, University of Chicago Press, Chicago, 1994.
                
                \bibitem{henry-labordereAnalysisGeometry2009}
                P.~{Henry-Labord{\`e}re}, Analysis, Geometry, and Modeling in Finance: Advanced Methods in Option Pricing, CRC Press, Boca Raton, 2009.
                
                \bibitem{buchholzLocalityStructure1982}
                D.~Buchholz, K.~Fredenhagen, Locality and the structure of particle states, Communications in Mathematical Physics 84~(1) (1982) 1--54.
                \newblock \href {https://doi.org/10.1007/BF01208370} {\path{doi:10.1007/BF01208370}}.
                
                \bibitem{nairQuantumField2005}
                V.~P. Nair, Quantum {{Field Theory}} a Modern Perspective, Graduate Texts in Contemporary Physics, Springer, New York, NY, 2005.
                
                \end{thebibliography}
\end{document}